\documentclass{aa}

\usepackage[utf8]{inputenc}
\usepackage{dcolumn}
\usepackage{graphicx}
\usepackage{natbib}
\usepackage{enumerate}
\usepackage{aas_macros}
\usepackage{float}
\usepackage{txfonts}
\usepackage{color}
\usepackage{grffile}
\usepackage{array,float}
\allowdisplaybreaks

\newcommand{\vlsr}{V$_{\rm{LSR}}$}
\makeatletter

\newcommand{\kms}{$\rm km\,s^{-1}$}
\newcommand{\Rmnum}[1]{\expandafter\@slowromancap\romannumeral #1@}

\newcolumntype{d}[1]{D{.}{\cdot}{#1}}
\newcolumntype{.}{D{.}{.}{-1}}
\usepackage{aas_macros}

\newcommand{\hi}{H~{\sc i}}
\newcommand{\hii}{H~{\sc ii}}
\newcommand{\uchii}{UC\,H~{\sc ii}}

\newcommand{\sex}{\texttt{SExtractor}}
\newcommand{\Disperse}{\texttt{DisPerSE}}

\newcommand{\mum}{$\mu$m}

\newcommand{\lsun}{L$_\odot$}
\newcommand{\msun}{M$_\odot$}

\begin{document}

   \title{ATLASGAL --- A Galaxy-wide sample of dense filamentary
   structures\thanks{The full version of Tables\,1 and 3 and Fig.\,9 are only
   available in electronic form at the CDS via anonymous ftp to cdsarc.u-strasbg.fr (130.79.125.5) or via http://cdsweb.u-strasbg.fr/cgi-bin/qcat?J/A\&A.
   The full catalogue and the corresponding data files are
available at
http://atlasgal.mpifr-bonn.mpg.de/cgi-bin/ATLASGAL\_FILAMENTS.cgi.}
   }

\authorrunning{Guang-Xing Li et al.}
\titlerunning{A Galaxy-wide sample of dense filamentary structures}

\author {Guang-Xing Li\inst{1,2}\thanks{E-mail:
gxli@usm.lmu.de}, J.\,S.\,Urquhart\inst{1,3}, S.\,Leurini\inst{1},
T.\,Csengeri\inst{1}, F.\,Wyrowski\inst{1},  K.\, M.\, Menten\inst{1} \and
F.\,Schuller\inst{4} }

  \institute {$^1$Max-Planck-Institut f\"ur Radioastronomie, Auf dem H\"ugel 69,
    D-53121 Bonn, Germany \\
  $^2$University Observatory Munich, Scheinerstr. 1, D-81679 M\"unchen, Germany\\
$^3$Centre for Astrophysics and Planetary Science, University of Kent, Canterbury, CT2 7NH\\
	$^4$European Southern Observatory, Alonso de Cordova 3107, Vitacura, Santiago, Chile
}

\date{Received xxx; accepted xxx}


\abstract {Filamentary structures are ubiquitous in the interstellar
medium.
Investigating their connection to the large-scale structure of the Galaxy and
their role in star formation is leading to a paradigm shift in our understanding
of star formation.}
{We study the properties of filamentary structures from the ATLASGAL survey,
which is the largest and most sensitive systematic ground-based survey of the
inner Galactic plane at submillimeter wavelengths.}
  {We use the \Disperse\ algorithm to identify spatially coherent structures
  located across the inner-Galaxy ($300\degr < \ell < 60\degr$ and $|b| < 1.5$).
  As a result we produce a catalogue of $\sim$1800 structures; these were then
  independently classified by the five lead authors into one of the following
  types: marginally resolved, elongated structures, filaments, network of
  filaments and complexes.  This resulted in the identification of 517
  filamentary structures. We determine their physical properties and investigate
  their overall Galactic distribution.
   }
 {We find that almost 70\% of the total 870\,\mum\ flux associated with these
 structures resides in filaments and networks of filaments and we estimate that
 they are likely to be associated with a similar fraction of the mass.
Correlating these structures with tracers of massive star formation we also find
that a similar fraction of the massive star forming clumps are associated with
filaments and networks of filaments, which highlights the importance of these
types of structures to star formation in the Galaxy.  We have determined
distances, masses and physical sizes for 241 of the filamentary structures. We
find a median distance of 3.8 kpc, a mean mass of a few
10$^3$\,\msun, a mean length of $\sim$6\,pc and a mass-to-length ratio of
($M/L$) $\sim$\,200-2000\,\msun\,pc$^{-1}$.
We  also find that these filamentary structures are tightly correlated with the
spiral arms in longitude and velocity, and that their semi-major axis is preferentially aligned
parallel to the Galactic mid-plane and  therefore with the direction of
large-scale Galactic magnetic field. We find many examples where the dense filaments
identified in ATLASGAL are associated with larger scale filamentary structures
($\sim$100\,pc), and argue that this is likely to be common, and as such these
may indicate a connection between large-scale Galactic dynamics and star
formation. }
 {We have produced a large and Galaxy-wide catalogue of dense  filamentary
 structures that are representative of a particular size and mass range not
 previously well studied in the literature. Analyses of the properties and
 distribution of these filaments reveals that they are correlated with the
 spiral arms and make a significant contribution to star formation in the Galaxy.
 Massive star formation is ongoing within $\sim$20\% of the filaments and is
 strongly correlated with the filaments with the largest mass-to-length ratios.
 The luminosity of the embedded sources has a similar distribution to the
 Galactic-wide samples of young massive stars and can therefore be considered to
 be representative.}

   \maketitle

\section{Introduction}

The presence of filamentary structures in interstellar clouds has been revealed
by observations at different wavelengths and in different tracers over the last
few decades \citep[e.g.][]{1979ApJS...41...87S,1987ApJS...63..645U}. However,
the importance of filaments in the process of star formation has only recently
been fully realised thanks to the {\it Herschel} satellite. The high
sensitivity, spatial dynamic range and angular resolution have revealed the
ubiquity of filamentary structures throughout the interstellar medium
\citep[e.g.][]{molinari2010,andre2010,arzoumanian2011}. Such filamentary
structures have been found in both low- and high-mass star-forming molecular
clouds, as well as in quiescent clouds \citep[e.g.][]{miville2010}. In  nearby
molecular clouds, a substantial fraction of pre-stellar cores are located on
filaments, suggesting that the formation of filaments is a step preceding star
formation \citep[see][and references therein]{andre2014}.

The mechanisms leading to their formation and
their link to the star formation process is not well understood. Filaments can
form through compression or shear at a scale that is comparable to the
filament length. This can be achieved with gravitational instabilities,
gravitational  collapse (e.g. \citealt{2004ApJ...616..288B,gomez2014}, Li et
al., 2015), supersonic turbulence
\citep{2001ApJ...553..227P,2008ApJ...674..316H}, and converging flows
\citep{2008ApJ...674..316H,vazquez2011}.
Turbulence can produce filaments through a combination of compression and
shear
\citep{2013A&A...556A.153H,2016MNRAS.457..375F,2015ApJ...807...67M,arzoumanian2011}.
 Their physical characteristics are
constrained by a relatively limited number of examples in the literature;
filamentary structures  have been reported showing a broad range of lengths,
from a few parsecs \citep{schisano2014} to large-scale structures of several
tens to hundreds of parsecs
\citep[e.g.][]{li2013,goodman2014,ragan2014,wang2015} with a variety of aspect
ratios, widths and masses.

To fully understand their origin, subsequent evolution, and thus their role
in star formation, systematic studies of large samples of filamentary structures
are necessary. Unbiased
surveys of the inner Galactic plane, and of entire star-forming regions at submillimetre and far-infrared wavelengths are
providing a rich database for such studies. Such surveys include the Hi-Gal
\citep{molinari2010} and Gould Belt projects, \citep{andre2010} both of which
cover the wavelength regime $70\,\mu$m--$500\,\mu$m with {\it Herschel}, as
well as the APEX Telescope Large Area Survey of the Galaxy (ATLASGAL) survey
 \citep{schuller2009_full}.

\begin{figure*}
\includegraphics[width=0.95 \textwidth]{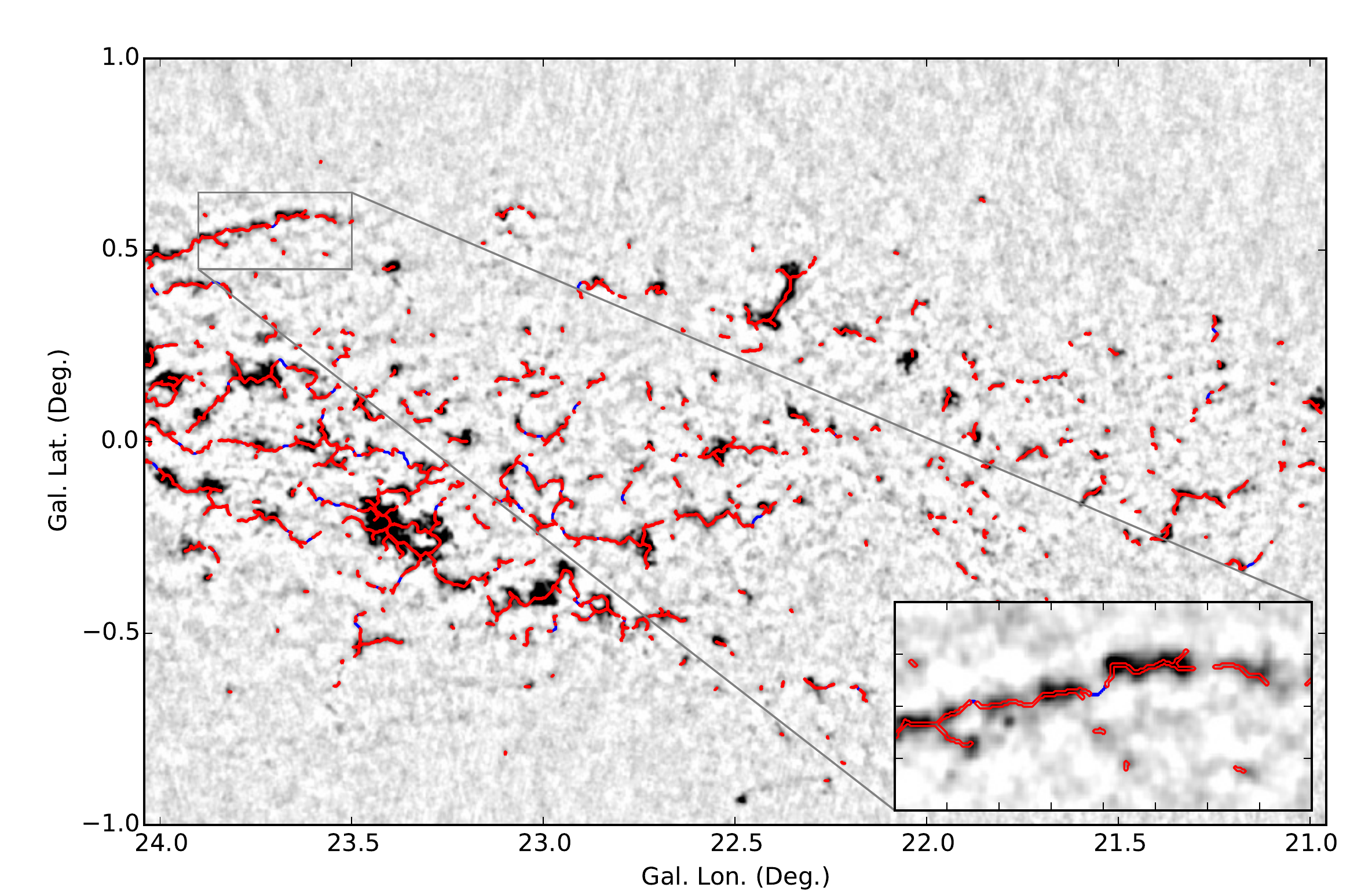}
\caption{Emission map of the $\ell = 22\degr$ field that has been used to optimise
the persistence and  robustness thresholds used by \Disperse. This map is overlaid
with the skeletons obtained using a high robustness threshold that trace the high-column
density regions shown in red (i.e. $p = 1.4$ and $r = 2.5$) and the inter-connecting
skeletons obtained using a lower robust threshold that traces the lower column density
connecting regions are shown in blue (i.e. $p = 1.4$ and $r = 1.95$).}\label{fig:combined}
\end{figure*}


In this paper we present the first systematic search for filamentary
structures across the inner Galactic plane at submillimetre wavelengths based on
the ATLASGAL survey.  ATLASGAL covered 420 square degrees of the inner
Galactic plane at 870\,$\mu$m with an angular resolution of 19.2\arcsec.
Emission from dust is optically thin at this wavelength and therefore ATLASGAL
is an excellent tracer of column density and total mass \citep{schuller2009_full}.
In the ATLASGAL survey data, diffuse emission is
filtered out, leaving only the compact high density structures. The ATLASGAL maps are
not affected by saturation.
These properties make the ATLASGAL emission maps an ideal place to conduct the
first Galactic-wide search for filaments. Furthermore, having a wealth of
ancillary data in hand from complementary ATLASGAL follow-up programmes
(\citet[e.g.][]{wienen2012,wienen2015a,giannetti2014}), we are able to assign
distance estimates to the extracted structures, allowing a direct estimate of their
physical properties.

 The structure of the paper is as follows: in Section\,\ref{sect:method} we
 describe the procedures used to identify coherent structures from the dust
 emission maps and describe how their properties are determined.
 The classification scheme is described in Section\,\ref{sect:classification}
 where we also discuss the differences in the structural and statistical
 properties of the various type of structures identified. We determine
 velocities, distances, masses, lengths and widths for the filamentary
 structures in Section\,\ref{sect:physical_properties}. In
 Section\,\ref{sect:discussion} we investigate their connection to larger scale
 structures such as giant molecular filaments (e.g. \citealt{li2013, ragan2014,
 goodman2014}) and the spiral arms.
 In this section we also investigate their association with infrared dark
 clouds, \hii\ region bubbles and massive star formation in an effort to link
 these structures to other catalogues and evaluate their contributions to star
 formation in the Galaxy. In Section\,\ref{sect:summary} we present a summary of
 this work and highlight our main findings.

\section{Identifying structures }\label{sect:method}
 The ATLASGAL survey \citep{schuller2009_full} covers $300\degr < \ell <
60\degr$ and $|b| < 1.5^{\circ}$. It has a typical noise level of 50-70 mJy
beam$^{-1}$ and is sensitive to gas with $\rm H_2$ column densities exceeding
$\sim 10^{22} \;\rm cm^{-2} $.
For our sample, the beam size of ATLASGAL (which is 19.2 arcsec) corresponds to
$\sim 0.4 \;\rm pc$ at a typical distance of $\sim 4\;\rm kpc$.
Submillimetre emission detected from the ground are dominated by
 fluctuations from the atmosphere. To remove these fluctuations,  median
values of the correlated emission between all bolometers are subtracted in the
data reduction process. A consequence of this is that uniform emission on scales
larger than $\sim 2.5$ arcmin are filtered out. However, structures larger than
this can be recovered provided that the structures are compact enough
\citep{schuller2009_full} and therefore filamentary structures will be preserved
as long as their widths are smaller than the scale above which filtering is
effective.

We aim to study filamentary structures in the ATLASGAL survey. This has
been achieved through two steps. First, we extract skeleton representations of
structures in the dust continuum emission maps with the \Disperse\ (Discrete
Persistent Extractor;
\citealt{sousbie2011}). Then, the extracted structures are
classified, and filamentary structures  are identified. In this section we
describe the extraction of structures. The classification of the extracted
structures into different categories will be
described in the following section.

\subsection{{ Extraction of skeletons}}
\label{sect:source_extraction}

\begin{figure*}
\centering
\includegraphics[width = 0.99\textwidth]{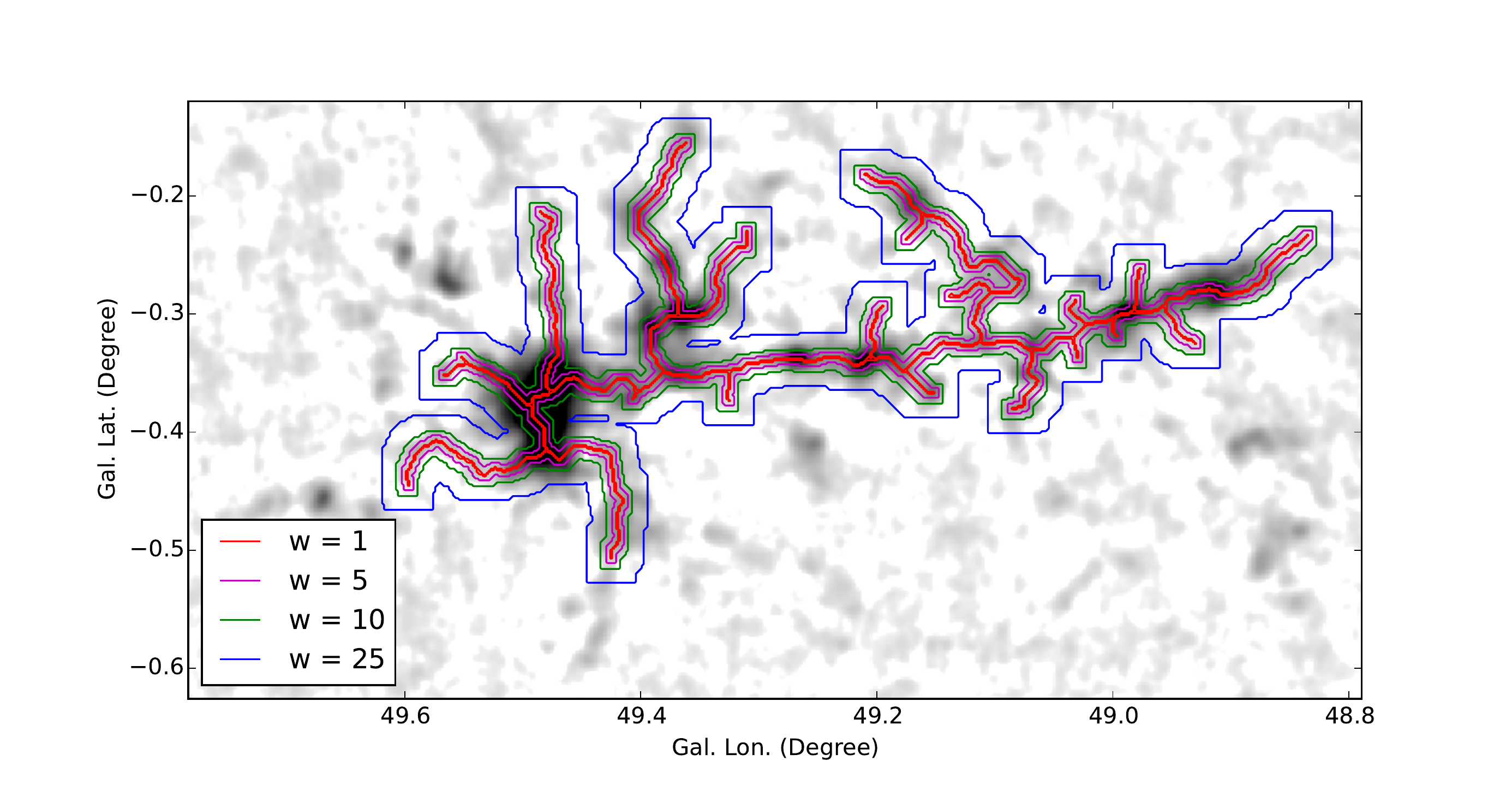}
\caption{Demonstration of the dilation operation applied to the { source
skeleton} identified towards the W51 molecular cloud. The grayscale image shows
the dust continuum emission mapped with ATLASGAL while surrounding contours show
how the size and shape of the { dilated skeleton} changes as it is dilated by
squares of increasing sizes.
The red contour shows a dilation by a square of 1 pixel and simply
traces the original skeleton as this also has a width of 1 pixel.
The pink, green, blue contours show the structure after the original
skeleton has been dilated with a square of width of $w$=5, 10 and 25\,pixels, respectively.}
\label{fig:snake:dilate}
\end{figure*}

We use the publicly-available \Disperse\ (Discrete Persistent
Extractor;
\citealt{sousbie2011})\footnote{\url{http://www2.iap.fr/users/sousbie/web/html/indexd41d.html?}}
source extraction algorithm to identify large spatially coherent structures that
are located within the ATLASGAL survey region. This algorithm is based on
discrete Morse theory and identifies persistent topological features such as
peaks, voids, walls, and in particular, filamentary structures from 2D or 3D
datasets. The resulting skeletons\footnote{A skeleton is a set of connecting points that trace the high-density ``crests'' of the emission map.} extracted by \Disperse\ are therefore
representations of the topological structure of the emission. \Disperse\ has
been successfully used to trace structures in  {\it Herschel} data (e.g.
\citealt{andre2010,hill2011,arzoumanian2011}) as well as on column density maps
produced by simulations (e.g. \citealt{smith2014b}). A
number of other algorithms have been used in recent studies (e.g.
\textit{getfilaments} --- \citealt{2013A&A...560A..63M}, \textit{CuteX} ---
\citealt{molinari2011} and Hessian matrix method --- \citealt{schisano2014}),
however, with the except of \textit{getfilaments}, none are currently publicly
available. We therefore decided to use \Disperse, as it is the most widely used
code for this purpose in the literature and has been shown to produce reliable
results (e.g. \citealt{andre2010,hill2011,arzoumanian2011}).

The two most important parameters of the program are the persistence ($p$) and
robustness ($r$) thresholds. The persistence threshold is a measure of the
absolute difference of the values of a pair of critical points, while the
robustness threshold is a measure of the contrast of a structure with respect to
the local background. In order to determine the optimal values for these two thresholds we tried different combinations on a test field. The region selected for these tests was
$\ell \sim$ 22.5\degr\ as it contains a large filamentary structure as well as a
number of molecular complexes and was therefore considered an ideal region to test the code.

{
We applied the \Disperse\ algorithm to ATLASGAL tiles with a pixel size of
6\arcsec.} Initial tests on the ATLASGAL emission maps revealed two
difficulties.
First, the noise varied over the field and is significantly higher in the upper and
lower thirds of the maps (i.e. $|b| > 1\degr$). \Disperse\ only allows a single
value for the thresholds and so results in large numbers of spurious structures
being detected in these noisy regions. This problem can be significantly reduced
using signal-to-noise ratio maps for the initial identification of structures.
{ This technique has been successfully applied in our previous works}
(e.g. \citealt{csengeri2014}). The second problem is related to the
ground-based nature of the survey; in order to remove the contribution of the
sky the median value of the correlated emission between all bolometers is
removed. Although this is a very effective way of removing the sky noise, it
also filters out coherent emission from large-scale structures. As a result,
filamentary structures might appear as chains of disconnected
fragments in the ATLASGAL emission maps, and \Disperse\ does not identify them
as a single structure.  In order to increase the sensitivity of the survey for
the diffuse emission, we smoothed the maps using a Gaussian kernel with
a FWHM of twice the ATLASGAL beam. This lowers the overall resolution by a
factor 2.2 ($\sim$42\arcsec, which corresponds to a FWHM of $\sim$7 pixels in
the maps) but significantly increases the sensitivity to the lower column
density regions by roughly a factor of 4. 

Our primary goal is to detect and quantify coherent dust condensations in the
ATLASGAL survey, and as a result a good skeleton of the emission has to satisfy
the following requirements: first, it should trace structures that appear to be
coherent, and second, noise should not be represented by the skeleton.
Investigating the performance of \Disperse\ with different combinations of
parameters we found that the skeleton from a single set of values for the
persistence and robustness parameters was not sufficient to obtain a good representation of the
emission seen in the maps. Either the parameters were too constraining and only
the highest column density regions were picked up or they were too
permissive, resulting in large numbers of spurious sources.
We therefore settled for two sets of parameters and
carefully filtered the results  to identify a reliable sample of coherent
structures.

The final skeleton used in the analysis is constructed as follows: first, we
make a catalogue of short skeletons that trace the high-column density dust
condensations using a high robustness parameter. This is used to make a primary
catalogue of reliable fragments. We then try to connect these fragments to
identify larger coherent structures using a lower robustness threshold, which is
more sensitive to the lower column density regions between the dense clumps. The
primary skeleton is produced with $p = 1.4$ and $r = 2.5$, in which only structures
 with significant contrast are included, and the secondary skeleton is
produced with $p = 1.4$ and $r=1.95$. We combined the results of these two
extractions to produce a reliable catalogue for further analysis. In
Fig.\,\ref{fig:combined} we show the result of this two stage procedure for the
test field. In this image the primary skeletons that trace the high-column
density fragments are shown in red while the lower column density
interconnecting fragments are shown in blue. The interconnecting fragments are
only included in the analysis if (a) lengths of the segments are shorter than 4
pixels, or (b) average flux of the pixels on the skeleton is larger than $4$
times the image rms.

We found that even with a relatively conservative threshold the output produced
by \Disperse\ still contained structures that have low signal-to-noise ratios.
Therefore to ensure the reliability of the catalogue produced from our analysis
we removed any structures that are approximately 4.5 times
smaller than the smoothed beam (e.g. $<$30 pixels) in order to focus the
catalogue on the filamentary structures; this has the consequence of excluding shorter and
perhaps more distant structures.

A visual inspection of the maps confirms that the method described above is
able to recover all of the most significant structures that would be
identified by eye and provides confidence that nothing important has been
missed. Although we have developed our method to ensure that all important
structures are recovered,  our catalogue is limited by the
sensitivity and resolution of the observations.

\section{Classification}
\label{sect:classification}

\subsection{Classification scheme}

Following the procedure presented in Section\,\ref{sect:source_extraction} we
have produced a preliminary catalogue of 1812 structures. They cover a range
of sizes, morphologies and complexity from relatively roundish featureless blobs to large star forming complexes. The
emission associated with the more complicated regions is often
blended and the underlying structure is unclear, making interpretation of any
derived properties difficult.

Based on a visual inspection of the observed properties of the catalogue
sources we were able to identify a relatively small number of common types of
objects that can be used to broadly classify the whole sample. Below we provide a list of the
six classes and a brief description of their observed properties:

\begin{enumerate}

\item { Marginally resolved clumps}:  structures that appear as
relatively isolated structures with simple morphologies and small aspect ratios ($<$\,2).
We also use this classification for small groups of isolated clumps that have no obvious
connecting emission.

\item Resolved elongated structures:  structures classified in this
group are dominated by elongated structures with aspect ratios of 2-3 and a
relatively small number of associated spine points ($<$\,100 pixels). These may
simply be unresolved filaments or fragments of larger filaments where the
connecting lower density gas does not have sufficient column density to have
been detected by ATLASGAL. This class also includes unresolved structures
that are extended in more than two directions; these typically consist of one or
two central bright clumps associated with three of four lower density strands
protruding radially away from the centre.

\item Filaments: elongated linear structures that are clearly resolved across
their lengths and widths and consisting of a single dominant filament with relatively few
sub-branches and typical aspect ratios larger than 3.

\item Networks of filaments: several filaments that seem to be connected to each
other. In many cases these networks are coincident with similar networks of
infrared-dark clouds (IRDC) that are seen in absorption at mid-infrared
wavelengths against bright Galactic background emission.

\item Complexes: they are regions of very bright extended and often complex
emission, and although \Disperse\ has connected the emission together, there is
significant blending of the emission features. Consequently, the derived
parameters are unlikely to be reliable. Many structures that fall into this category are
already well-known complexes such as W43 and W49.

\item Unclassified: this is a category for structures that cannot be
classified as any of the types described above. An example of a structure that
is included in this group is the Galactic centre region where the source
density is so high that the algorithm is unable to identify individual
structures and effectively groups all of the emission into a single enormous structure.
This category is therefore not considered a distinct structure type and
is excluded from further analysis.

\end{enumerate}

In Fig.\,\ref{fig:structure_types} we present a typical example of each of
these classification types to illustrate some of the features discussed above.
For each structure type we present the dust emission map and images of the
mid-infrared emission. In many cases corresponding structures are seen in the
infrared images and these can therefore provide strong support for the
structures identified by \Disperse\ and can assist in the classification,
particularly when the dust emission is weak.

Cloud structures are complicated and difficult to categorise as they have
irregular morphologies,  and their boundaries are not well
defined (e.g. \citealt{goldsmith2008, williams2000}). For these reasons a
definitive classification of different types of structures
is often not possible. The
descriptions provided above are quite loose and the resulting classification can
be somewhat subjective. To mitigate this, classifications have been made
independently by five members of the team using the criteria given above. Only structures that have been given
the same classification by three or more of the five leading authors are considered
to be bona fide examples of each type. A similar method has been applied in a
number of other studies, which have resulted in the production of reliable and
representative samples of different types of Galactic structures (e.g. giant
molecular filaments --- \citealt{ragan2014}, \hii\ region bubbles ---
\citealt{churchwell2006} and extended green objects ---
\citealt{cyganowski2008}).

\setlength{\tabcolsep}{2pt}

\begin{table*}

\begin{center}\caption{\label{tbl:cattable} Catalogue
of ATLASGAL structures extracted by \Disperse\ and identified as filaments by visual inspection of the emission maps.
The columns are as follows: (1) name derived from Galactic coordinates of the maximum intensity
in the structure; (2) level of agreement; (3)-(6) Galactic coordinates of the emission centroid and
the angular extent of the structure; (7)-(10) semi-major and semi-minor size,
aspect ratio and source position angle measured anti-clockwise from Galactic mid-plane;
(11) the width estimated from the normalized intensity distribution (e.g. Fig.\,3); (12) integrated flux densities determined from
the flux of all pixels found within the \emph{skeleton box}; (13) number of connecting pixels
that form the spine of the skeleton; (14) number of associated clumps identified in the ATLASGAL CSC.}
\begin{minipage}{\linewidth}
\small
\centering
\begin{tabular}{lc.........r.c}
  \hline \hline
  \multicolumn{1}{c}{Catalogue} &  \multicolumn{1}{c}{Agreement}
  &  \multicolumn{1}{c}{$\ell$} &  \multicolumn{1}{c}{d$\ell$} &  \multicolumn{1}{c}{$b$} &  \multicolumn{1}{c}{d$b$} &
  \multicolumn{1}{c}{$\sigma_{\rm{maj}}$} &  \multicolumn{1}{c}{$\sigma_{\rm{min}}$} &   \multicolumn{1}{c}{Aspect} & \multicolumn{1}{c}{$\theta$} &\multicolumn{1}{c}{Width}   &
  \multicolumn{1}{c}{$S_{\rm{int}}$} &  \multicolumn{1}{c}{nPixels}
&  \multicolumn{1}{c}{nCSC} \\

   \multicolumn{1}{c}{name} &  \multicolumn{1}{c}{flag}
  &  \multicolumn{1}{c}{(\degr)} &  \multicolumn{1}{c}{(\degr)} &  \multicolumn{1}{c}{(\degr)} &  \multicolumn{1}{c}{(\degr)} &
  \multicolumn{1}{c}{(\arcmin)} &  \multicolumn{1}{c}{(\arcmin)} &
  \multicolumn{1}{c}{ratio} &\multicolumn{1}{c}{(\degr)} &
  \multicolumn{1}{c}{(\arcsec)}  &
  \multicolumn{1}{c}{(Jy)} &  \multicolumn{1}{c}{} & \multicolumn{1}{c}{} \\


  \multicolumn{1}{c}{(1)} &  \multicolumn{1}{c}{(2)} &  \multicolumn{1}{c}{(3)} &  \multicolumn{1}{c}{(4)} &
  \multicolumn{1}{c}{(5)} &  \multicolumn{1}{c}{(6)} &  \multicolumn{1}{c}{(7)} &  \multicolumn{1}{c}{(8)} &
  \multicolumn{1}{c}{(9)} &  \multicolumn{1}{c}{(10)} &  \multicolumn{1}{c}{(11)} & \multicolumn{1}{c}{(12)} &  \multicolumn{1}{c}{(13)} &  \multicolumn{1}{c}{(14)}   \\
  \hline
  G008.255+0.165  &       4       &       8.244   &       0.912   &       0.168   &       0.878   &       1.3     &       0.8     &       1.5     &       -86.6   &       32.1    &       15.4    &       74      &       3       \\
  G008.675$-$0.689        &       5       &       8.667   &       0.947   &       -0.692  &       0.865   &       2.1     &       0.5     &       4.0     &       -68.3   &       45.5    &       16.8    &       69      &       3       \\
  G008.800$-$0.359        &       4       &       8.807   &       0.895   &       -0.370  &       0.923   &       1.8     &       1.2     &       1.5     &       2.1     &       30.1    &       12.0    &       116     &       4       \\
  G008.914$-$0.318        &       4       &       8.914   &       0.978   &       -0.326  &       0.862   &       3.2     &       0.5     &       6.7     &       85.5    &       30.3    &       8.3     &       95      &       2       \\
  G009.230+0.157  &       4       &       9.234   &       0.898   &       0.146   &       0.868   &       1.3     &       0.6     &       2.3     &       -67.2   &       32.0    &       4.8     &       50      &       1       \\
  G009.280$-$0.152        &       4       &       9.280   &       0.870   &       -0.157  &       0.880   &       0.8     &       0.3     &       2.5     &       37.0    &       28.8    &       7.6     &       33      &       1       \\
  G009.541$-$0.664        &       4       &       9.544   &       0.928   &       -0.657  &       0.883   &       1.9     &       0.9     &       2.2     &       78.4    &       32.0    &       13.0    &       82      &       3       \\
  G009.853$-$0.739        &       4       &       9.812   &       1.077   &       -0.733  &       0.908   &       3.3     &       0.6     &       6.0     &       -72.8   &       35.1    &       62.3    &       171     &       7       \\
  G009.970$-$0.024        &       5       &       9.964   &       0.942   &       -0.026  &       0.865   &       1.6     &       0.4     &       3.7     &       -75.5   &       30.4    &       12.8    &       79      &       3       \\
  G010.531$-$0.024        &       3       &       10.538  &       0.928   &       -0.032  &       0.860   &       2.3     &       0.7     &       3.4     &       -82.6   &       31.0    &       6.0     &       66      &       2       \\
  \hline
\end{tabular}\\
\end{minipage}\\
Notes: Only a small portion of the data is provided here,
the full table is only  available in electronic form at the CDS via anonymous ftp to
cdsarc.u-strasbg.fr (130.79.125.5) or via http://cdsweb.u-strasbg.fr/cgi-bin/qcat?J/A+A/.
\end{center}
\end{table*}

\subsection{Estimating sizes and total flux of structures}
\label{sect:dilation}

As discussed in the previous subsection, the algorithm traces the high column
density structures by connecting regions of emission seen in the dust maps;
the resulting skeletons essentially form the backbone of the structures and are
referred to as the {\em source skeletons}. Although these source skeletons are
useful and convenient to trace {the underlying structures}, they are not
sufficient for estimating the angular sizes and the total flux associated with each structure.
To achieve this we use an image processing operation called
{\em{dilation}}\footnote{\url{http://en.wikipedia.org/wiki/Dilation_(morphology)}};
this is a morphological operation that can be applied to binary images by
expanding shapes contained in the input image with another structural element.

The structural element used in dilation is a box with both sizes of the same
length. Our starting point is the source skeleton. When the square has a size of 1 pixel we are essentially just
tracing the skeleton, but as we increase the size of the square we start to
dilate the  skeleton and incorporate more of the surrounding
emission.
In Fig.\,\ref{fig:snake:dilate} we present an example of this method to illustrate
the results of this process for different  size squares. In general, a dilation
operation with a square of size $N$ (pixels) generates a \emph{skeleton box}
of width $N$ (pixels). The challenge is to determine the optimum  size of the
square, in pixels, to use so that the majority of the flux associated with a
structure is captured while avoiding possible contamination by
unassociated nearby structures.

\begin{figure}
\centering
\includegraphics[width = 0.5 \textwidth]{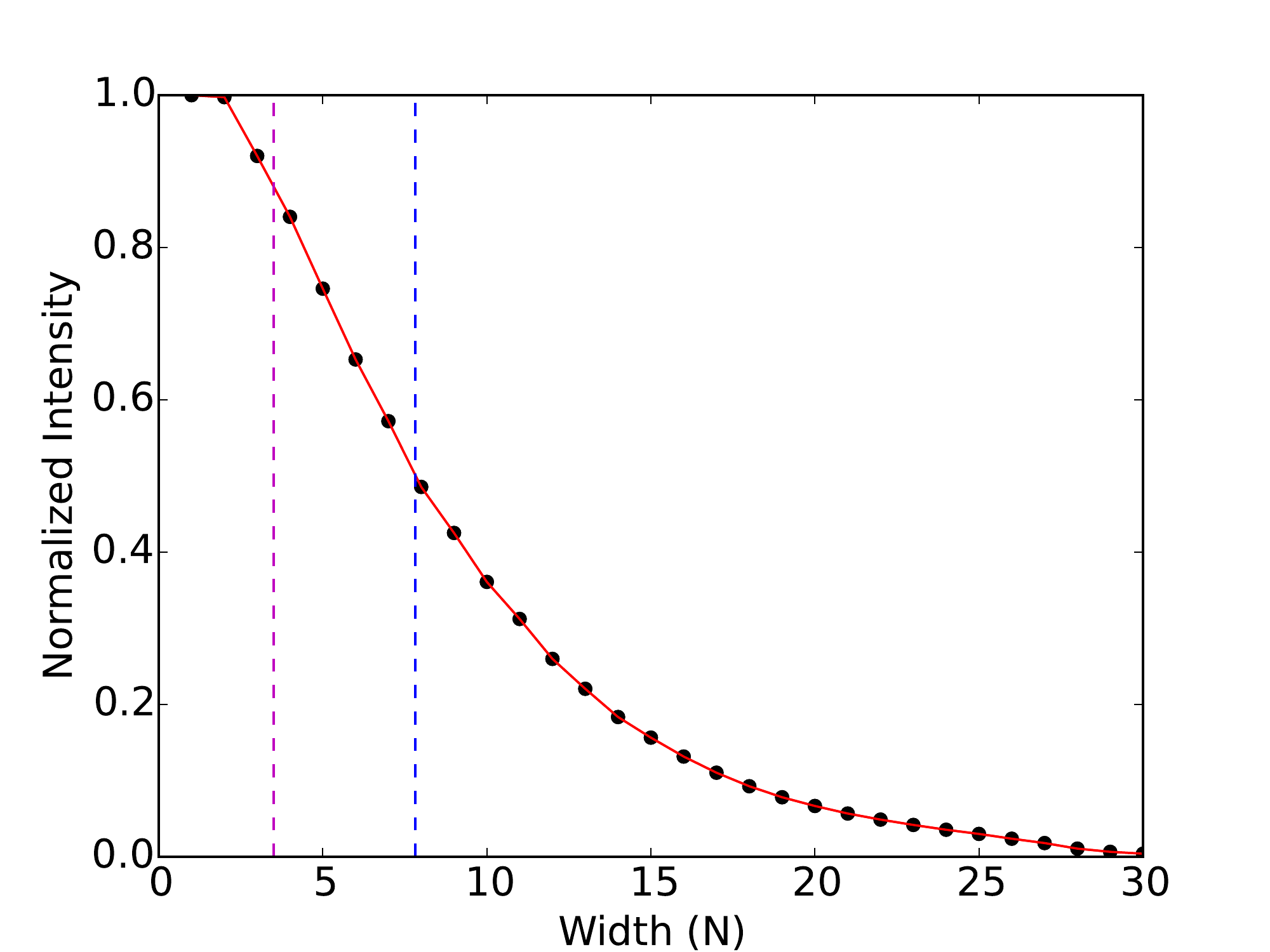} \\
\caption{Example of the normalized mean intensity as a function of the
width of the dilated skeleton for W51. The filled circles show the measurements
while the solid red line simply traces their distribution. The magenta and blue dashed vertical lines, respectively, indicate the angular resolution of the original ATLASGAL maps (19.2\arcsec\  corresponding to $\sim$3.2 pixels) and the width of the structure, which is defined as the size at which the intensity
reaches half of its maximum. }
\label{fig:mass:width}
\end{figure}

Two quantities can be readily defined with the help of the image dilation. Here,
we define the integrated flux, $S_{\rm{int}}$, to the sum of all pixels contained in a skeleton box of width $N$ as

\begin{equation}
S_{\rm int}(N) = \int_{w = N} I(x, y) \; {\rm d} x\; {\rm d } y\;,
\end{equation}

\noindent where the integration is carried out inside the skeleton, and ${\rm
d} x$ and ${\rm d} y$ are measured in pixels. Similarly, we can define the total
number of pixels inside the skeleton of width $N$, which is denoted as $N_{\rm pixel}(N)$. The normalised intensity is then defined as:

\begin{equation}
I(N) = \frac{{\rm d} S_{\rm int}(N)}{{\rm d} N_{\rm pixel}(N)}\;.
\end{equation}

An important point to note is that while the identification of the structures
has been performed on smoothed maps to improve the sensitivity to lower density
connecting fragments, we have used the original higher resolution ATLASGAL
emission maps to estimate their associated properties. In
Fig.\,\ref{fig:mass:width} we present the normalised pixel intensity as a
function of increasing dilation width. It is clear from this plot that the peak
flux density decreases with increasing distance from the emission ridge, which
is traced by the skeleton. From inspection of the distribution shown in
Fig.\,\ref{fig:mass:width} and the example shown in Fig.\,\ref{fig:snake:dilate}
we estimate the optimum value for the skeleton box width is approximately 25
pixels; this width ensures that the vast majority of the fluxes are captured and
that fluxes from pixels above 10\% of the peak intensity are included.  This
is also the maximum angular scale that the ATLASGAL survey is sensitive to. We
have therefore used this value to  derive the dilated skeletons, which are
subsequently used to estimate the total flux and to calculate the structure
sizes and their orientations.

\begin{figure*}
\begin{center}
\includegraphics[width = 0.95 \textwidth]{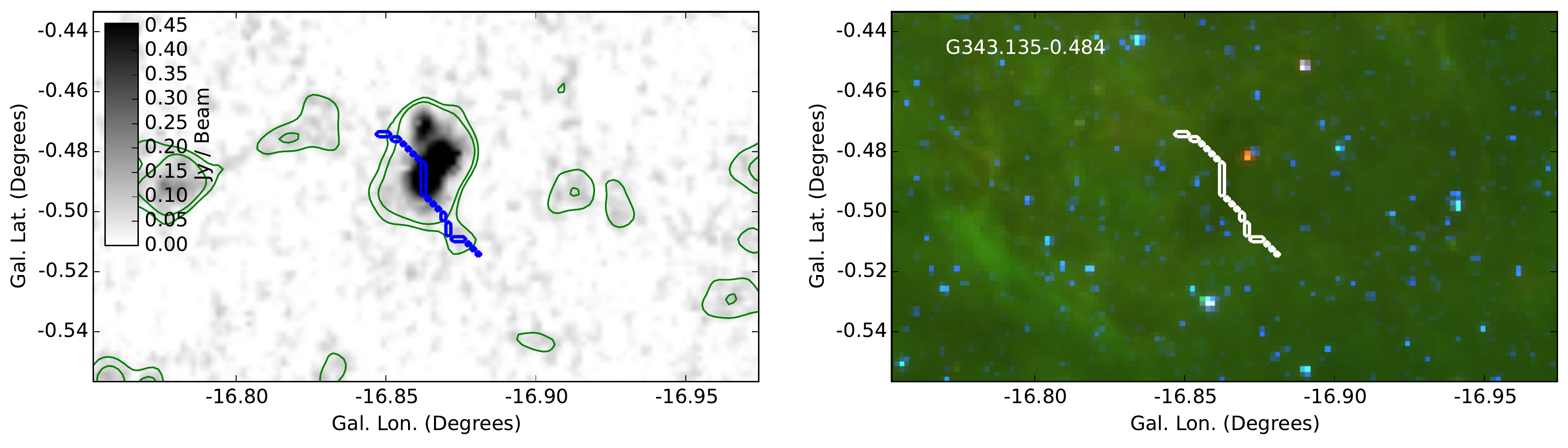}
\includegraphics[width = 0.95 \textwidth]{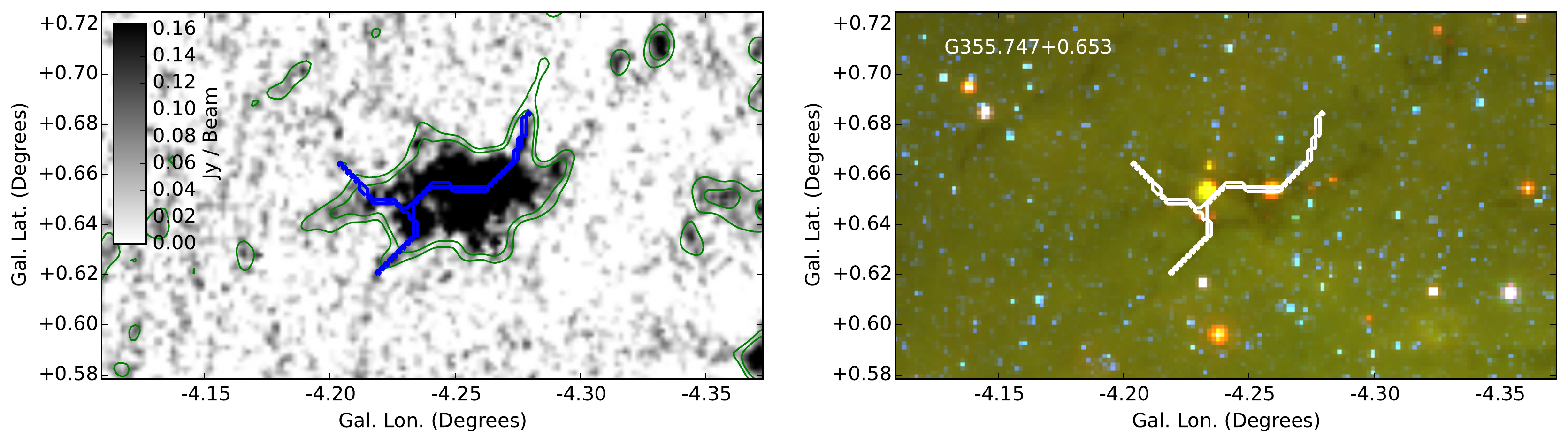}
\includegraphics[width = 0.95 \textwidth]{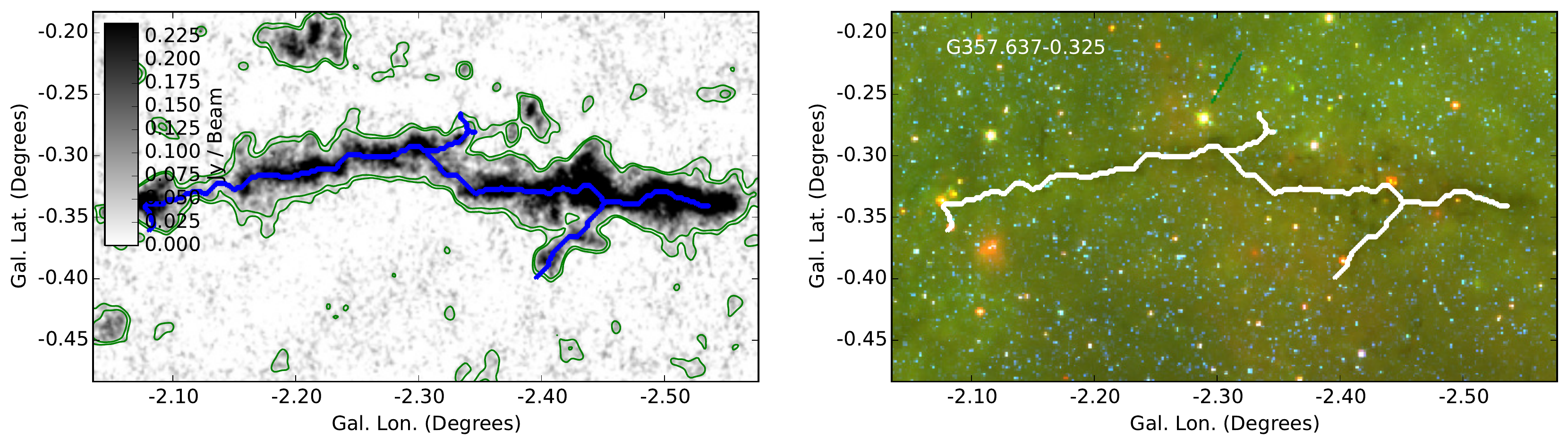}
\caption{Examples of the different structure types identified. The dust
continuum images are presented in the left hand panels while in the right hand panels we
present three colour composite mid-infrared images; they are produced using the
4.5, 8.0\,\mum\ IRAC bands taken from the GLIMPSE survey (\citealt{benjamin2003,churchwell2009})
and the 24\,\mum\ MIPS band image extracted from the MIPSGAL survey (\citealt{carey2005}).
The green contours show the 3 and 5$\sigma$ noise levels. The blue and white
lines overlaid on the left and right panels trace the skeletons of
the structures identified by the \Disperse\ algorithm.
In the upper, middle and lower panels we show an example of an marginally
resolved clumps, a marginally resolved elongated structure and a filament,
respectively.}
\label{fig:structure_types}
\end{center}
\end{figure*}

\setcounter{figure}{3}

\begin{figure*}
\begin{center}
\includegraphics[width = 0.95 \textwidth]{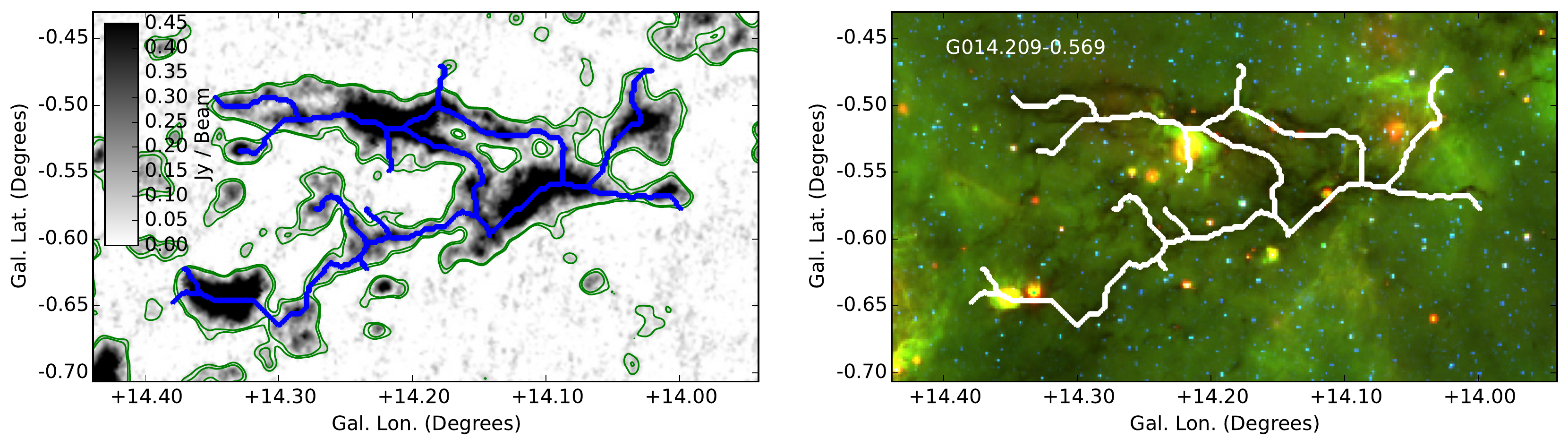}
\includegraphics[width = 0.95 \textwidth]{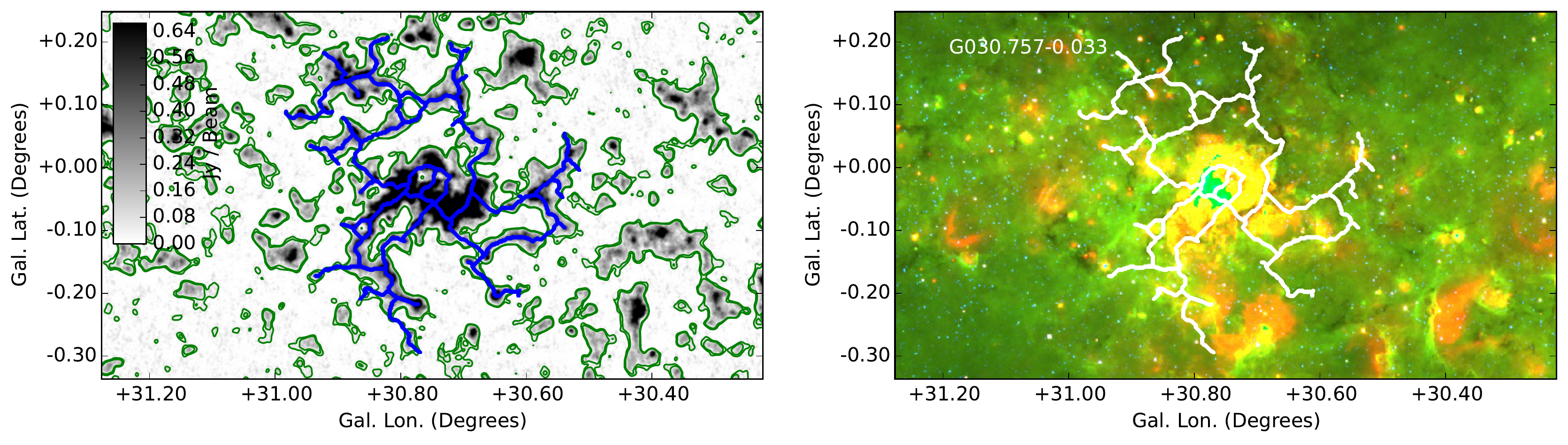}
\includegraphics[width = 0.95 \textwidth]{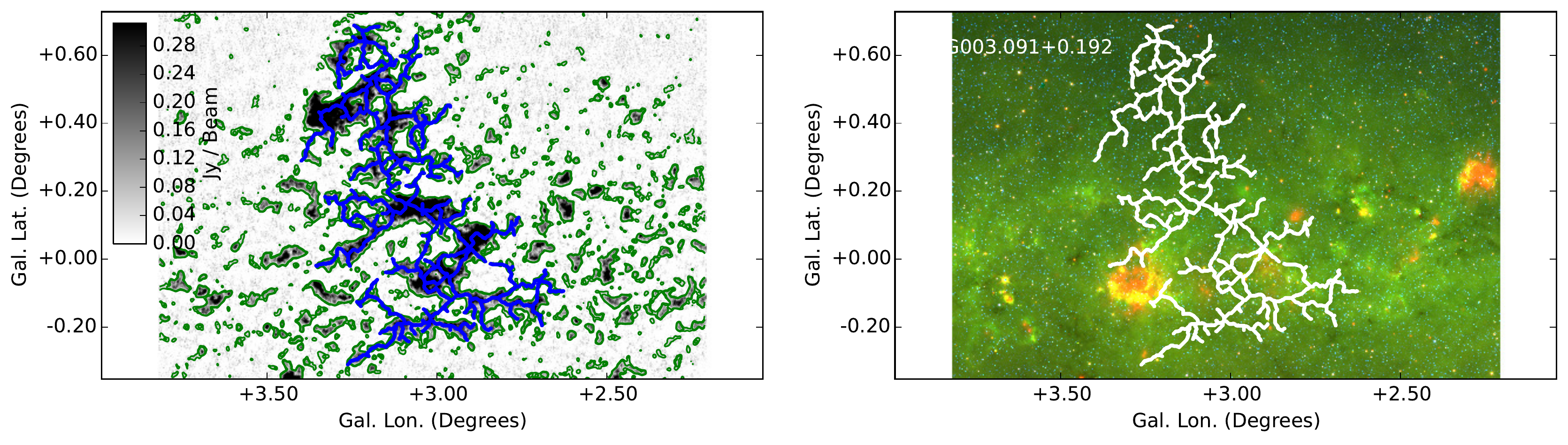}
\caption{Continued. In the upper, middle we show an example of an network of
filaments, a complex and an unclassified structure, respectively. Structures
that are assigned as unclassified are excluded from further considerations.}
\end{center}
\end{figure*}

We estimate the width of each structure taking the point where the averaged
intensity  drops to half of its maximum (see the blue vertical line shown in
Fig.\,\ref{fig:mass:width}). Note that our definition of the width differs from
the definition commonly used in analysing \emph{Herschel} observations
\citep[e.g.][]{arzoumanian2011,2010A&A...518L.103M} where they fit analytical formulae
 to the filament profiles obtained by stacking the maps based on the
skeletons. We choose our dilation-based definition since we are interested in
how gas concentrates around the skeleton, rather that studying the profiles in
detail. Properties of the detected structures are presented in Table
\ref{tbl:cattable}.

\subsection{Catalogue of { identified structures}}

Of the 1812 structures identified we find agreement for 1339 structures, which
is approximately three-quarters of the sample and demonstrates the difficulties involved in
classifying these types of objects. The situation is a little better
than the statistics would suggest as many of the disagreements are for
structures that fall between two similar types of structures e.g.
when does an elongated structure become a filament and when does a filament with a
number of sub-branches become a network? In this paper we have chosen to focus
on the structures where the morphology is relatively unambiguous.

\setlength{\tabcolsep}{6pt}


In estimating the sizes, orientation and total flux associated with each
structure we take all the pixels inside the $w = 25\;\rm pixels$ mask into
account (see Section\,\ref{sect:dilation}) with a SNR (signal-to-noise ratio)
$>$3.
The size of the major and minor axes and their orientation is determined by diagonalizing
the tensor of second moments of the position coordinates weighted by the pixel
intensity. We define the major and minor sizes as the larger and smaller
eigenvalue of the tensor, respectively, while the orientation as the angle
between the larger eigenvector and the Galactic mid-plane.

In Table\,\ref{tbl:cattable} we present the observed properties of the 517 filamentary structures identified. The structure names given in Col.\,1 are based on the Galactic coordinates of
flux weighted centroid position. In Cols.\,2-5 we give the coordinates of the
centroid position and angular size of each structure while in Cols.\,6-7 we
give semi-major and semi-minor axis lengths that represent the standard deviation of
the pixel co-ordinate values about the centroid position, weighted by the pixel
values. The aspect ratio and the orientation of the structure, $\theta$, is
presented in Col.\,8 and Col.\,9 while in Col.\,10 we give the width in arcsecs.
In Cols.\,11 and 12 we give the integrated
870\,\mum\ flux and number of pixels that
define the spine of the structures. Finally, in Col.\,13 we give the
number of associated ATLASGAL Compact Source Catalogue (CSC; \citealt{contreras2013,urquhart2014c}), which will be discussed in more detail later in this section.

\subsection{Properties of the various kinds of structures}
\label{sect:properties_}

\begin{figure}
\centering
\includegraphics[width = 0.5 \textwidth]{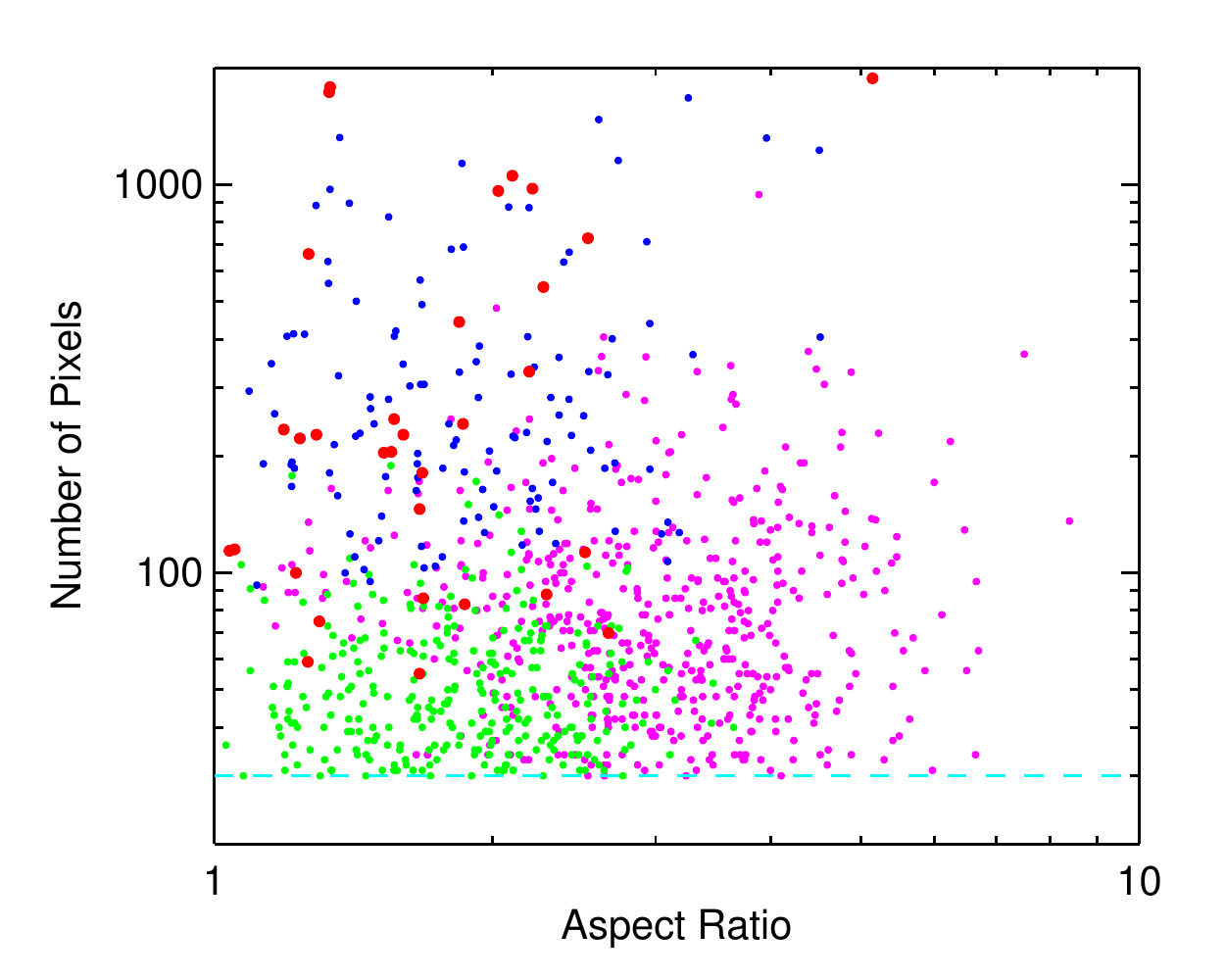} \\
\includegraphics[width = 0.5 \textwidth]{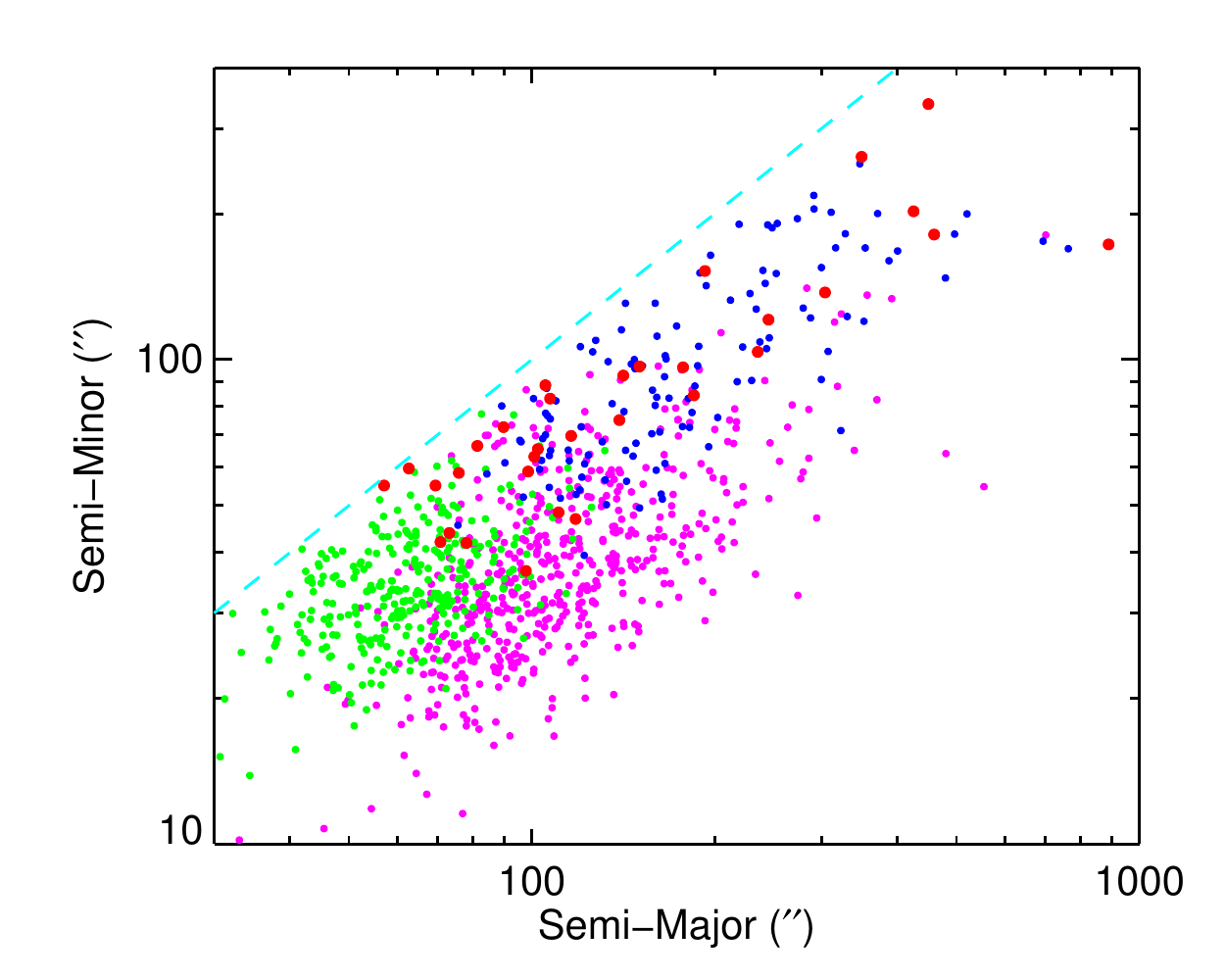} \\
\caption{Observed properties of the different structure types identified in
Section\,\ref{sect:classification}.
These plots show the distributions of angular sizes,
number of associated spine points and shapes of the elongated structures,
filaments, networks of filaments and complexes, which are shown as
filled green, magenta, blue and red circles, respectively.
For clarity we have excluded the { marginally resolved} structures as there
is significant overlap with the elongated structures. The horizontal dashed line shown
in the upper panel indicates the threshold of thirty pixels required for inclusion in
the catalogue while the diagonal dashed line shown in the lower panel shows the line of
equality where the semi-major and semi-minor axes are equal.}
\label{fig:structure_properties}
\end{figure}

The classification has been performed from a visual examination of the dust and
mid-infrared images, however, we might expect the different structure types
to have significantly different properties. In Fig.\,\ref{fig:structure_properties}
we present plots showing the distributions of the number of pixels as a function
of the aspect ratio (upper panel) and the semi-major and semi-minor axes (lower panel).
The subsamples are well separated in the parameter
space although we note there are still some significant overlaps. The complexes
and networks appear to be well mixed with similar aspect ratios and numbers of
pixels, and it might transpire that they are in fact similar kinds of
structures with the complexes being more distant and active regions that are
simply less well resolved in the ATLASGAL beam. These two types of structures
are, however, relatively well separated from the filaments and elongated
structures, which are themselves well separated from each other. Despite the
difficulties associated with classifying these structures we do find that
the different types of structures have fairly distinct properties and this
provides strong reassurance that both the classification scheme and the method
used are reliable.

\setlength{\tabcolsep}{2pt}

\begin{table*}

\begin{center}\caption{Summary of structure types and their statistical
properties.
The flux ratio is estimated by considering only the flux associated with
structures where agreement is found. The fluxes are normalised with
respect to the total flux in all the structures where agreement is found for the five main structure types. When
determining the statistics for the association with the massive star forming { (MSF)}
clumps we only consider structures outside of the inner 10\degr\ of the
Galactic centre where the matching between ATLASGAL and massive star formation tracers is complete (for details see \citealt{urquhart2014b}).}
\label{tbl:statistical_properties}
\begin{minipage}{\linewidth}
\small
\begin{tabular}{l..........}
\hline \hline
\multicolumn{1}{c}{Structure}&  \multicolumn{1}{c}{Number}& \multicolumn{1}{c}{Fraction} &\multicolumn{1}{c}{Fraction} & \multicolumn{1}{c}{Aspect} &	\multicolumn{1}{c}{Mean number}    &\multicolumn{1}{c}{Total number}  &\multicolumn{1}{c}{Mean number } &\multicolumn{1}{c}{Total number} &\multicolumn{1}{c}{Total number}&\multicolumn{1}{c}{Fraction}\\

\multicolumn{1}{c}{type}&  \multicolumn{1}{c}{}&\multicolumn{1}{c}{of
structures}&\multicolumn{1}{c}{of total flux}&\multicolumn{1}{c}{ratio}&
\multicolumn{1}{c}{of pixels}   &\multicolumn{1}{c}{of CSC assoc.}
&\multicolumn{1}{c}{of CSC assoc.}&\multicolumn{1}{c}{of MSF assoc.}
&\multicolumn{1}{c}{of MSF clumps}&\multicolumn{1}{c}{of MSF clumps}
\\

\multicolumn{1}{c}{(1)}&  \multicolumn{1}{c}{(2)}&\multicolumn{1}{c}{(3)}&\multicolumn{1}{c}{(4)}&\multicolumn{1}{c}{(5)}&	\multicolumn{1}{c}{(6)}   &\multicolumn{1}{c}{(7)}  &\multicolumn{1}{c}{(8)}&\multicolumn{1}{c}{(9)} &\multicolumn{1}{c}{(10)}&\multicolumn{1}{c}{(11)} \\

\hline

Clumps	&	370	&	0.28	&	0.06	&	2.09	&	44	&	325	&	0.9	&	11	&	12	&	0.02
\\
Elongated	&	301	&	0.22	&	0.11	&	1.91	&	54	&	499	&	1.7	&	33	&	36	&	0.07	\\
Filament	&	517	&	0.39	&	0.26	&	3.18	&	93	&	1495	&	2.9	&	43	&	63	&	0.12	\\
Network	&	119	&	0.09	&	0.37	&	1.97	&	357	&	1609	&	13.5	&	29	&	80	&	0.15	\\
Complex	&	32	&	0.02	&	0.21	&	1.81	&	444	&	574	&	17.9	&	9	&	32	&	0.06	\\
\hline\\
\end{tabular}\\

\end{minipage}

\end{center}
\end{table*}
\setlength{\tabcolsep}{6pt}

In Table\,\ref{tbl:statistical_properties} we present the number
of structures identified for each structure type along with the fraction
of the total flux normalized with respect to the total flux in all the
detected structures. Here we only included the 1339
structures for which we have agreement. A number of other statistical
properties are also included. This table provides some immediate insights into the relative importance of the various kinds of
structures identified. The combination of the { marginally resolved} and
elongated structures contribute approximately 50\% of the structures identified,
but they only contribute a relatively small fraction of the total flux ($\sim$17\%).
The next most numerous structure type is the filaments, which contribute
$\sim$40\% of the total number of structures, however, these contribute a
disproportionate amount of the total flux given their number ($\sim$26\%).

The majority of the flux ($\sim$37\%) is associated with the filamentary networks, which
themselves contribute approximately 10\% of the structures. Combined with filaments
 we find that these kinds of structures
 make up approximately 50\% of all  identified sources and are
associated with 60\% of the total flux. If we assume that the different types of
structures have a similar distance and temperature distribution, and there is no
reason to think otherwise, then we can use the integrated fluxes as a proxy
for the masses. Although this is likely to be somewhat imprecise it provides a
useful ballpark estimate for the fraction of mass associated with each structure
type.
This crude analysis suggests that filaments and filamentary
networks may also contribute roughly 60\% of the mass associated with the
 identified structures. Furthermore, due to the limited resolution and
 sensitivity of the ATLASGAL survey it is likely that the fraction of { filaments} we have found
 is a lower limit. This highlights the ubiquitous nature of filamentary
 structures and their potential importance to current and future star formation
 in the Galaxy.

Complexes make up only 2\% of the structures identified, however, given their
size they still contribute a significant amount of the total flux, and therefore also the
mass across the inner Galaxy ($\sim$22\%). Many of the well know star formation
complexes belong to this group, such as W43, W49 and RCW106/G333. Altogether
with the network of filaments we find that they contribute only 11\% of the total number of structures but due to their sizes they are associated
with $\sim$60\% of the total flux and likely to contain a similar fraction of
the total mass found in these 1339 structures. This would suggest that much
of the dense mass, and presumably also the star formation, is concentrated in a relatively small
number of very large and massive structures. This is consistent with the
findings of \citet{murray2010} and \citet{urquhart2014b} that a significant
fraction (30-50\%) of all of the Galactic star formation is concentrated in the
most luminous 20-30 regions located in the Galactic mid-plane.

In Table\,\ref{tbl:statistical_properties} we also give the aspect ratio and the
mean number of associated spine pixels. The aspect ratio is simply the ratio of
the semi-major and semi-minor axes. As one would expect filaments have
significantly larger aspect ratios than any of the other structure types,
which is also clearly seen in Fig.\,\ref{fig:structure_properties}. For the
elongated structures and { filaments} the number of pixels can be used as a
reasonable approximation of their observed length. Here again there is a clear
trend for increasing numbers of pixels as we move from elongated structures to
filaments and this is roughly correlated with the increasing aspect ratio for
these structure types.
However, this trend breaks down for networks and complexes as these tend to
be associated with multiple intersecting filamentary structures. The structural
complexity makes it difficult to provide reliable estimates of the
lengths of these structures.

\begin{figure}
\centering
\includegraphics[width = 0.5 \textwidth]{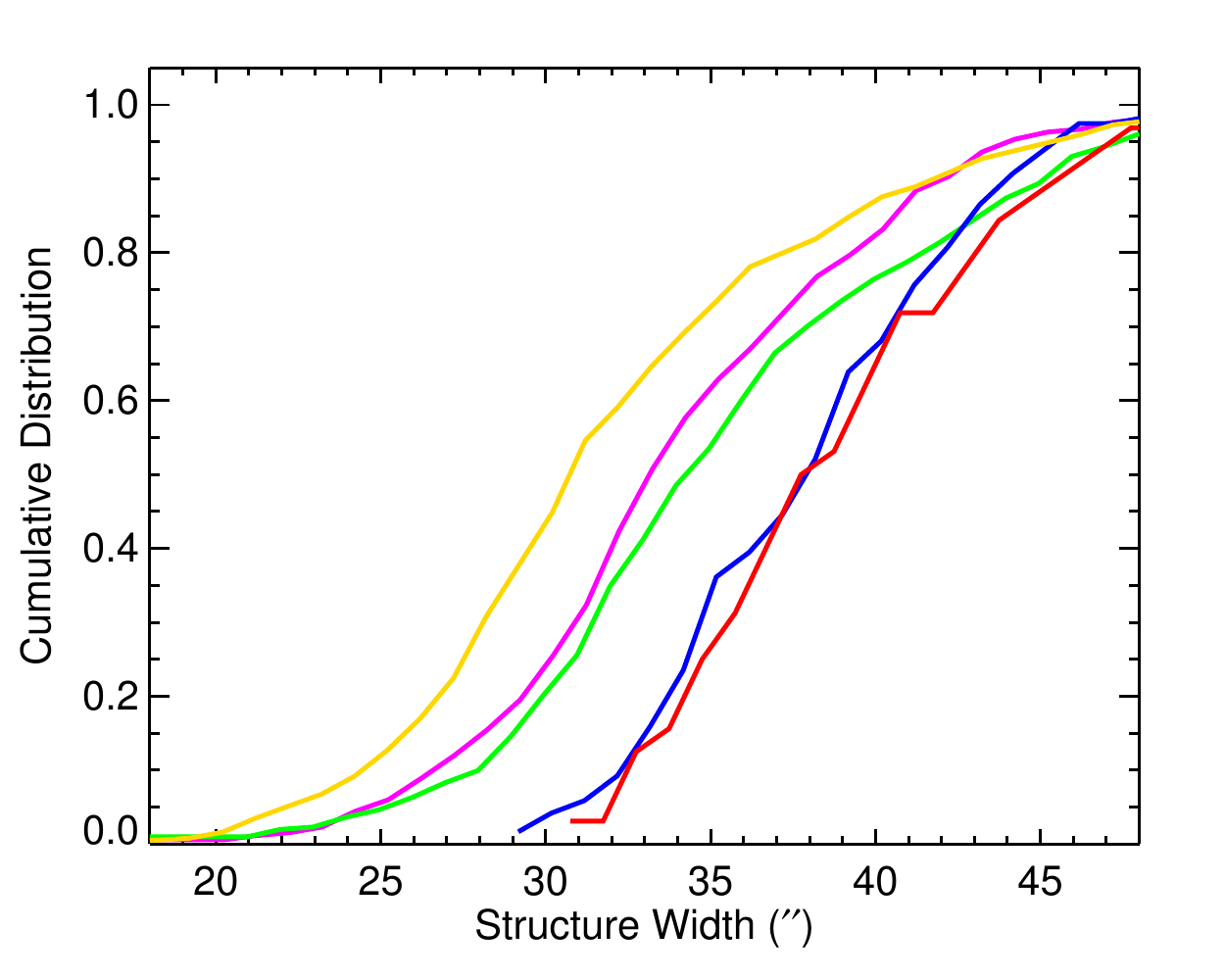} \\
\caption{Cumulative distribution of the pixel widths of each type of structure.
The yellow, green, magenta, blue and red curves show the distribution of
marginally resolved, elongated structures, filaments, networks of filaments and
complexes, respectively.}
\label{fig:width}
\end{figure}

In Fig.\,\ref{fig:width} we show the cumulative distribution of the widths of
the various structures. The properties have been extracted from the original
maps and it is therefore clear from this plot that the widths of majority of structures are resolved by
ATLASGAL. We also note that, with the exception of the marginally resolved
structures, the filaments have significantly narrower widths than the other
structures.
We will discuss the widths of the filaments in more detail in
Section\,\ref{sec:sizes}.

\begin{figure}
\centering
\includegraphics[width = 0.5 \textwidth]{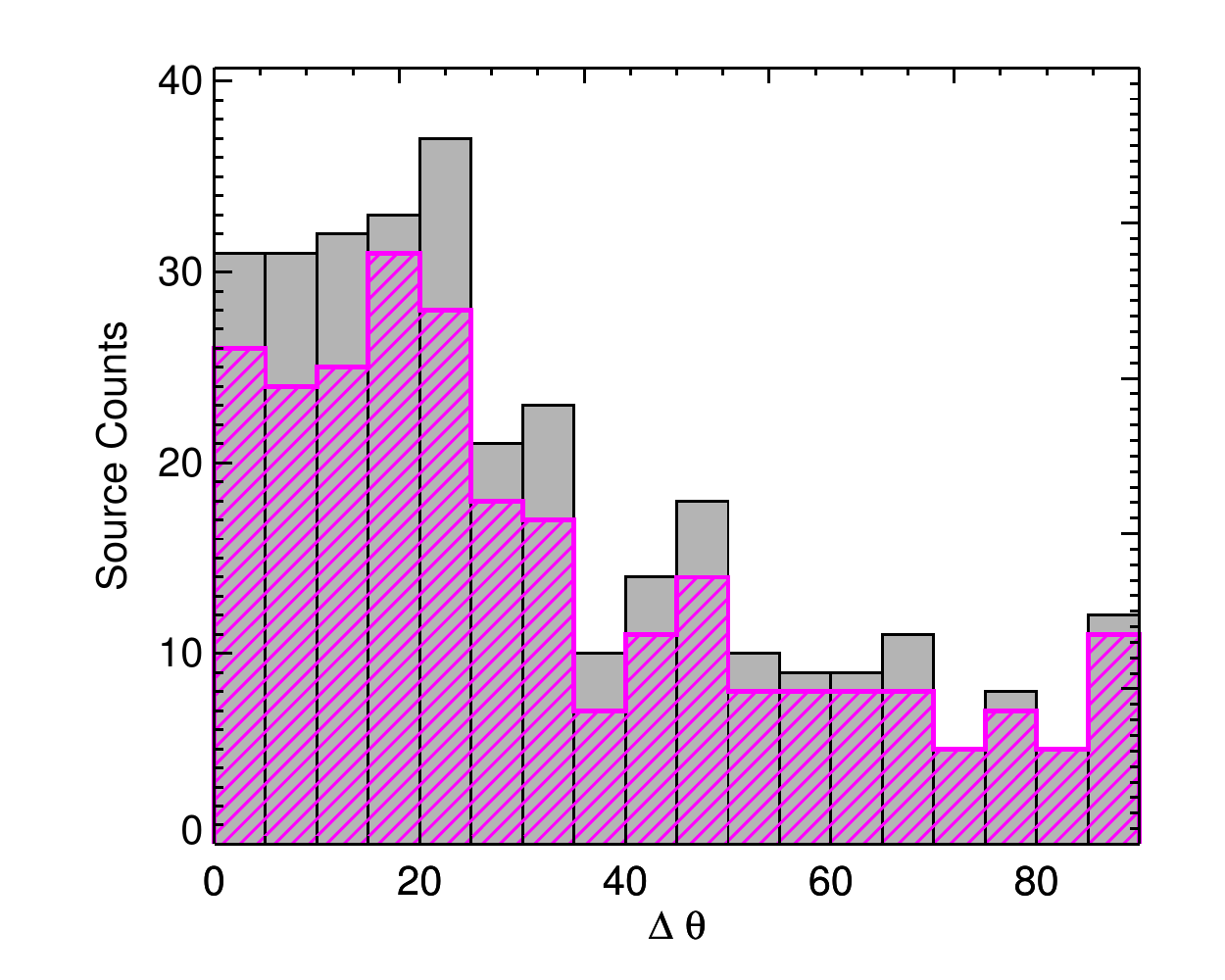} \\
\caption{\label{fig:pa_distribution} Angle
between the semi-major axis of the filaments and the Galactic
plane. The results of all structures with an aspect ratio greater than 3 are shown as a grey histogram while those classified as filaments are shown in magenta.}
\end{figure}

In Fig.\,\ref{fig:pa_distribution} we show the distribution of  the angle
between the semi-major axis of the filaments and the Galactic
plane for all structures with an aspect ratio larger than 3, the vast majority
of which have been classified as filaments. We have chosen to restrict this analysis to
the more elongated structures to minimise the uncertainty in the position angle.
This plot reveals a broad but significant peak in the structure orientation with
respect to the Galactic mid-plane with an angle of between 0 and 20\degr. This suggests
that the elongated structures are preferentially aligned parallel to the plane
of the Galaxy. This correlation is independent of their Galactic location i.e.
we find no correlation between the orientation of the  and Galactic longitude
or angular separation from the mid-plane (i.e. $|b|$). We will investigate this
correlation in more detail in Section\,\ref{sect:orientation}.

\subsection{Association with ATLASGAL Compact Source Catalogue}
\label{sect:csc_associations}


In the following, we match the sample with the ATLASGAL Compact Source Catalogue
(CSC;
\citealt{contreras2013,urquhart2014c})\footnote{http://atlasgal.mpifr-bonn.mpg.de/cgi-bin/ATLASGAL\_DATABASE.cgi.},
which was produced using the \sex\ algorithm (\citealt{bertin1996}). A useful
output produced by this extraction code are image masks. They are the same size
and dimensions as the input emission maps but each pixel is given an integer
value that relates each pixel to a source in the catalogue. These masks
therefore contain the emission footprint of each source and by mapping the spine
positions onto these maps we can unambiguously identify associations between the
CSC and structures identified by  \Disperse, which in turn can provide some
insight into source sub-structure. Furthermore, making the link between the
structures and the CSC allows us to assign velocities and distances to many of
the structures that have been determined from previous work by the ATLASGAL team
(e.g. \citealt{wienen2012,wienen2015a,urquhart2014b} and {\color{black}K\"onig
et al. 2015}) and from the literature (e.g.
\citealt{bowers2015,green2011b,dunham2011b,schlingman2011,roman2009}).

\begin{figure}
\centering
\includegraphics[width = 0.5 \textwidth]{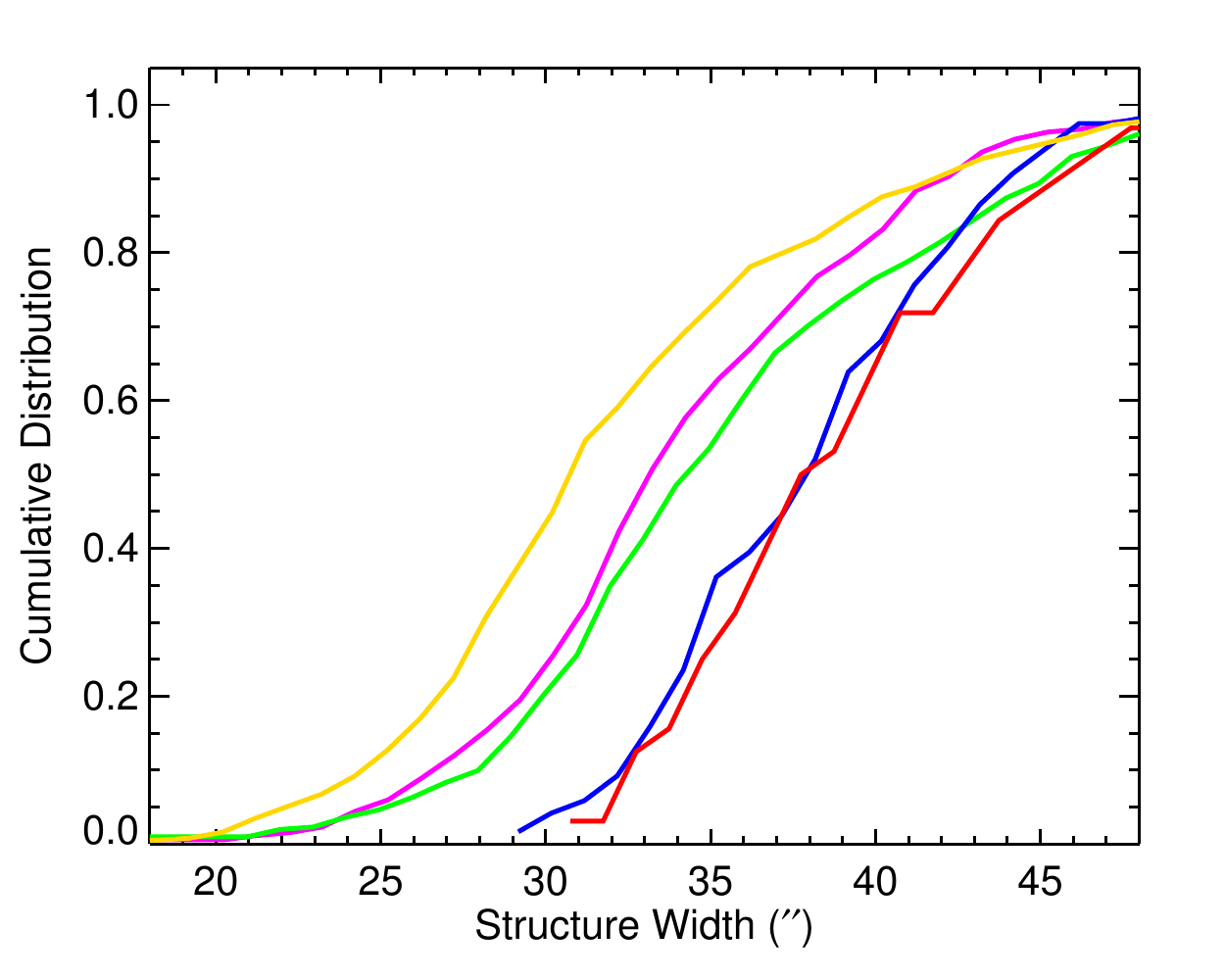} \\
\caption{Cumulative distribution showing the fraction of the various types of
structures with a particular integer number of CSC or few associations. The
green, magenta, red and blue show the distributions of the elongated structures and
filaments and complexes and networks of filaments, respectively.}
\label{fig:csc_distribution}
\end{figure}

 The compact source catalogue derived from the ATALSGAL survey provided an
unbiased  view of the distribution of dense gas (i.e.  $ > \sim 10^{22}$\,cm$^{-2}$)
in the Galaxy \citep{schuller2009_full}. We find
that roughly two-thirds of the structures where agreement is found are
associated with one or more ATLASGAL CSC, with an average of four clumps
being associated with each structure (4502 ATLASGAL CSC associated with 1042 structures).
A full breakdown of the structures and the number of associations is given in
Table\,\ref{tbl:statistical_properties}. In Fig.\,\ref{fig:csc_distribution} we
present a cumulative distribution plot showing the fraction of structures as a
function of the number of CSC associations. As one would expect the mean number
of CSC associations increases with the complexity of the structure, with
{ marginally resolved} structures being associated with an average of $\sim$1
compact clump while filaments are associated with an average of 3 clumps and
complexes with an average of $\sim$20 dense clumps; {these values are relatively unaffected if
we exclude the Galactic centre region (i.e. |$\ell$| $<$10\degr).}

Previous work has been done to match dense clumps in the ATLASGAL CSC with
signposts of massive star formation. In a recent series of papers
\citet{urquhart2013a,urquhart2013b,urquhart2014b} identified a large sample
of massive star forming (MSF) clumps by associating ATLASGAL clumps with
methanol masers identified by the methanol multibeam (MMB; \citealt{green2009}),
which are considered to be an excellent tracers of massive star formation (e.g.
\citealt{walsh1998,minier2003}), mid-infrared bright massive young stellar
objects (MYSOs) identified by the Red MSX Source (RMS) survey
(\citealt{lumsden2013}) and compact and ultracompact \hii\ regions identified by
both the RMS and CORNISH (\citealt{hoare2012,purcell2013}) surveys. In total
some $\sim$1300 MSF clumps have been identified, however, two of the three
surveys used for the matching exclude the inner ten degrees of the Galactic
plane. The MSF associated sample of clumps also includes the ATLASGAL-Extension
region ($\ell=280-300\degr$ and $b=-2$ to 1\degr), which is not considered in
this work. The total number of MSF in the area of interest considered here is
981, of which we find 520 are associated with structures where agreement is
found; this corresponds to approximately 53\% of the MSF associated clumps.

Comparing the fraction of MSF clumps associated with each type of structure
(Columns 4 and 11 in Table\,\ref{tbl:statistical_properties}) we find it is tightly correlated with the fraction of the total flux
associated with these structure types.
Both the flux and fraction of
MSF clumps agree within a few percent. It would therefore appear that the number
of MSF sites associated with any type of structure is roughly proportional to
the amount of dense material associated with them, which  indicates that
these gas condensations share a similar structure at the clump scale.

\begin{figure*}
\begin{center}
\includegraphics[width = 0.95
\textwidth]{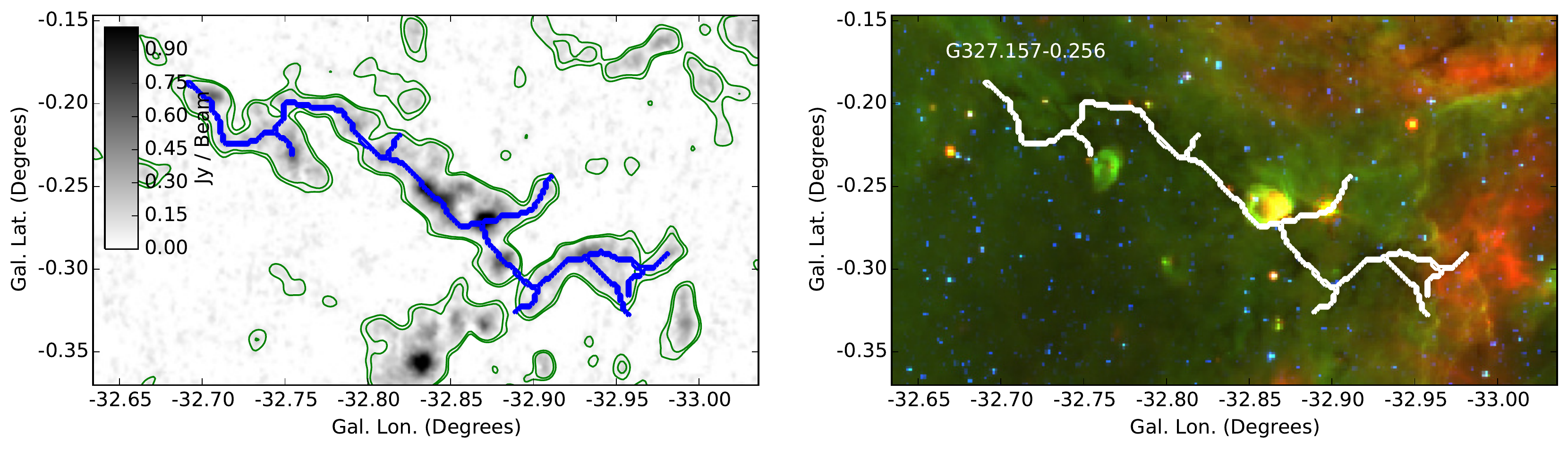}
\includegraphics[width = 0.95
\textwidth]{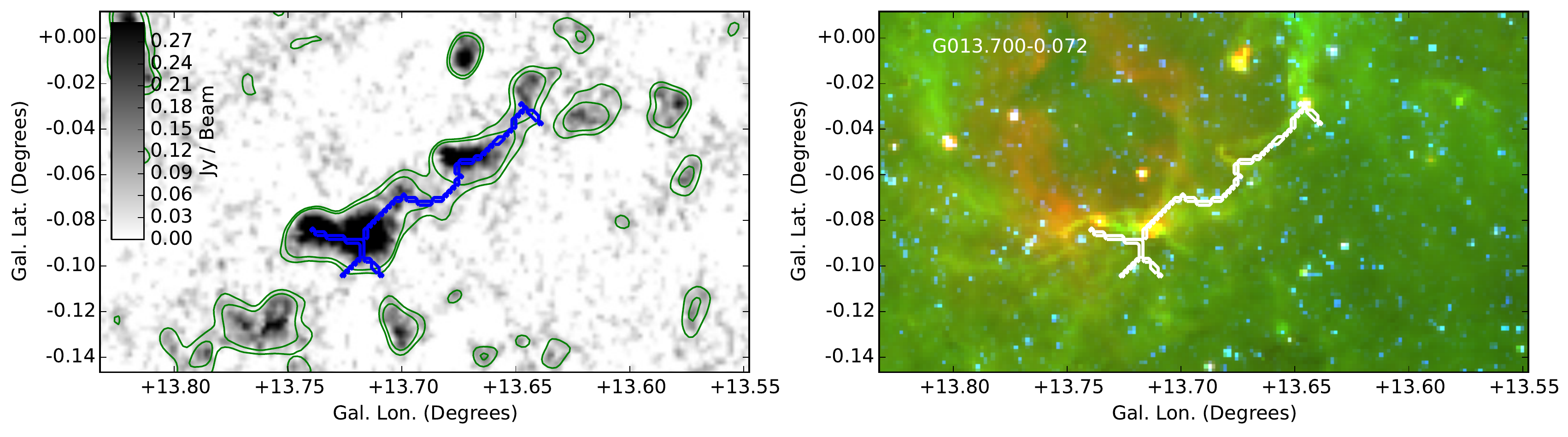}
\includegraphics[width = 0.95
\textwidth]{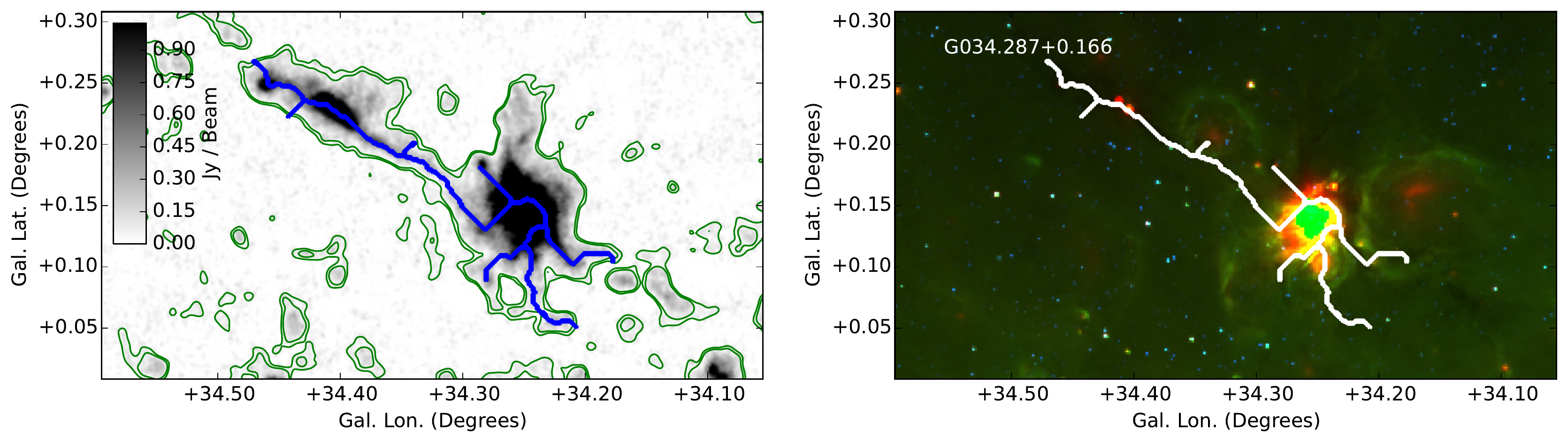}
\caption{Examples of filaments. For image details see caption of Fig.\,\ref{fig:structure_types}.}
\label{fig:filaments}
\end{center}
\end{figure*}

\section{Properties of the filaments}
\label{sect:physical_properties}


\setlength{\tabcolsep}{3pt}

\begin{table*}

\begin{center}\caption{Physical properties of the 279 ATLASGAL filaments for which a velocity has been determined. }
\label{tbl:atlasgal_filaments}
\begin{minipage}{\linewidth}
\small
\begin{tabular}{l.c.c.......}
\hline \hline
\multicolumn{1}{c}{Structure}&\multicolumn{2}{c}{\vlsr} & \multicolumn{2}{c}{Distance} &	\multicolumn{1}{c}{Length}    &\multicolumn{1}{c}{Width}    &\multicolumn{1}{c}{Log(Total Mass)}  &\multicolumn{1}{c}{Log(Mean $N_{\rm{H_2}}$)}&\multicolumn{1}{c}{Log(Peak $N_{\rm{H_2}}$)}  &\multicolumn{1}{c}{IRDC}&\multicolumn{1}{c}{Bubble} \\

\multicolumn{1}{c}{name}& \multicolumn{1}{c}{(\kms)}&\multicolumn{1}{c}{(Ref.)}&\multicolumn{1}{c}{(kpc)}&\multicolumn{1}{c}{(Ref.)}&	\multicolumn{1}{c}{(pc)}   &\multicolumn{1}{c}{(pc)}   &\multicolumn{1}{c}{(\msun)}  &\multicolumn{1}{c}{(cm$^{-2}$)}&\multicolumn{1}{c}{(cm$^{-2}$)}&\multicolumn{1}{c}{assoc.}&\multicolumn{1}{c}{assoc.}  \\

\multicolumn{1}{c}{(1)}&  \multicolumn{1}{c}{(3)}&\multicolumn{1}{c}{(4)}&\multicolumn{1}{c}{(5)}&	\multicolumn{1}{c}{(6)}   &\multicolumn{1}{c}{(7)}   &\multicolumn{1}{c}{(8)}  &\multicolumn{1}{c}{(9)}&\multicolumn{1}{c}{(10)}&\multicolumn{1}{c}{(11)}&\multicolumn{1}{c}{(12)}&\multicolumn{1}{c}{(13)}  \\
\hline
G001.653+0.203  &       163.2   &       1       &       \multicolumn{1}{c}{$\cdots$}    &       \multicolumn{1}{c}{$\cdots$}    &       \multicolumn{1}{c}{$\cdots$}    &       \multicolumn{1}{c}{$\cdots$}    &       \multicolumn{1}{c}{$\cdots$}    &       21.52   &       21.74   &       \multicolumn{1}{c}{$\cdots$}    &       \multicolumn{1}{c}{$\cdots$}    \\
G003.456+0.006  &       5.8     &       1       &       \multicolumn{1}{c}{$\cdots$}    &       \multicolumn{1}{c}{$\cdots$}    &       \multicolumn{1}{c}{$\cdots$}    &       \multicolumn{1}{c}{$\cdots$}    &       \multicolumn{1}{c}{$\cdots$}    &       21.60   &       21.76   &       \multicolumn{1}{c}{$\cdots$}    &       \multicolumn{1}{c}{$\cdots$}    \\
G005.101$-$0.104        &       45.4    &       2       &       2.8     &       1       &       5.03    &       0.40    &       3.176   &       21.56   &       21.76   &       \multicolumn{1}{c}{$\cdots$}    &       \multicolumn{1}{c}{$\cdots$}    \\
G005.357+0.101  &       11.1    &       2, 3    &       3.0     &       1       &       2.58    &       0.41    &       2.802   &       21.63   &       21.88   &       \multicolumn{1}{c}{$\cdots$}    &       \multicolumn{1}{c}{$\cdots$}    \\
G005.938$-$1.274        &       11.1    &       3       &       2.9     &       Near    &       4.50    &       0.13    &       3.495   &       21.87   &       22.08   &       \multicolumn{1}{c}{$\cdots$}    &       \multicolumn{1}{c}{$\cdots$}    \\
G006.104$-$0.623        &       15.7    &       2, 3    &       3.7     &       1       &       6.10    &       0.62    &       3.705   &       21.71   &       21.87   &       \multicolumn{1}{c}{$\cdots$}    &       \multicolumn{1}{c}{$\cdots$}    \\
G006.589$-$0.106        &       12.8    &       3       &       3.4     &       1       &       5.48    &       0.57    &       3.849   &       21.82   &       22.13   &       \multicolumn{1}{c}{$\cdots$}    &       \multicolumn{1}{c}{$\cdots$}    \\
G008.000$-$0.272        &       39.9    &       4       &       11.8    &       1       &       9.13    &       1.60    &       4.373   &       21.74   &       22.04   &       \multicolumn{1}{c}{$\cdots$}    &       \multicolumn{1}{c}{$\cdots$}    \\
G008.003$-$0.507        &       130.4   &       4       &       8.4     &       Tangent &       6.41    &       1.00    &       3.560   &       21.65   &       21.78   &       \multicolumn{1}{c}{$\cdots$}    &       \multicolumn{1}{c}{$\cdots$}    \\
G008.255+0.165  &       18.8    &       4       &       3.0     &       2, 3    &       2.25    &       0.38    &       3.090   &       21.59   &       21.93   &       \multicolumn{1}{c}{$\cdots$}    &       \multicolumn{1}{c}{$\cdots$}    \\
\hline
\end{tabular}\\
Velocity references: (1) \citet{purcell2012}, (2) \citet{jackson2013}, (3)
\citet{wienen2012}, (4) \citet{dunham2011}, (5) \citet{shirley2013}, (6)
\citet{bronfman1996}, (7) \citet{csengeri2015a}, (8) \citet{urquhart2011}, (9)
\citet{urquhart_13co_north}, (10) \citet{dempsey2013}, (11) {\color{black}
Wienen et al. submitted}, (12) \citet{2016MNRAS.456.2885R}, (13)
\citet{urquhart_13co_south}, (14) \citet{jackson2008}, (15) {\color{black}
Wyrowski et al. in prep.}, (16) \citet{urquhart2014a}, (17) {\color{black}
Urquhart et al. in prep.}\\
Distance references: (1) \citet{wienen2015a}, (2) \citet{ellsworth2013}, (3)
\citet{dunham2011}, (4) {\color{black} Urquhart et al. in prep.}, (5)
\citet{urquhart2014c}, (6) \citet{roman2009}, (7) \citet{battisti2014}, (8)
\citet{urquhart2014a}, (9) \citet{davies2012}, (10) \citet{reid2014}.\\
Notes: Only a small portion of the data is provided here,
the full table is only  available in electronic form at the CDS via anonymous ftp to
cdsarc.u-strasbg.fr (130.79.125.5) or via http://cdsweb.u-strasbg.fr/cgi-bin/qcat?J/A+A/.
\end{minipage}

\end{center}
\end{table*}
\setlength{\tabcolsep}{6pt}

 The primary motivation for this paper is to produce a representative
catalogue of filamentary structures and therefore the rest of the paper we will
focus exclusively on the 517 of these structures identified in the previous
section; for brevity we will refer to these as filaments but note that these may
be physically distinct structures from other ``filaments'' discussed in the
literature \footnote{ The full catalogue and the corresponding data files are available
at \url{http://atlasgal.mpifr-bonn.mpg.de/cgi-bin/ATLASGAL_FILAMENTS.cgi}. }.
Several examples of filaments are presented in Fig.\,\ref{fig:filaments}. We do
not consider the networks of filaments as they are rather more complicated and
their properties are more difficult to determine and somewhat less reliable. In Table\,\ref{tbl:cattable} we present the structure names and physical properties derived in this section. Of these, 130
filaments were agreed on by three reviewers, 160 were agreed on by 4 reviewers,
and 227 filaments were unanimously agreed upon. These can be considered as a
crude confidence flag and so we include this information in
Table\,\ref{tbl:cattable} where the integer values of 3, 4 and 5 indicate the
level of agreement for the filaments with a 5 considered to have the highest
reliability.

 Given the sensitivity and limited spatial resolution of the ATLASGAL survey
this catalogue of filamentary structures is unlikely to be complete. However,
given the unbiased nature of the ATLASGAL survey and the systematic way the
filaments have been identified this sample is likely to be representative of a
particular size and mass not previously well studied in literature. This sample therefore
represents a significant advance on what was previously known (see
Sect.\,\ref{sect:gmf} for a thorough discussion.)



\subsection{Integrated flux distribution}

\begin{figure}
\centering
\includegraphics[width = 0.49 \textwidth]{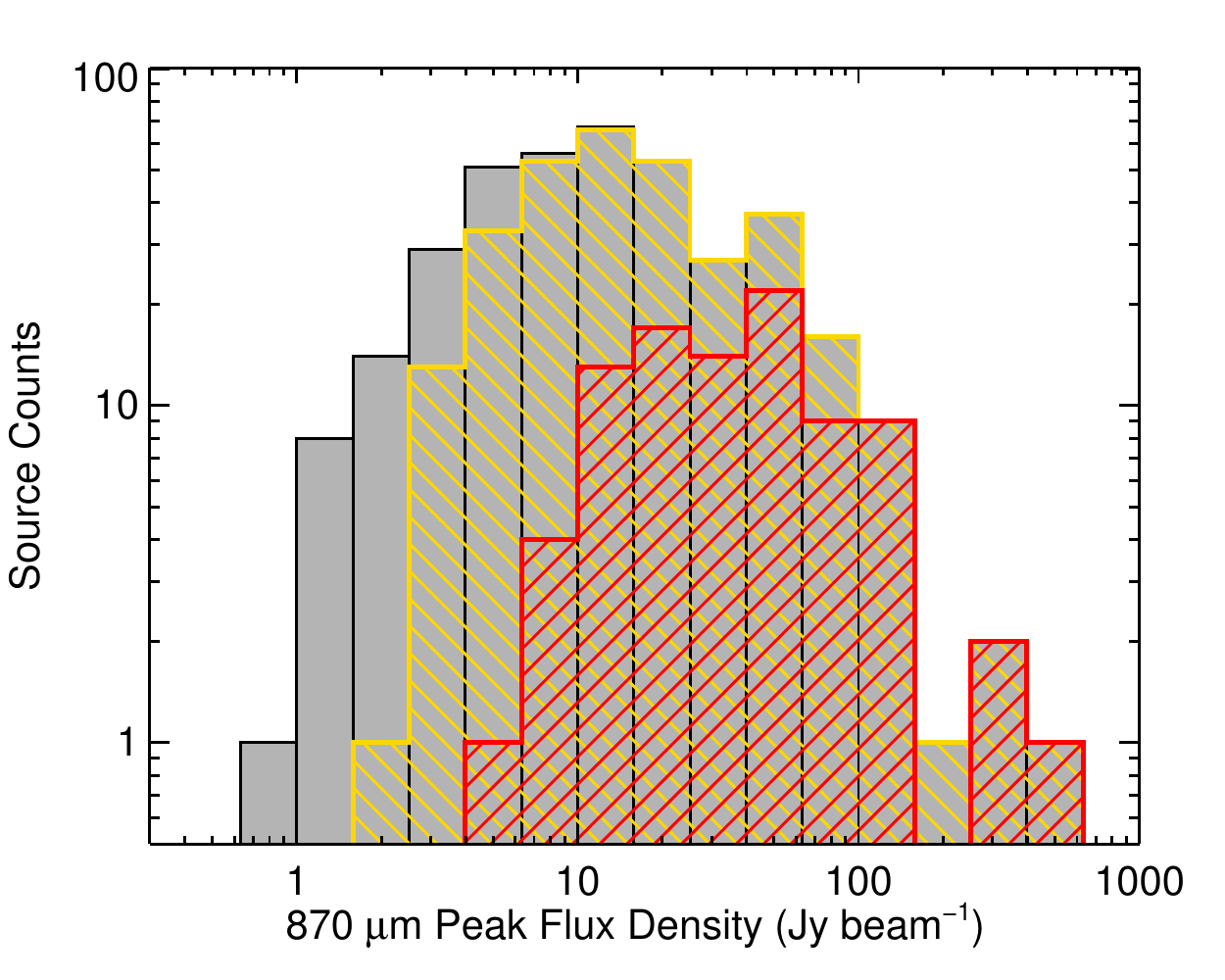} \\
\caption{Integrated flux distribution of all filaments is shown
by the grey histogram while the yellow hatched histogram shows only those
associated with one or more ATLASGAL CSC sources. The red hatched histogram shows the
flux distribution of the filaments that are associated with massive star forming clumps.}
\label{fig:filaments_flux_dist}
\end{figure}

The filaments are associated with 1495 CSC sources with an average of
three clumps being associated with each filament. There are, however, some 96
filaments that do not have any counterparts in the CSC. In
Fig.\,\ref{fig:filaments_flux_dist} we present the integrated flux distribution
for the whole filament sample (grey histogram) and those associated with one or
more ATLASGAL CSC sources (yellow hatched histogram). Is it clear from this plot that
the filaments that are not associated with any compact sources have
significantly lower fluxes. It is likely these have only become detectable in the
smoothed maps used with \Disperse\ to identify large structures.

A consequence of this is that there is little complementary data
available for these weaker filament. However, these may be more distant
examples of these types of structures or may simply be lower density nearby
filaments. The  mean aspect ratios of the filaments associated and unassociated
with CSC sources is not significantly different. However, the latter are associated with
approximately half as many spine pixels and so are generally smaller, which is
consistent with what we would expect if they are more distant examples of the
same type of structures. If this is indeed the case then the analysis of the
421 filaments that are associated with CSC sources should produce statistically
robust results, as these structures are expected to be representative of
the whole sample.

\subsection{Radial velocities}
\label{sect:vlsr}

In the following analysis we will only consider the sample of 421 filaments
associated with ATLASGAL clumps as velocities and distances are available for a
large fraction of these from both our own programmes of follow-up molecular line
observations (\citealt{wienen2012,giannetti2014,wienen2015a,2015arXiv151105138C}), and
from the results of other large-scale line surveys previously mentioned. These
observations are directed towards the peak intensity region found towards the

 of the dense clumps.
This data is heterogeneous as transitions from various molecules with different
critical densities are used to determine the velocities. However, a large
proportion have high critical densities such as CS\,(2-1) and NH$_3$\,(1,1) that
are able to unambiguously identify the correct radial velocities of the clumps;
when available they are preferred over CO measurements that are also sensitive
to low density diffuse clouds, which can lead to multiple velocity components
being detected along the line of sight.

We have matched the results of these line observations with compact sources
identified in the CSC and by association we are able to obtain velocity
measurements to 310 filaments. Furthermore, since many of the filaments (230)
are associated with multiple clumps with independent velocity measurements we
are able to distinguish between filaments that have a coherent velocity
 from those that are likely the result from blending of different
structures along the line of sight. Looking at these velocity measurements we
find there are 31 structures with a standard deviation $>$ 10\,\kms, the
majority of these are likely to be chance alignment and can be rejected from the
catalogue. Although some of these filaments may be
broadly coherent in velocity with relatively little line of sight contamination,
we prefer to be conservative and exclude them. It is likely that there are still
some structures in the sample that are the result of chance alignments,
however, statistical analysis suggests that this is likely to be a relatively small
fraction. Poisson counting statistics suggests this number is
$\sim$11 or 4\% of the remaining sample of 279 filaments for which we have a
velocity, which means that they are unlikely to have a
significant impact on our analysis.

\begin{figure*}
\centering
\includegraphics[width = 0.99\linewidth,  trim= 0.5cm 0.5cm 1cm 0.5cm, clip]{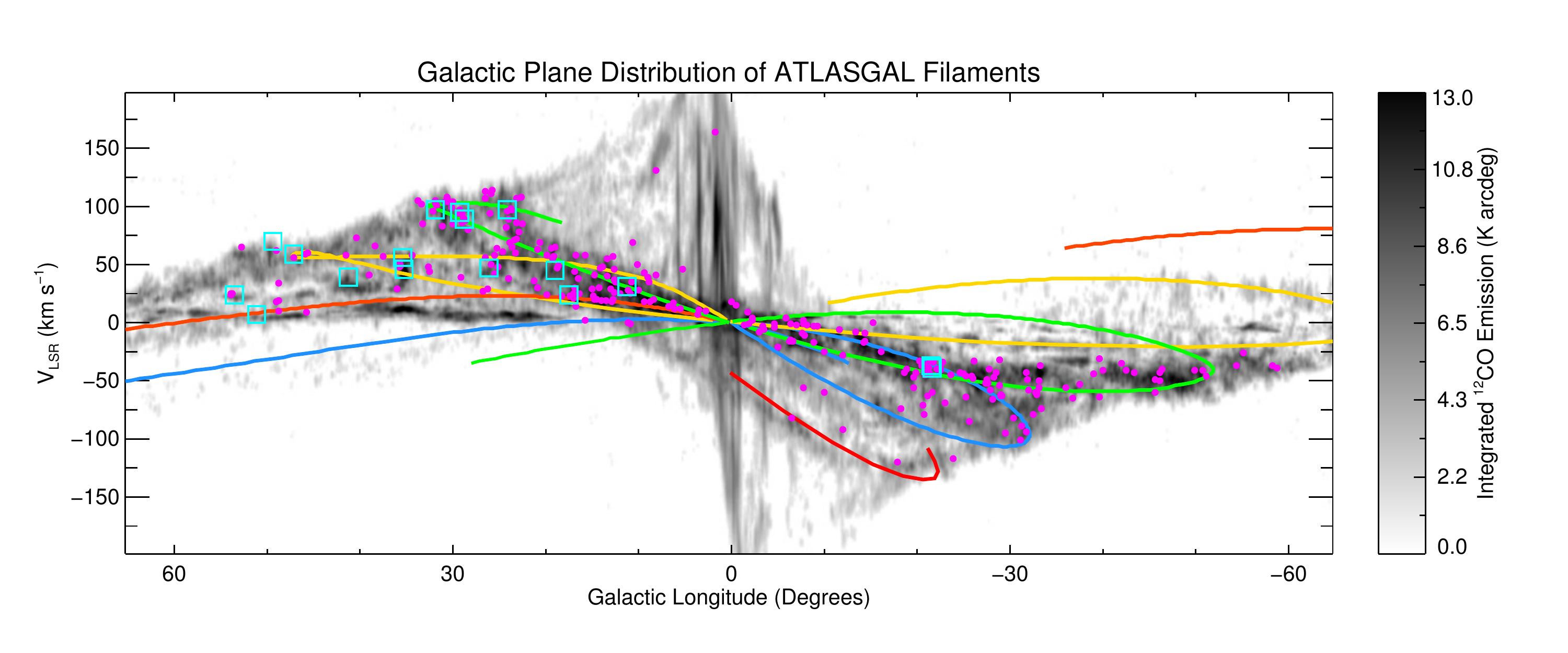} \\
\caption{Galactic longitude-velocity distribution of all filaments. The background image shows the
distribution of molecular gas as traced by the integrated $^{12}$CO\,(1-0) emission
(\citealt{dame2001}); the colour bar on the right shows the relative intensity of the emission.
The magenta circles mark the positions of the ATLASGAL filaments while the cyan squares show the
positions of other filamentary structures reported in the literatures (see
Section\,\ref{sect:gmf} for details).
The coloured lines trace the location of the spiral arms taken from the model by
\citet{tayor1993} and updated by \citet{cordes2004}.
The blue, green, yellow and orange curves trace the Norma, Scutum-Centaurus,
Sagittarius and Perseus arms, which are the four main spiral arms, while the red
curve tracers the Near 3-kpc expanding arm (\citealt{bronfman2000}).}
\label{fig:lv_distribution}
\end{figure*}

In Fig.\,\ref{fig:lv_distribution} we present a plot of the Galactic longitude
and velocity distribution of the filaments overlaid on the large-scale
distribution of molecular gas as traced by the integrated $^{12}$CO map of
\citet{dame2001}, along with the positions of the four main spiral arms as
determined by \citet{tayor1993} and the 3-kpc expanding arm from
\citet{bronfman2000}.
The positions of the filaments appear to be tightly correlated with the loci of
the spiral arms. For comparison we overlay a number of large-scale filamentary
structures reported in the literature on Fig.\,\ref{fig:lv_distribution}. We
will investigate this correlation in more detail in
Section\,\ref{sect:correlation_spiral_arms}.

There are 111 filaments for which we have no velocity information. Of these, 54
 are located towards the Galactic Centre (i.e. $350\degr < \ell <
10\degr$), which is an area that many molecular line studies have avoided,
partly due to complexity of emission in this region, and partly because
kinematically derived distances are unreliable. Of the remaining filaments all but 8
are located in the southern Galactic plane where the coverage by molecular line
surveys is significantly poorer. Of the 315 filaments located outside the
Galactic centre region we have assigned velocities to 258 ($\sim$82\%).

\subsection{Distances}

After excluding the filaments with a velocity dispersion greater than 10\,\kms\
our sample is reduced to 279 filaments with available velocity information. In
order to estimate values of the physical properties such as masses and sizes of
the filaments we first need to determine their distances. There has been a number of
 recent dedicated Galactic plane survey datasets that have focused on
determining distances to star forming regions (e.g. ATLASGAL ---
\citealt[][Urquhart et al. 2016, in prep.]{wienen2015a}; the Red MSX Survey
(RMS) --- \citealt{urquhart2014a}; Bolocam Galactic Plane Survey (BGPS) ---
\citealt{dunham2011, bowers2015, battisti2014}; \hii\ region Discovery Survey
(HRDS) --- \citealt{bania2010} and the Galactic Ring Survey (GRS;
\citealt{jackson2006}) --- \citealt{roman2009}). The majority of these use the
radial velocities of sources, obtained from molecular line observations, with a
Galactic rotation curve to estimate the kinematic distances. For sources located
within the Solar Circle the rotation curves return two possible distances,
equally spaced on either side of the tangent position; they are commonly
referred to as the \emph{near} and \emph{far} kinematic distances. This distance
ambiguity is normally resolved by comparing the source velocity with the \hi\
profile taken towards the source; this technique is known as the \hi\
self-absorption method (for a more detailed discussion of this method see
\citealt{jackson2003} and \citealt{roman2009}). This is not the only method but
it is the most commonly used and the only one relevant for the studies we have
used to assign distances to our sample of filaments.

\begin{figure}
\centering
\includegraphics[width = 0.49 \textwidth]{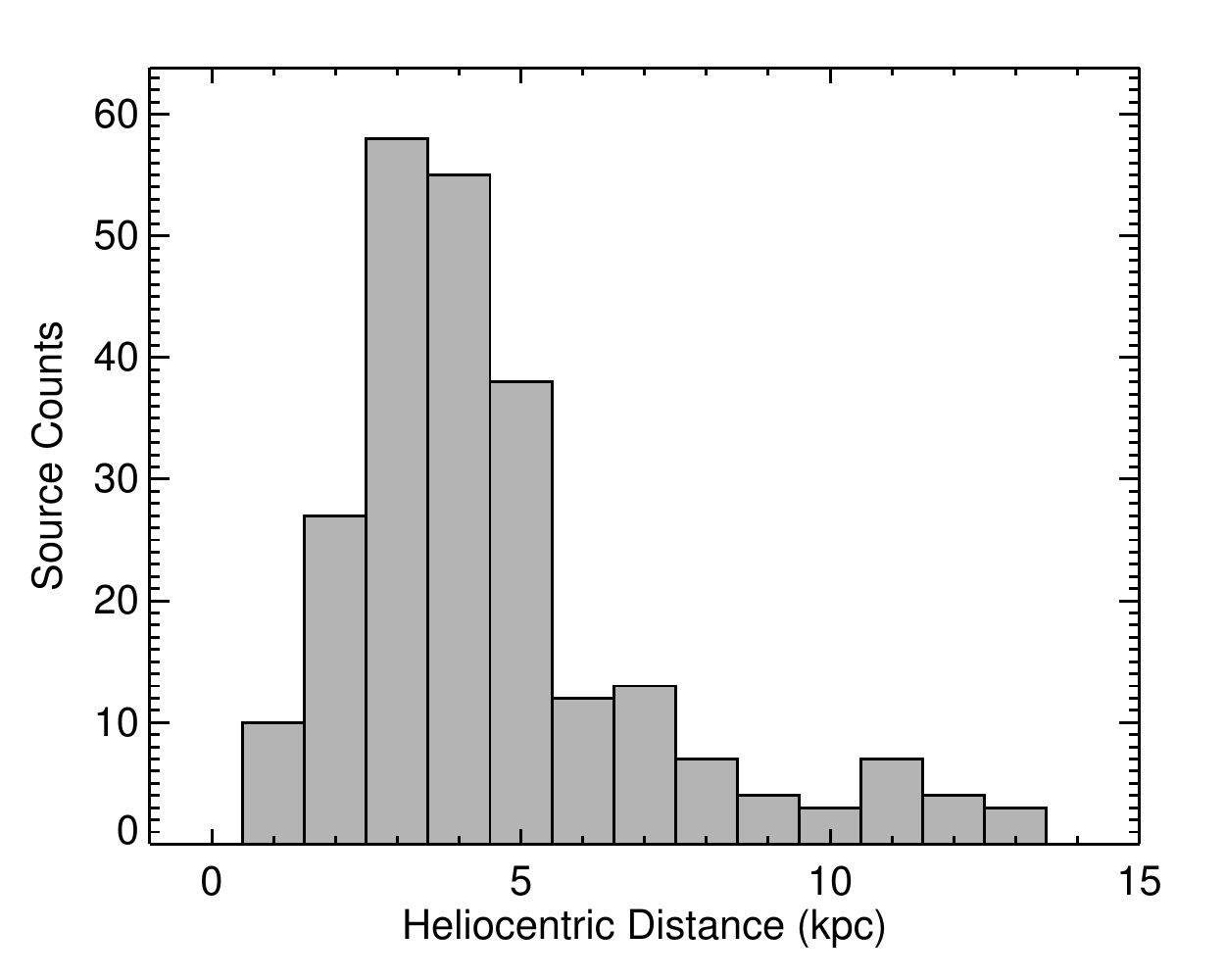} \\
\caption{Distance distribution. The bin size used is 1\,kpc.}
\label{fig:distance_dist}
\end{figure}

We estimate the systemic velocity from the average velocity of their associated clumps and use this to determine the near
and far distances using the \citet{brand1993} rotation model. We then resolve
any distance ambiguities using the information given in the literature. We do
not assign kinematic distances to structures with radial velocities close to the
solar velocity (i.e. |\vlsr| $<$ 10\,\kms) as they are unreliable; this
affects {30} filaments, however, maser parallax measurements are
available for 4 of these filaments and reliable distances for another 2 of these
can be found in the literature. We have been able to determine
distances to 241 filaments from the various \hi\ studies; the assigned distance
and literature reference are given in Cols.\,4 and 5 of Table\,\ref{tbl:atlasgal_filaments}. In
Fig.\,\ref{fig:distance_dist} we present a histogram showing the distribution of
heliocentric distances.

\begin{figure}
\begin{center}

\includegraphics[width=0.49\textwidth, trim= 20 10 20 20] {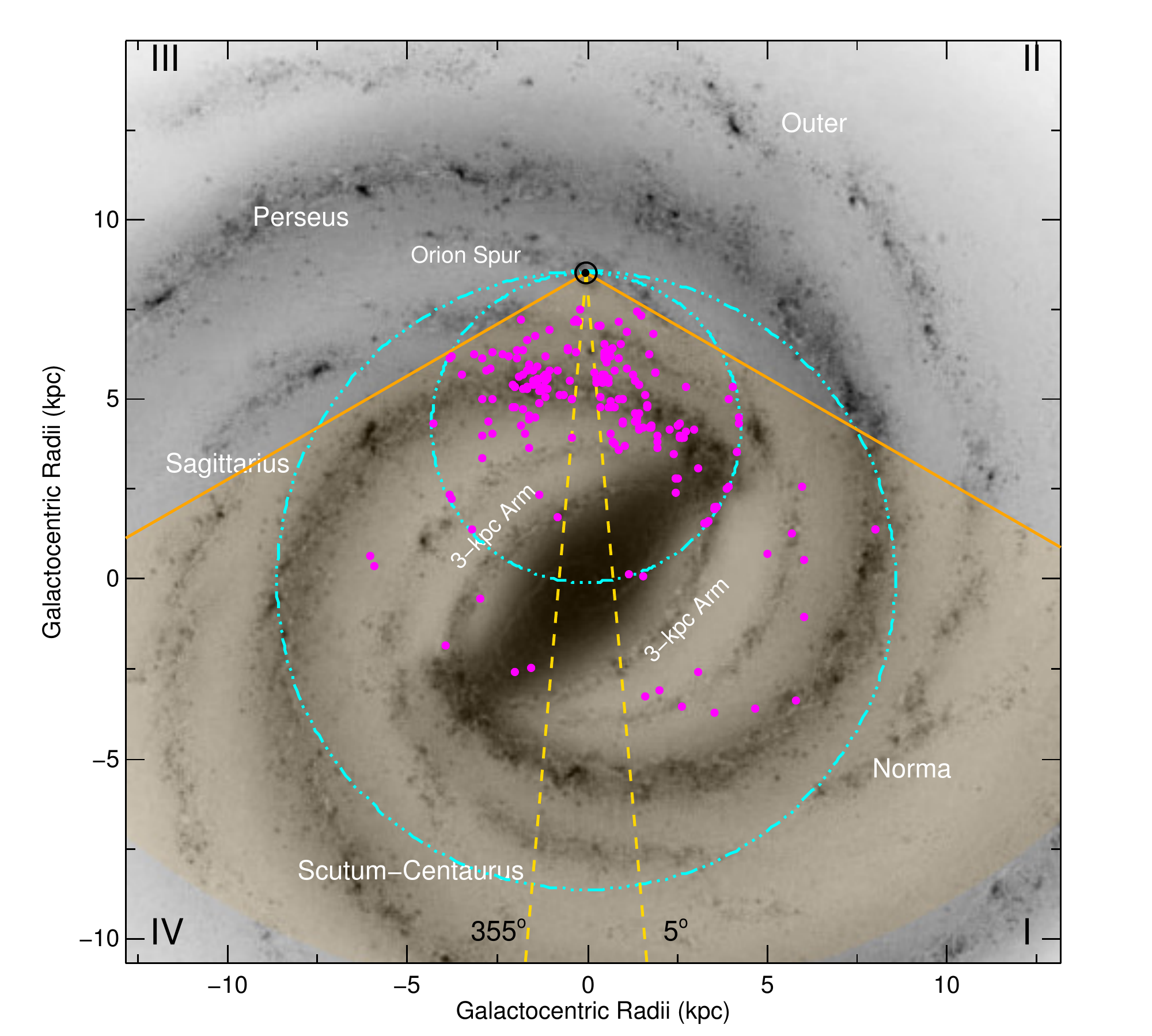}

\caption{Galactic distribution of all ATLASGAL filaments for which a distance
has been determined.
The orange shaded area indicates the region of the Galactic plane covered by the ATLASGAL
survey to a distance of 20\,kpc, within which the survey is complete for compact clumps
with masses $>$1000\,\msun. The background image is a schematic of the Galactic disk as viewed
from the Northern Galactic Pole (courtesy of NASA/JPL-Caltech/R. Hurt (SSC/Caltech)).
The Sun is located at the apex of the wedge and is indicated by the $\odot$ symbol.
The smaller of the two cyan dot-dashed circles represent the locus of tangent points,
while the larger circle traces the Solar Circle. The spiral arms are labelled in white and
Galactic quadrants are given by the roman numerals in the corners of the image.
The yellow lines shows the innermost region towards the Galactic centre where distances are not reliable.}
\label{fig:galactic_distribution}
\end{center}
\end{figure}

In Fig.\,\ref{fig:galactic_distribution} we show the Galactic distribution of the
filaments overlaid on a schematic diagram of the Milky Way that includes many of
the key elements of Galactic structure, such as the location of the spiral arms
and the Galactic long and short bars (\citealt{churchwell2009}). Comparing the
distribution of the filaments with the spiral arms it is clear that the majority
are associated with the near side of the Scutum-Centaurus arm that
fills a large proportion of inner Galactic plane and has a heliocentric distance
of $\sim$4\,kpc. There is reasonable correlation between the filaments not
associated with the Scutum-Centaurus arm and the other arms, however, the
sampling is too sparse to draw any conclusions from this correlation. It is a
little surprising that there are so few associations with the Sagittarius arm
since it is the closest arm to us and covers a larger $\ell$ range. This is
likely to be related to the removal to large angular scales in the data reduction process, decreasing our sensitivity to
nearby extended low-density structures (as discussed in
Section\,\ref{sect:source_extraction}).
 We also note that while this spiral arm covers a large portion of the sky it falls
outside the dense ring of material found towards the inner part of the Galaxy
(i.e. the 4-5\,kpc ring where $\sim$70\%
of the molecular gas in the Galaxy is found; \citealt{jackson2006})
and therefore this part of the Sagittarius arm has a lower mass surface
density so perhaps we might expect to find few filaments anyway.
We will investigate the Galactic distribution in more detail in
Section\,\ref{sect:correlation_spiral_arms}.

\subsection{Lengths and widths}
\label{sec:sizes}

\begin{figure}
\centering
\includegraphics[width = 0.49 \textwidth]{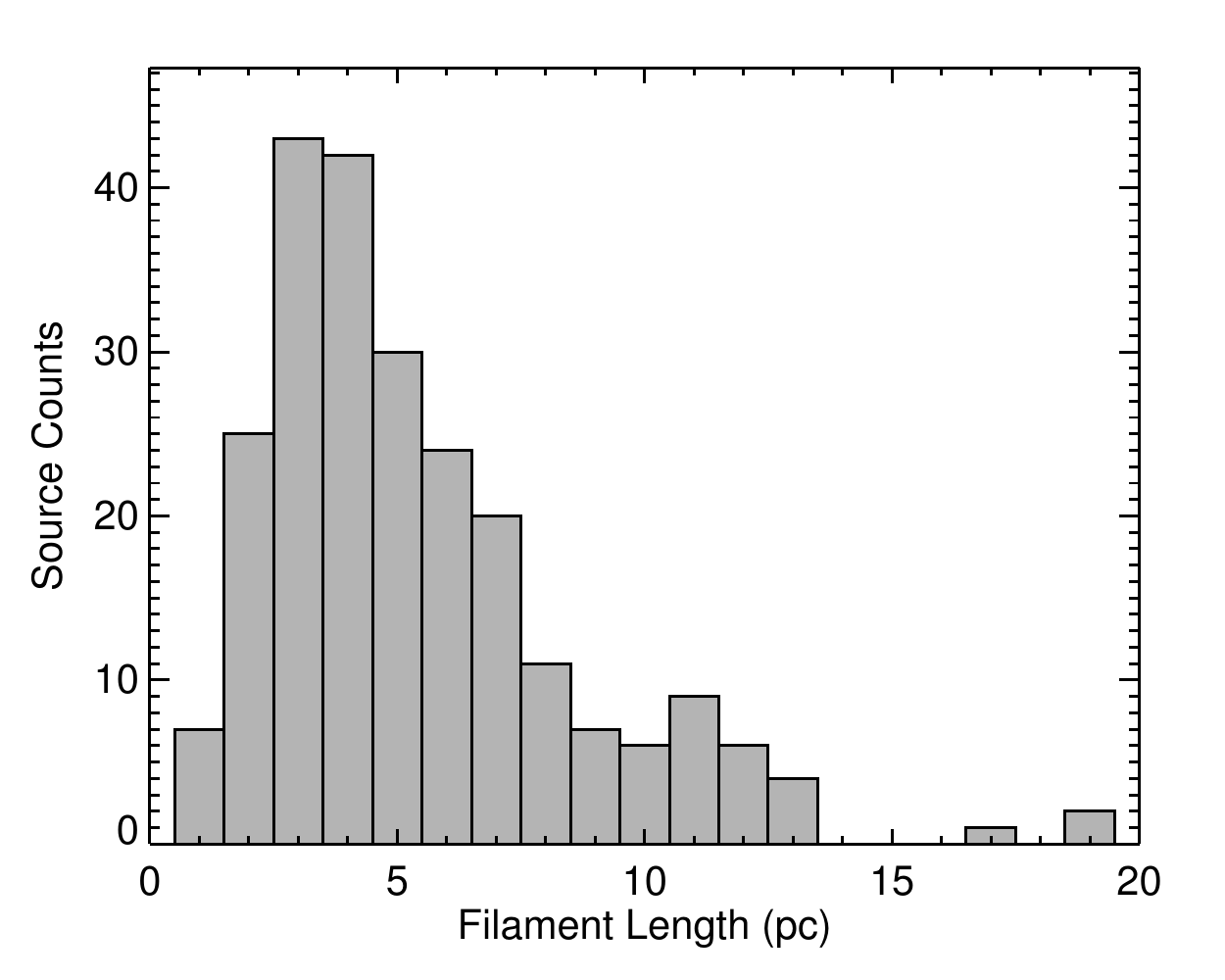} \\
\caption{Distribution of filament lengths as estimated from the angular size of the major axis. The bin size used is 1\,pc. }
\label{fig:size}
\end{figure}

There are two possible methods we can use to estimate the length of a
filament: the number of points that form the source skeleton or the length of
the major axis. The former is likely to overestimate the length for structures
 with numerous sub-branches, while the latter is likely to underestimate the true
length as it makes no allowance for any curvature that might be present in the
source structure. We have opted to use the major axis as a measure of the
physical length of filaments as this is easy to define and the results
relatively easy to reproduce. Furthermore, although sizes are likely to be
underestimated this is likely to be less of a problem for structures with
large aspect ratios. In Fig.\,\ref{fig:size} we show the length distribution of the
filaments and give the statistical breakdown in Table\,\ref{tbl:derived_para}.
Considering only filaments with an aspect ratio greater than 3 we estimate the
range of lengths to be between $\sim$2 and 20\,pc with a median value of
3.8\,pc. This is similar to the lengths reported in a recent
\emph{Herschel}-HIGAL study of the $\ell= 216.5-225.5\degr$ region presented by
\citet{schisano2014}. The lengths
of many filaments in our sample are significantly
smaller than some of the large-scale filaments that have recently been reported
in the literature (e.g. \citealt{ragan2014,li2013,jackson2010}), which have
lengths of several tens to hundreds of parsecs (this will be discussed in detail
in Section\,\ref{sect:gmf}).

\begin{figure}
\centering
\includegraphics[width = 0.49 \textwidth]{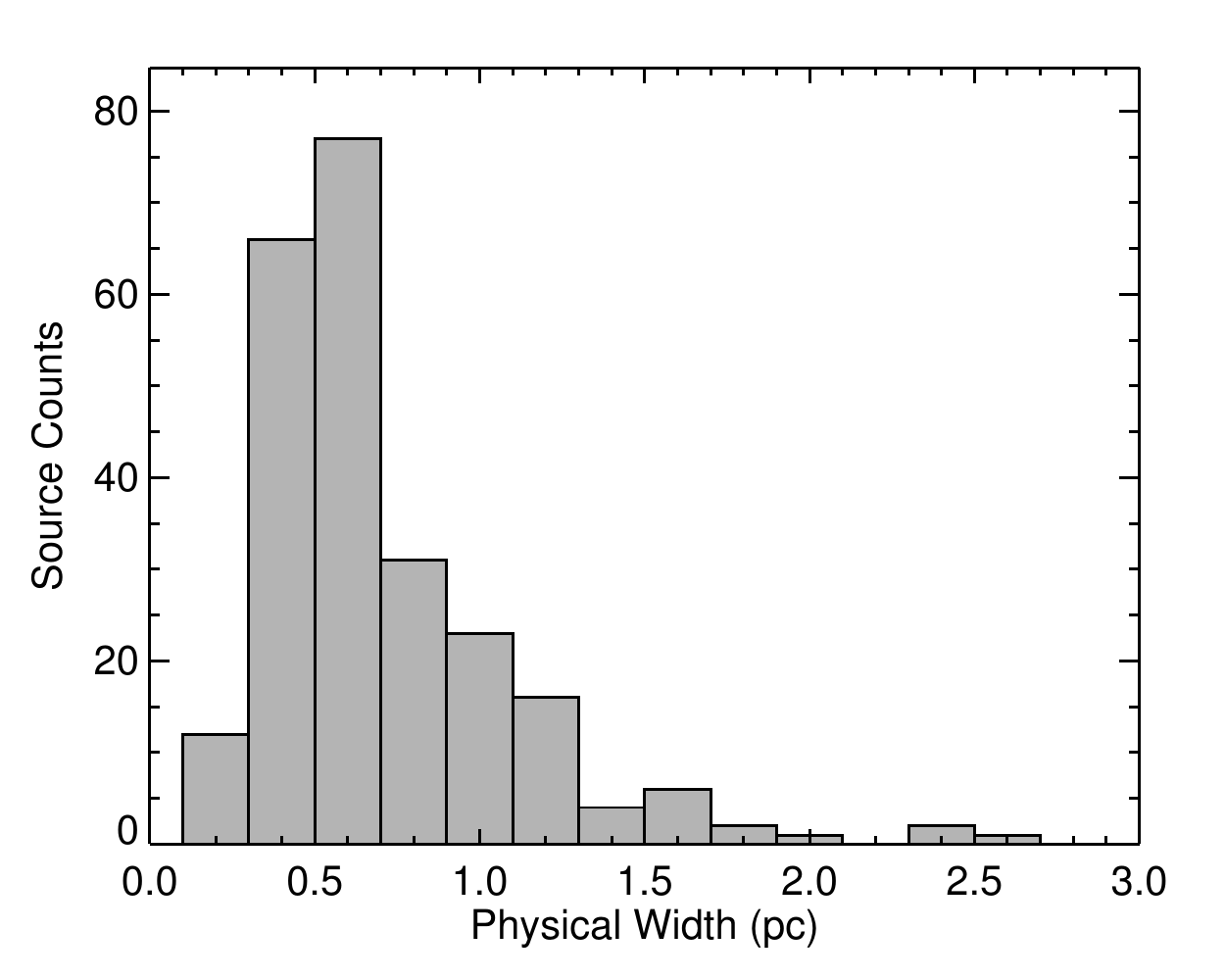} \\
\caption{Width distribution for 241 resolved filaments. The widths are determined from the
original resolution maps, which have a resolution of 19.2\arcsec\ derived from
the APEX beam.
The bin size used is 0.2\,pc.}
\label{fig:width_dist}
\end{figure}

In Fig.\,\ref{fig:width_dist} we show the distribution of widths for all
resolved filaments. The median width is $\sim$0.5\,pc, which is significantly
larger than the widths measured from the \emph{Herschel} Gould Belt Survey
(\citealt{andre2014}) where filament  inner widths of $\sim$0.1\,pc are
typically reported for nearby clouds (e.g.
\citealt{arzoumanian2011,2012A&A...548L...6H}),  and  we have made no attempt to
 separate  out the contribution from a possible power-low component
\citep{2013A&A...550A..38P}.
 Our measurements are in
line with what has been found for other filaments found towards the inner Galaxy,
(e.g. `Nessie' has a width of 0.5\,pc --- \citealt{jackson2010}; and the
`Snake' has a width of 0.8\,pc --- \citealt{wang2014}), the widths of $\sim$0.3
reported by \citealt{schisano2014} and the range reported by
\citet[][0.6-3\,pc]{wang2015} for a sample of 9 \emph{large-scale filaments}.
However, we note that the values calculated here show a distance-dependence,
i.e.
we tend to find filaments with larger widths at larger distances (cf
\citealt{schisano2014}).

We also note that while the major axis length is likely
to underestimate the length of the filaments the minor axis is likely to
overestimate their widths. The widths presented here provides an estimate of how
concentrated the emission is with regards to the skeleton, but these parameters
are not particularly well constrained by our data and should be considered to be upper
limits and used with caution.

\subsection{Masses, column densities and mass-to-length ratios}

We estimate the mass of the filaments using the total integrated flux from the
dilation operation described in Section\,\ref{sect:dilation} and following the method described by
\citet{hildebrand1983}

\begin{equation}
M_{\rm{filament}} \, = \,\frac{D^2 \, S_\nu \, R}{B_\nu(T_{\rm{dust}}) \, \kappa_\nu},
\end{equation}

\noindent where $S_\nu$ is the integrated 870\,$\mu$m flux, $D$ is the
heliocentric distance to the source, $R$ is the gas-to-dust mass ratio (assumed
to be 100), $B_\nu$ is the Planck function for a dust temperature
$T_{\rm{dust}}$, and $\kappa_\nu$ is the dust absorption coefficient taken as
1.85\,cm$^2$\,g$^{-1}$; this values was interpolated by
\citet{schuller2009_full} from Table\,1, Col.\,5 of \citet{ossenkopf1994}. In
this calculation we are assuming that the total filament mass is proportional to
the total flux density integrated over the source. In previous studies we have
used temperatures of 18\,K and 20\,K when estimating the masses (e.g.
\citealt{csengeri2014,urquhart2014c}). However, these were based on temperature
measurements made towards the peaks of the submillimetre emission, which are
often associated with star formation and therefore tend to be a little warmer
than the temperature of their extended envelope. Here we assume a dust
temperature of 15\,K, which is more typical of the temperatures reported for
IRDCs and quiescent clumps (e.g. \citealt{wienen2012,urquhart2015b}). Using this
temperature the equation simplifies to

\begin{equation}
\left(\frac{M_{\rm{filament}}}{\rm{M}_\odot}\right) \, = \,8.63 \left(\frac{D}{\rm{kpc}}\right)^2 \, \left(\frac{S_{\rm{int}}}{\rm{Jy}}\right).
\end{equation}

\begin{figure}
\centering
\includegraphics[width = 0.49 \textwidth]{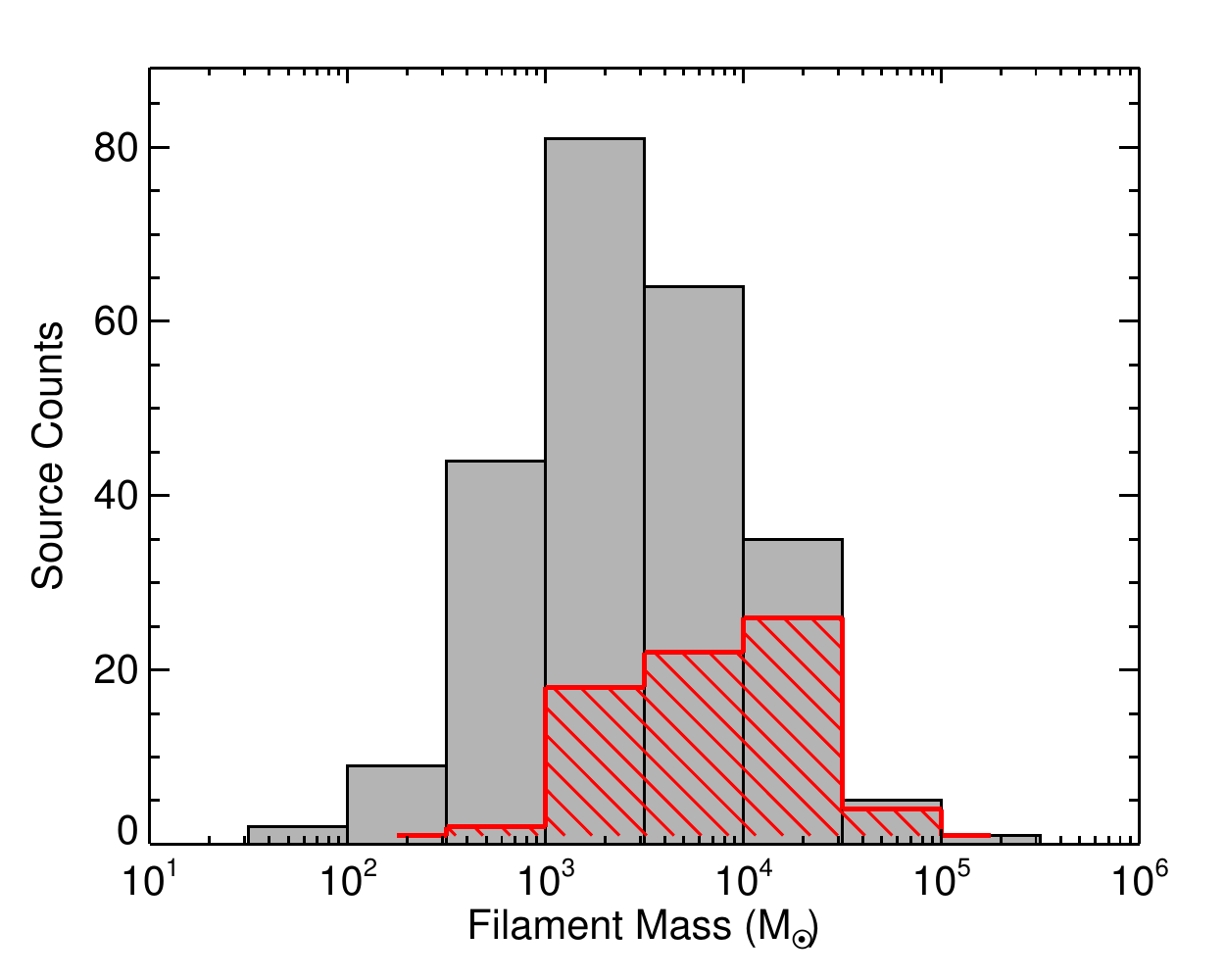} \\
\caption{The mass distribution of all filaments is shown in grey and is
overlaid with the mass distribution of filaments associated with MSF (massive
star forming) clumps in red. The bin size used is 0.5\,dex. }
\label{fig:mass_dist}
\end{figure}

In Fig.\,\ref{fig:mass_dist} we show the filament mass distribution. The
filaments have masses from $\sim$100\,\msun\ to 10$^5$\,\msun\ with typical
values of a few 1000\,\msun, which is typical of the mass reservoir often
associated with massive star formation. We also show the distribution of the
filaments that are associated with massive star formation
(Section\,\ref{sect:csc_associations}). We use the Kolmogorov-Smirnov test
(\citealt{press1992}) to determine if the mass distribution of the MSF and
non-MSF filaments are significantly different from each other. This test starts
with the null hypothesis that both samples are drawn from the same parent
population and this is rejected if this probability is less than 0.0013 (i.e.
more than 3$\sigma$); this is known as the $p$-value.
Comparing the masses are able to reject the null hypothesis with $p$-value $\ll$
0.0013 and so the difference in mass is statistically significant.

To estimate the column density, following
previous works \citep[e.g.][]{schuller2009_full,csengeri2014}, we use

\begin{equation}
N({\rm H_2}) = \frac{S_{\rm \nu}\; R}{B_{\rm \nu}(T_{\rm dust})\; \Omega\;
\kappa_{\nu}\; \mu_{\rm H_2}\; m_{\rm H}}\;,
\end{equation}

\noindent where $\Omega$ is the beam solid angle, $\mu_{\rm H_2}$ is the mean
molecular weight of the interstellar medium with respect to hydrogen molecules,
which is equal to 2.8 \citep{2008A&A...487..993K}, and $m_{\rm H}$ is the mass
of an hydrogen atom. The other parameters are as previously described. Again assuming
a dust temperature of 15\,K, the conversion between flux and column density can be expressed as

\begin{equation}\label{eq:conversion}
\left(\frac{N({\rm H_2})}{{\rm{cm}}^{-2}}\right) = 3.92\times 10^{22} \;
\frac{S_{\nu}}{\rm Jy / beam}\;.
\end{equation}

\begin{figure}
\centering
\includegraphics[width = 0.49 \textwidth]{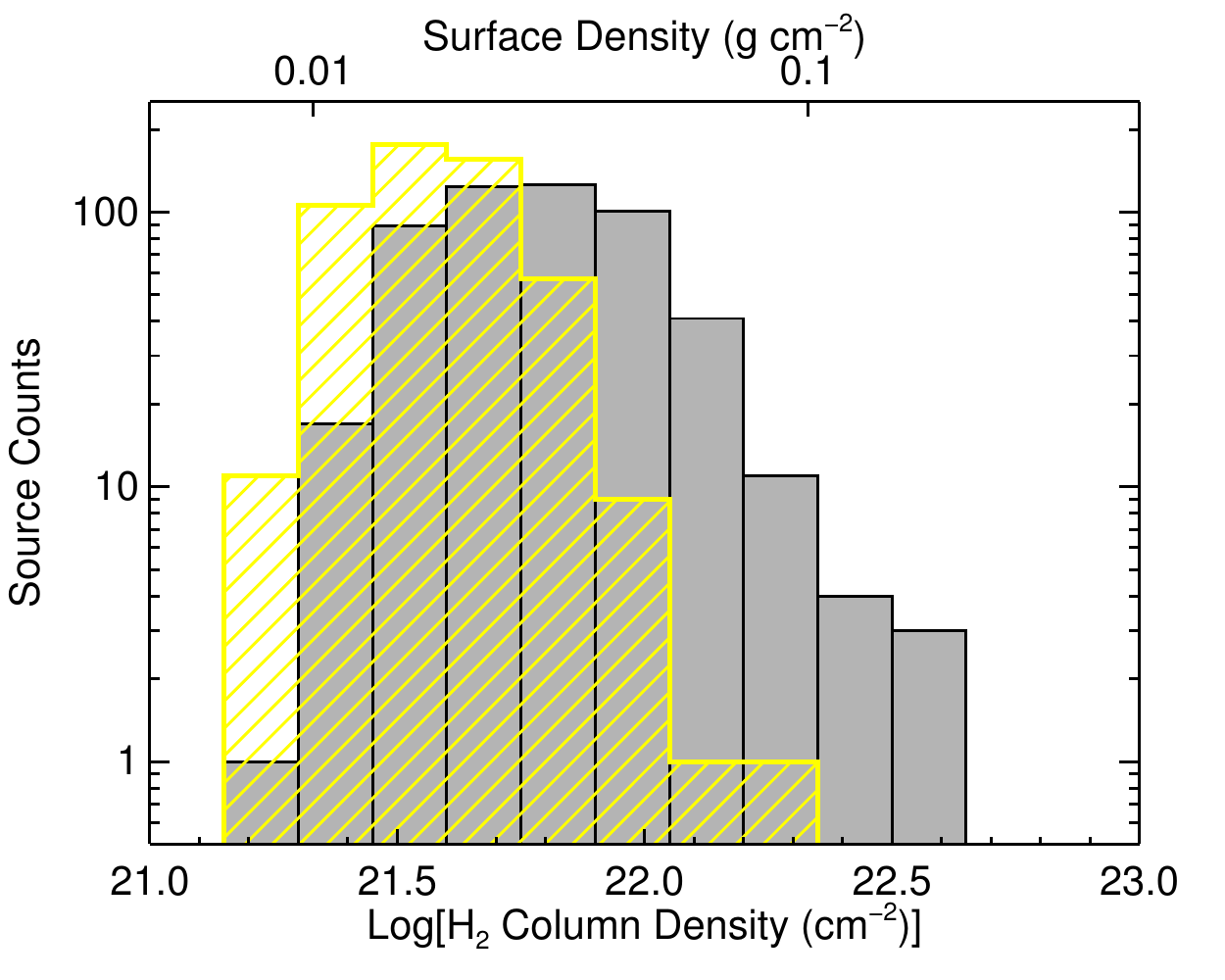} \\
\caption{Log-normal column density distribution. The filled grey histogram shows
the peak column density of the skeleton pixels while the hatched yellow histogram shows the mean column density
of all pixels associated with each filament. The bin size used is 0.15\,dex. }
\label{fig:colden_dist}
\end{figure}

In Fig.\,\ref{fig:colden_dist} we show peak and mean column density
distributions. The peak column densities of the filament are determined by
fitting a log-normal function to the distributions of column densities on pixels
that form the spine of the source skeleton; and the mean column densities of the
filaments are determined by fitting a log-normal function to the distribution of
column densities of all pixels found within the dilated skeleton (i.e.  $w =
25$\,pixel). The average column densities of the filaments range between
$\sim$2-$13\times 10^{21}$\,cm$^{-2}$ with mean value of $3.55\times
10^{21}$\,cm$^{-2}$. This is almost an order of magnitude higher than the mean
value determined for the  catalogue of filaments reported by
\citet[][1.7-4.8$\times 10^{20}$\,cm$^{-2}$]{schisano2014} and explains how they
were able to identify a similar number of filaments from a relatively small
fraction of the Galactic plane.


Allowing for a difference in temperature of $\pm$5\,K, a distance uncertainty of
$\sim$10\%, the uncertainty on the flux calibration (15\%;
\citealt{schuller2009_full}) and the fact that the dust to gas ratio and
$\kappa_{\nu}$ are poorly constrained we estimate the uncertainties for the mass and column density are likely to be a factor of a few. However, these uncertainties are
unlikely to have a significant impact on the overall distribution or the
statistical analysis of these parameters as they are systematic and affect all
structures equally.

\begin{figure}
\begin{center}
\includegraphics[width = 0.49 \textwidth]{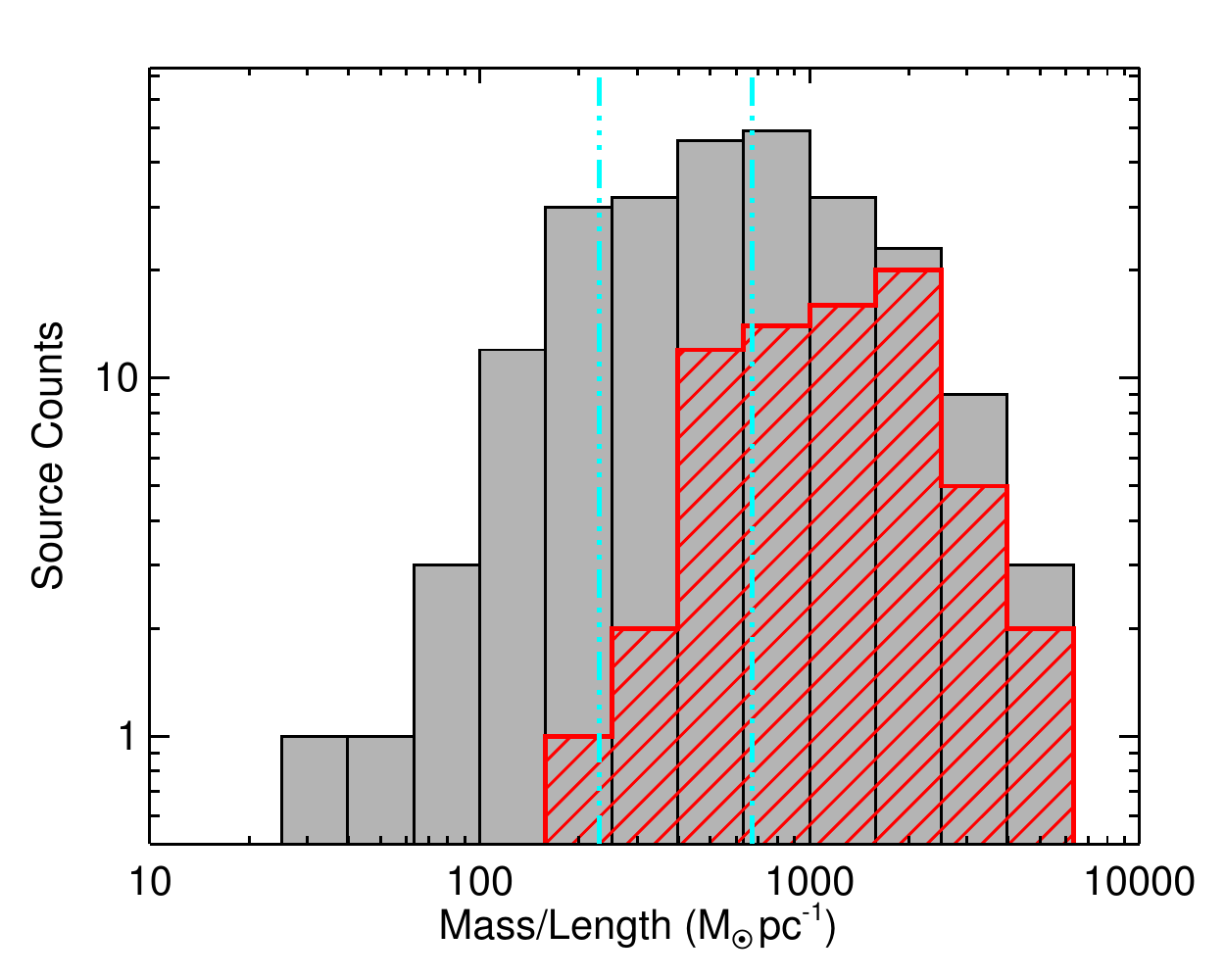}
\includegraphics[width = 0.49 \textwidth]{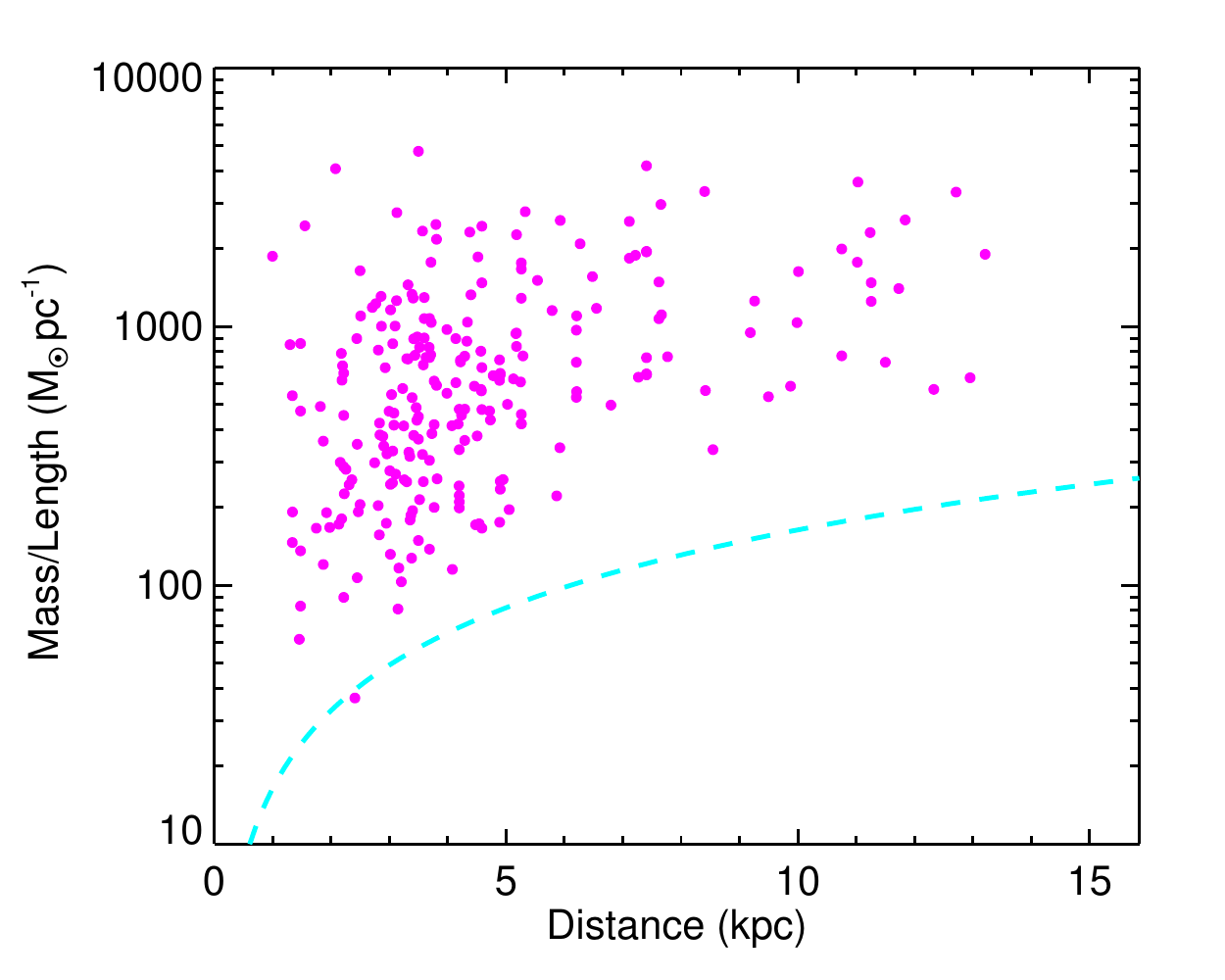}

\caption{Upper panel: Mass-to-length ($M/L$) ratio for all filaments and for those associated
with massive star forming clumps are shown as grey and red hatched histograms,
respectively. The vertical cyan lines show the critical $M/L$ thresholds for filaments with
line-widths of 0.7 and 1.2\,\kms; this corresponds to masses per unit length of 230 and 670\,\msun\,pc$^{-1}$, respectively. Filaments above these thresholds are unstable to radial collapse (see Sect.\,\ref{star_formation_engines} for details). The bin size is 0.2\,dex. Lower panel: The mass-to-length ratio is plotted as a function of heliocentric distance; the dashed cyan curve in this plot indicates the sensitivity of the ATLASGAL survey.}
\label{fig:ml_plot}
\end{center}
\end{figure}

In the upper and lower panels of Fig.\,\ref{fig:ml_plot} we plot the
distribution of mass-to-length ratio of the filaments and this ratio as a
function of distance. The majority of the filaments have mass-to-length ratios
of  $M/L \sim$\,200-2000 \ \msun\,pc$^{-1}$ and is relatively independent of
distance.

\begin{table*}

\begin{center}\caption{Statistical properties of the filaments.}
\label{tbl:derived_para}
\begin{minipage}{\linewidth}
\small
\begin{tabular}{l.......}
\hline \hline
  \multicolumn{1}{l}{Parameter}&  \multicolumn{1}{c}{Number}&	\multicolumn{1}{c}{Mean}  &	\multicolumn{1}{c}{Standard Error} &\multicolumn{1}{c}{Standard Deviation} &	\multicolumn{1}{c}{Median} & \multicolumn{1}{c}{Min}& \multicolumn{1}{c}{Max}\\
\hline
Distance (kpc) &          173&4.73&0.20 & 2.67 & 3.72 & 1.00 & 12.95\\
Length (pc) &          241&5.72&0.29 & 4.44 & 4.49 & 1.01 & 35.22\\
FWHM Width (pc) &     241&0.66&0.03 & 0.39 & 0.54 & 0.13 & 2.51\\
Mass (\msun) &          241&6137.93&685.31 & 10638.86 & 2736.38 & 62.52 &
116603.13\\
Log[Peak H$_2$ Column Density (cm$^{-2}$)] &           517&21.79&0.01 & 0.22 & 21.77 & 21.30 & 22.56\\
Log[Mean H$_2$ Column Density (cm$^{-2}$)] &              517&21.57&0.01 & 0.15 & 21.57 & 21.24 & 22.22\\
Log[RMS Integrated Luminosity (\lsun)] &           89&4.53&3.76 & 4.74 & 4.15 &
3.04 & 5.46\\
$M/L$ All Filaments (\msun\,pc$^{-1}$) &          241&851.85&51.50 & 799.47 &
611.27 & 36.72 & 4761.01\\
$M/L$ MSF Filaments (\msun\,pc$^{-1}$) &
72&1424.13&108.67 & 922.11 & 1288.36 & 214.59 & 4761.01\\
\hline\\
\end{tabular}\\

\end{minipage}

\end{center}
\end{table*}

\section{Discussion}
\label{sect:discussion}

\subsection{Filamentary structures in the Galaxy}
\label{sect:gmf}

\setlength{\tabcolsep}{6pt}

\begin{table*}


\begin{center}
\caption{Properties of giant molecular filaments.
This table combines many of the parameters given in Table\,2 and 3 presented by
\citet{ragan2014} and includes a few similar structures reported in the
literature. We identify structures that are coincident with filaments reported
in this paper by appending a $\star$ to the literature name in Col.\,1.
\label{tbl:literature_filaments}}
\begin{minipage}{\linewidth}
\small
\centering
\begin{tabular}{l......ccc}
\hline \hline
\multicolumn{1}{c}{Literature name}&  \multicolumn{1}{c}{$\ell$}& \multicolumn{1}{c}{$b$} &\multicolumn{1}{c}{\vlsr} & \multicolumn{1}{c}{Distance} &	\multicolumn{1}{c}{Length}  &\multicolumn{1}{c}{Log(Total Mass)}  &\multicolumn{1}{c}{Associations}  &\multicolumn{1}{c}{Arm} &\multicolumn{1}{c}{References} \\

\multicolumn{1}{c}{}&  \multicolumn{1}{c}{(\degr)}&\multicolumn{1}{c}{(\degr)}&\multicolumn{1}{c}{(\kms)}&\multicolumn{1}{c}{(kpc)}&	\multicolumn{1}{c}{(pc)}   &\multicolumn{1}{c}{(\msun)}  &\multicolumn{1}{c}{}&\multicolumn{1}{c}{}&\multicolumn{1}{c}{}  \\


\hline

GMF 18.0$-$16.8$\star$	&	17.3	&	0.6	&	23.0	&	2.2	&	88	&	5.18	&	M16, W37 &	Sag.	&	1	\\
GMF 20.0$-$17.9$\star$	&	18.9	&	-3.3	&	43.5	&	3.4	&	170	&	5.60	&	 W39	& Scu. &	1	\\
GMF 26.7$-$25.4	&	26.0	&	1.4	&	46.0	&	3.1	&	123	&	5.30	& \multicolumn{1}{c}{$\cdots$}	& Scu.	&	1	\\
GMF 38.1$-$32.4a$\star$	&	35.2	&	0.1	&	55.0	&	3.5	&	234	&	5.85	&	W44	&\multicolumn{1}{c}{$\cdots$}	& 1	\\
GMF 38.1$-$32.4b$\star$	&	35.1	&	-0.4	&	44.5	&	2.9	&	79	&	4.89	&	\multicolumn{1}{c}{$\cdots$}&\multicolumn{1}{c}{$\cdots$}	&	1	\\
GMF 41.0$-$41.3	&	41.1	&	-0.0	&	38.0	&	2.7	&	51	&	4.69	&	\multicolumn{1}{c}{$\cdots$} & \multicolumn{1}{c}{$\cdots$}	&	1	\\
GMF 54.0$-$52.0$\star$	&	53.4	&	0.0	&	23.0	&	2.0	&	68	&	4.83	&	W52	&\multicolumn{1}{c}{$\cdots$}	& 1	\\
 \hline
Nessie$\star$	&	338.4	&	-0.5	&	-38.0	&	3.1	&	81	&	5.00
&\multicolumn{1}{c}{$\cdots$} &	Scu.	&	2	\\
Nessie Extended$\star$	&	338.3	&	-0.5	&	-40.0	&	3.1	&	162	&	5.30	& \multicolumn{1}{c}{$\cdots$}&	Scu.	&	3	\\
G32.02+0.06$\star$	&	31.8	&	0.1	&	96.0	&	5.6	&	80	&	5.30	&	\multicolumn{1}{c}{$\cdots$} & \multicolumn{1}{c}{$\cdots$}	&	1, 4	\\
\hline
G11.11$-$0.12$\star$	&	11.2	&	-0.1	&	30.0	&	3.6	&	28	&	\multicolumn{1}{c}{$\cdots$}	&	The Snake	& \multicolumn{1}{c}{$\cdots$} &	5	\\
Molecular Wisp	&	51.0	&	0.7	&	6.2	&	9.8	&	500	&	5.00	&	G52L nebula	&Per.&	6	\\
\hline
CFG024.00$+$0.48$\star$ & 24.00 & 0.48 & 96.0 & 5.2 & 82 & 4.9 & \multicolumn{1}{c}{$\cdots$} & & 7\\
CFG028.68$-$0.28 & 28.68 & -0.28 & 88.2 & 4.89 & 60 & 4.7 & \multicolumn{1}{c}{$\cdots$} & Scu.& 7 \\
CFG029.18$-$0.34 & 29.18 & 0.34 & 93.8 & 3.5 & 99&  4.7 & \multicolumn{1}{c}{$\cdots$} & Scu.&7 \\
CFG047.06$+$0.26$\star$ & 47.06 & 0.26 & 57.5 & 4.44 & 73 & 4.2 & \multicolumn{1}{c}{$\cdots$} &Sag. &7 \\
CFG049.21$-$0.34$\star$ & 49.21 & -0.34 & 68.5 & 5.41 & 85 & 4.9 & W51 &Sag.&
7\\
\hline

BC 029.94-0.30 & 26.94 & -0.30	 &	 58 & 4.6 & 13 &	3.23 & \ldots &
Scu. & 8 \\
BC 025.24-0.45$\star$ & 25.24 & -0.45	 &	57.3  & 4.3 & 57 &	3.99 & N40 &
Scu. & 8 \\
BC 018.88-0.09 & 18.88 & -0.09	 &	46 & 3.7 & 45 &	3.90 & \ldots &
Scu. & 8 \\
BC 004.14-0.02 & 04.14 & -0.02	 &	8  & 3.1 & 37 &	4.04 & \ldots &
Scu. & 8\\
BC 335.31-0.29 & 335.31 & -0.29	 &	-42  & 3.2 & 34 &	4.00 & \ldots &
Scu. & 8\\
BC 332.21-0.04 & 332.21 & -0.04	 &	-49  & 3.3 & 52 &	4.20 & \ldots &
Scu. & 8\\

\hline
\end{tabular}\\

\end{minipage}
\end{center}
References: (1) \citet{ragan2014}, (2) \citet{jackson2010}, (3)
\citet{goodman2014}, (4) \citet{battersby2014b}, (5) \citet{wang2014}, (6)
\citet{li2013}, (7) \citet{wang2015}. (8) \citet{zucker2015}. \\
Notes: The prefix of CF on the source name of the \citet{wang2015} sample of
filaments corresponds to \emph{cold filament}. We have omitted the following
four filaments that appear in the sample presented by \citet{wang2015}:
CFG026.38$+$0.79 is the dense part of the larger structure GMF 26.7$-$25.4
identified by \citet{ragan2014} and is therefore not considered to be an
independent entity; CFG064.27$-$0.42 is located outside the region of interest
for this study; CFG011.11$-$0.12 and CFG338.47$-$0.43 are part of the Snake and
Nessie filament for which more detailed studies are available.

\end{table*}
\setlength{\tabcolsep}{6pt}

The masses and lengths of the filaments detected in this work range from a few
hundred to $\sim 10^4$\,\msun\ and from $\sim$2-20\,pc, respectively. The sizes
and masses are similar to the sizes and masses of many filaments reported in the
literature (e.g. IRDC\,18223 --- \citealt{tackenberg2014}; G351 --- \citealt{leurini2011}, both of which appear in our catalogue). However, there have also been a number
of much larger and more massive filamentary structures recently reported in the
literature; these can be several hundreds of parsecs in length and have masses
$\sim$10$^5$\,\msun\ (\citealt{ragan2014, jackson2010, battersby2014b, li2013,
wang2014}). Many of these have been initially found from visual inspection of
mid-infrared images  with the aim  of identifying coherent filamentary patterns
or using the $^{13}$CO (1-0) GRS data.
How are these structures of different sizes related?

To address this question we have compiled a list of some of the largest
structures reported in the literature. The names, positions and physical
properties for these structures are given in
Table\,\ref{tbl:literature_filaments}. This sample includes 7 large filaments
identified by \citet{ragan2014} with lengths from $\sim 50$ to $250$\,pc and 5
filaments identified more recently by \citet{wang2015}. To this sample we also
add some of the most well-known filaments: G11.11$-$0.12 (the `Snake';
\citealt{wang2014}), `Nessie' (\citealt{jackson2010}) and G32.02+0.06
(\citealt{battersby2014b}) all of which have sizes of $\sim$100\,pc. We also
include the truly giant filamentary structure  identified by \citet{li2013};
this has a length of $\sim$500\,pc and consists of two connected molecular
clouds (G052.42+0.74 and G051.69+0.074) that are located on the periphery of the
G52L nebula (\citealt{bania2012}).

\begin{figure*}
\centering
\includegraphics[width = .9 \textwidth]{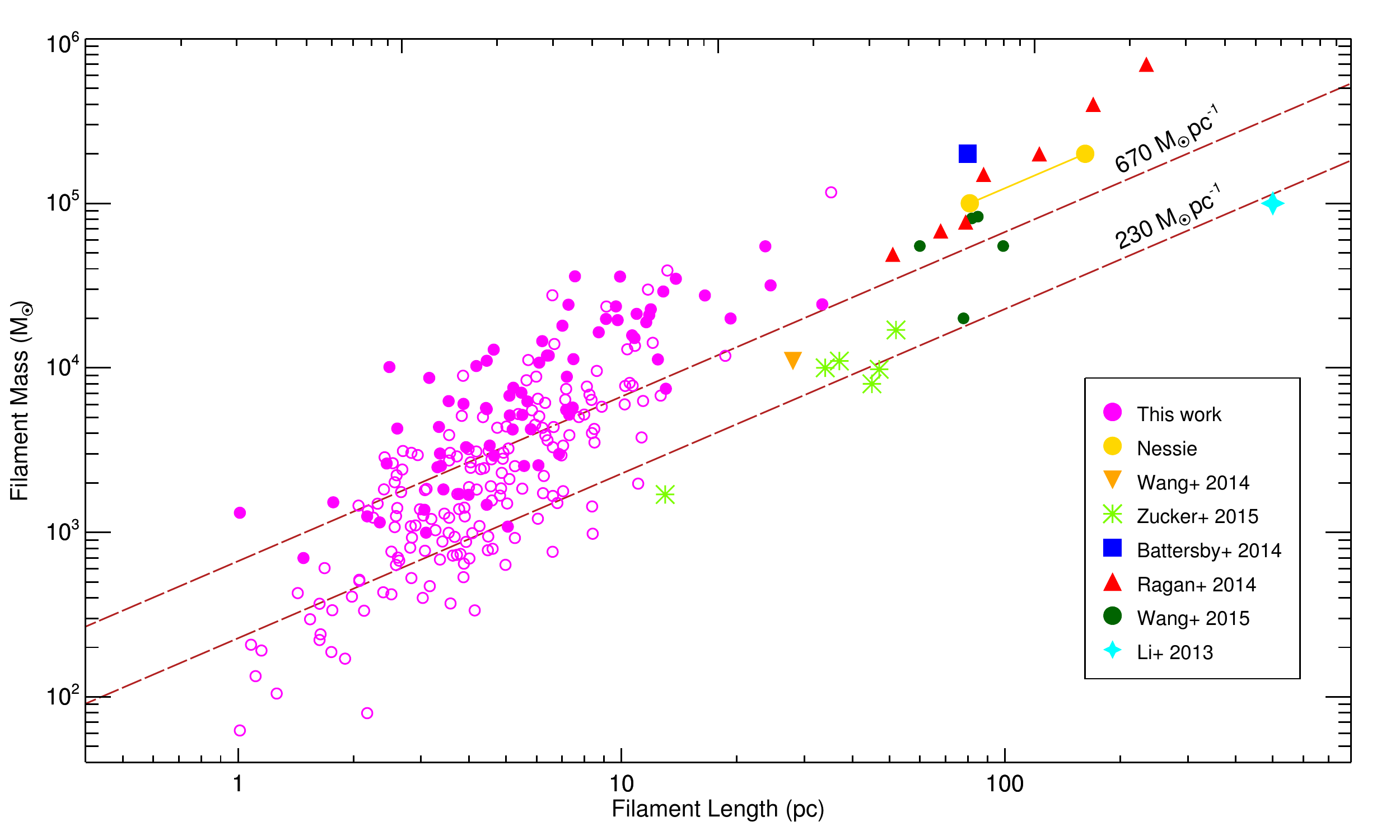}
\caption{Size-mass diagram of the filamentary structures discussed in
Section\,\ref{sect:gmf}. {\bf The values are taken from the original
publications.}
This diagram illustrates the differences discussed in the text between the filaments detected
in this work and giant filamentary structures reported in the literature.
The ``Nessie'' filament is represented by two connected circles; the left circle corresponds
to the structure identified by \citet{jackson2010} while the right circle
corresponds to the large structure identified by \citet{goodman2014}. The solid magenta circles identify filaments associated with massive star forming tracers while the open circles identify the non-MSF filaments. The red dashed lines correspond to the critical mass per unit length above which the filament becomes unstable and will start to collapse along their radial axis (\citealt{andre2014}); the upper and lower line corresponds to velocity dispersions of 1.2 and 0.7\,\kms, respectively, and assume an isothermal temperature of 10\,K.}
\label{fig:fila:comp}
\end{figure*}

\citet{ragan2014} refers to these structures as \emph{giant molecular filaments}
(GMF) and we will adopt this terminology for structures with lengths of
$\sim$50-200\,pc. The largest filament has been identified by \citet{li2013}.
These authors refer to it as a \emph{molecular wisp}. Since only a few studies
of this kind of structures have been conducted (\citealt{ragan2014,wang2015}),
which are limited to relatively nearby structures locate in the northern
Galactic plane, this sample of GMFs is unlikely to be complete and we have not attempted to compile an exhaustive list from the literature. However, this sample does contain some
of the most well-known filaments and is representative enough to investigate the
connection between structures of different size-scales and allow their
properties to be compared.

In Fig.\,\ref{fig:fila:comp} we present a size-mass diagram showing the
distributions of the filaments identified here and the compilation of larger
structures presented in Table\,\ref{tbl:literature_filaments}. The GMF
identified by \citet{ragan2014} and \citet{wang2015} are clearly much larger and
more massive than the filaments presented here. However, the dense gas fractions
of these structures is relatively low (e.g. 2-12\% --- \citealt{ragan2014}) so
it may be that we are tracing the dense fragments or sub-regions of these much
larger structures. Comparing the positions of our sample of filaments with
the GMFs we are able to find matches for 11 of the GMFs, and in a few of these
cases we find multiple filaments matched to a single GMF (e.g.
G016.763+00.020, G016.891+00.706, G016.919+00.268, G017.190+00.798 are
associated with GMF\,20.0$-$17.9 identified by \citealt{ragan2014}).

We do not find a match for four of the GMFs: GMF\,41.0$-$41.3, GMF\,26.7$-$25.4,
CFG028.68$-$0.28 and CFG029.18$-$0.34. However, we note that GMF\,41.0$-$41.3
has the lowest dense gas fraction and has a total dense gas mass almost an order
of magnitude lower than any of the other GMFs. We also note that approximately
50\% of GMF\,26.7$-$25.4 extends out of the region covered by ATLASGAL
($0.5\degr < b < 2.2\degr$). Furthermore, inspection of the images presented by
\citet[][their figure\,2]{wang2015} reveals that CFG029.18$-$0.34 appears to be
quite diffuse and CFG028.68$-$0.28 is located in a region of extended background
emission and is not so easily identified (see a combined Planck-ATLASGAL map
of this region in \citealt{csengeri2015a}).
We also do not find a counterpart of the molecular wisp identified by
\citet{li2013}, however, in spite of its size this structure has a similar total
mass as those identified by \citet{ragan2014} and is likely to have a
significantly lower column density; this source is significantly offset from the
other GMF shown in Fig.\,\ref{fig:fila:comp}.

{
In Fig.\,\ref{fig:fila:comp} we also include 6 more large filaments identified
by \citet{zucker2015}. This work follows from a study by \citet{goodman2014}
where they effectively doubled the size of the Nessie filament and speculate
that this structure may constitute one of many potential $bones$ of the Galaxy.
Among the 6 new objects, one is too close to the Galactic centre region -- a
region that we did not study because of the complexity of the emission.  We
compared positions and velocities of the remaining five objects and found that
BC 025.24-0.45 is matched to our filament G025.342-0.382. We have inspected
the ATLASGAL images of these new objects, and found that two of the objects (BC
026.94-0.30, BC 018.88-0.09) appear as chains of disconnected clumps, and are
not identified as filaments by \Disperse. This might because these objects are
extremely thin, and thus are not well resolved by ATLASGAL. The remaining two
of their objects (BC 335.31-0.29 and BC 332.21-0.04)
are located within crowed regions of emission. As ATLASGAL is sensitive to all
emission along the line of sight, these structures are more readily identified
in mid-infrared extinction.  }


Comparing our sample of filaments with the GMFs reported in the literature we
have found a match between our sample and 11 of the 16 larger scale filamentary
structures discussed in the previous paragraphs; these are identified by a star
appended to the literature name given to the various structures in
Table\,\ref{tbl:literature_filaments}. This supports the hypothesis that what we
have detected is the dense fragments of much larger filamentary structures, and
although the matched sample is relatively modest, it opens up the possibility
that many more of our filaments may also be dense fragments of much larger
structures.
The distributions of our filaments and the GMFs shown in
Fig.\,\ref{fig:fila:comp} is consistent with this possibility. A similar
conclusion was reached by \citet{schisano2014} from a comparison of the
structures identified in $Herschel$ maps. Relatively few large-scale structures
have been identified to date but this is likely to increase in the future with
the completion of a number of molecular line surveys (e.g. The Mopra Southern
Galactic Plane Survey (\citealt{burton2013}) and the James Clerk Maxwell
Telescopes CO Heterodyne Inner Milky Way Plane Survey
(CHIMPS; \citealt[][]{2016MNRAS.456.2885R}) at which point this hypothesis can
be more robustly tested.

\subsection{Orientation with respect to the Galactic Mid-Plane}
\label{sect:orientation}

In Section\,\ref{sect:properties_} we found that the filaments
 are preferentially aligned parallel with the Galactic mid-plane. This
suggests there is a connection between the dynamics of the Galactic disk and the
structure of the molecular material found within it. This connection is found
both at larger scale, with the recent discovery of $\geq$ 100\,pc molecular
structures (\citealt{li2013,ragan2014,goodman2014}), and small clumps
(\citealt{urquhart2015b}) all of which are found to have their major axis
preferentially aligned parallel to the Galactic mid-plane. This preferred
orientation therefore seems to apply to size scales from sub-parsec to $\geq$
100\,pc and therefore extends over two orders of magnitude. A similar
phenomenon can be seen in optical images of nearby face-on spiral galaxies such
as M31 and M51 (\citealt{schinnerer2013}), where filamentary dust lanes are
 roughly parallel to the spiral arms. The alignment between cold gas
condensations and spiral arms is also nicely reproduced in simulations
\citep{dobbs2013,smith2014a}. Following the discussion presented in the previous
subsection it seems likely that the filaments we have identified are the
high-density regions of these larger scale structures.

There are a number of explanations for the observed preferential
alignment of these elongated structures with the Galactic mid-plane. These
include magnetic fields and Galactic shear.
\citet{li2014} present evidence that the B-field in clouds is also aligned with the local large-scale
B-field in the diffuse ISM and that the latter is further aligned with the
Galactic plane \citep{han2002,jansson2012}.

This alignment is similar to what we
see in our filaments.
Furthermore, when these results are combined with recent observation of the
alignment of the B-field in the diffuse ISM and the Galactic plane \citep{li2015}
it suggests a picture where large-scale Galactic magnetic fields influence the dynamics of molecular gas over a wide range of
scales.  This is consistent with high resolution observations of the CO (2-1)
emission towards six of the most massive clouds in M31 reported by
\citet{li2011}. Results from the Planck satellite
\citep{2016A&A...586A.138P} also suggest that on large scales, filamentary
structures tend to align with the magnetic field \footnote{ In
\citet{2016A&A...586A.138P}, the orientation of structures is defined from the
{\it local} density gradient. In this work, we are concerned with the
{\it global} orientation of the structures. The
density threshold of $\log10({\rm N_H/cm^{-2}} ) \approx 21.7$  does not
contradict our results.}. In general, the mean magnetic field is both well
defined and is highly correlated with the spiral arms. Although it is clear that
filamentary structures are preferentially orientated parallel to the Galactic
mid-plane and with the large-scale magnetic field, it is unclear if
\emph{dynamically speaking}, the magnetic field plays a role in filament
formation and evolution, or that they share a common cause.

The discovery of filamentary structures that are larger than the thickness of
the molecular disk \citep{li2013,goodman2014,ragan2014} provides some support
for Galactic shear playing an important role in the formation of these
large-scale structures. The importance of Galactic shear in star
formation has been largely overlooked, probably due to the difficulty in
observing structures that are larger than the cloud scale \citep{li2013}.
However, it is much easier to study the impact of shear on cloud structures in
other galaxies such as M51 \citep{schinnerer2013}. Indeed, recent results indicate that shear
does have significant impacts on cloud dynamics and it can disrupt
the cloud and suppress star formation \citep{meidt2015, colombo2014}. This
agrees with  earlier theoretical studies
\citep{2006ApJ...646..213K,2002ApJ...570..132K,1986ApJ...311..554E} as well as
observational arguments \citep{2005MNRAS.361L..20S,2010ApJ...724.1503W}.
Although it is rather speculative at this stage, it does seem plausible that
the alignment between the filaments and the Galactic disk may be the result of
shear.

\subsection{Correlation of filaments with the spiral arms}
\label{sect:correlation_spiral_arms}

Fig.\,\ref{fig:lv_distribution} reveals a tight correlation between the spiral
arms and the filaments. To quantify this correlation we follow the analysis
presented by \citet{stark2006} and estimate the absolute difference between the
radial velocity of the source and the velocity of the nearest spiral arm along
its line of sight; they refer to this as the ``concentration statistic''
|$\Delta v$|. We use the velocities of the same spiral arms presented in
Fig.\,\ref{fig:lv_distribution} to determine this statistic for the filaments
and compare their distribution with a number of other samples. For all of the
samples considered we exclude the regions within 10\degr\ of the Galactic Centre
(i.e. $350\degr < \ell < 10\degr$) as the velocities of the spiral arms are
not well constrained in this region.

\begin{figure}
\centering
\includegraphics[width = 0.49\textwidth,  trim= 0 0 0 0, clip]{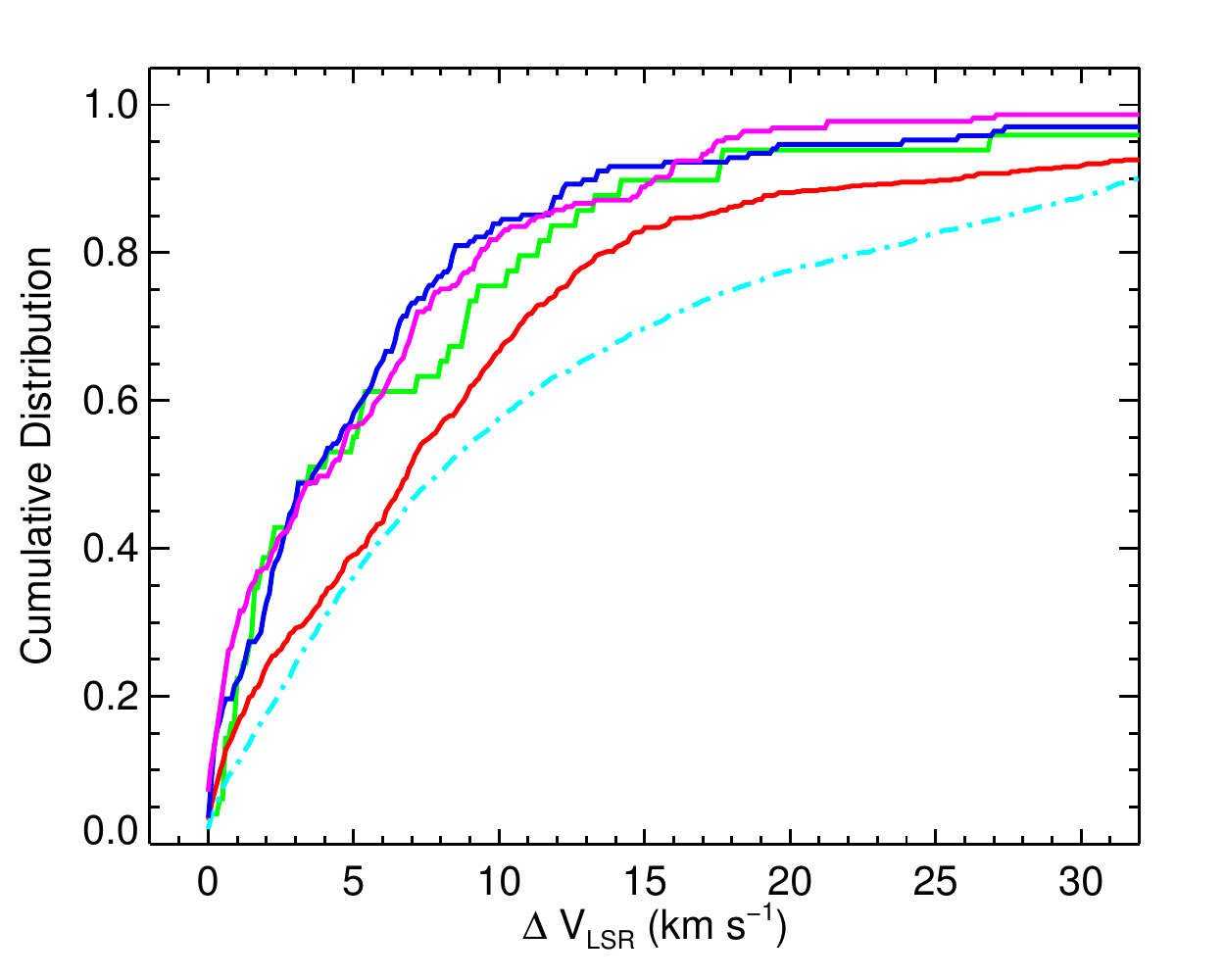} \\
\caption{Cumulative distribution of separation in velocity of filaments and molecular
clouds and their nearest spiral arm long the line of sight (|$\Delta V$|).
The spiral arm velocities are those shown in Fig.\,\ref{fig:lv_distribution}.
The |$\Delta V$| distribution of the filaments is shown by the magenta
curve, which is compared to \hii\ regions shown in blue, the 50 most
massive GRS clouds shown in green and the rest of the GRS lower mass
cloud population, which is shown in red. The cyan curve traces the
|$\Delta V$| distribution determined from a Monte Carlo simulated sample of clouds with random velocities.}
\label{fig:lv_correlation}
\end{figure}

In Fig.\,\ref{fig:lv_correlation} we present the result of this analysis in the
form of a cumulative distribution of the velocity differences. In this plot we
also show the results of the same analysis performed using the sample of compact
\hii\ regions identified by \citet{urquhart2013b} from the CORNISH survey, the GRS molecular cloud catalogue
(\citealt{roman2009}) and a sample with a random set of velocities produced by a
Monte Carlo simulation;  these were produced following the method outlined
in Sect.\,4 of \citealt{stark2006}. For the GRS clouds we plot the 50 most
massive clouds separately from the rest of the sample.

Observations of nearby spiral galaxies have revealed that young \hii\ regions
are amongst the most reliable tracers of their spiral structures (e.g.
\citealt{schinnerer2013}). The velocity differences between the
CORNISH sample of compact \hii\ regions and the spiral arms should therefore
provide a good reference point for comparison. It is clear from
Fig.\,\ref{fig:lv_correlation} that the distributions of the filaments (magenta) and the \hii\ regions (blue) are in excellent agreement with each other ($p$-value in a KS test is 0.11) and with the distribution of the 50 most massive GRS clouds (green) ($p$-value is 0.14). All three samples are very tightly correlated with
the spiral arms with over 80\% having offsets $<$10\,\kms. They are all
significantly different from the distribution of lower mass clouds in the GRS
sample (Kolmogorov-Smirnov test $p$-value $<$ $1.8\times10^{-4}$ for all
samples), which is likely to include a number of inter-arm clouds ($\sim$10\%; \citealt{stark2006}). The
difference in the distribution of the most massive clouds and the rest of the
population is in excellent agreement with the findings of \citet{stark2006}.
This is in spite of the fact that we have used a different catalogue of
molecular clouds and spiral arm loci, although the spiral arm loci derived by different models
are in broad agreement. We note however, despite the differences in the
distributions of the filaments, \hii\ regions, massive and less massive clumps
we find them all to be significantly more tightly correlated with the spiral
arms than that of the randomly sampled clouds produced by the Monte Carlo
simulation (cyan) (with all $p$-values $<2\times10^{-10}$) showing the
correlations are real and could not be produced by chance.

We note that \citet{ragan2014} reported that most of their GMFs were located in
the inter-arm regions; this led them to speculate that they may be analogs to
the spurs observed in nearby spiral galaxies. We find that many of the
filaments we have identified, which includes many identified by
\citet{ragan2014}, appear to be tightly correlated with the spiral arm loci (see Figs.\,\ref{fig:lv_distribution} and
\ref{fig:lv_correlation}). However, we note that along many lines of sight towards the inner Galaxy the
spiral arms overlap in velocity and this is likely to lead to blending of
different physical structures. This makes it more difficult to
identify large-scale structures associated with spiral arms than in the
inter-arm regions, and might explain why \citet{ragan2014} found the GMF to be
located in the inter-arm regions. This is rather speculative and requires a
more detailed investigation to verify and will be explored in a subsequent
paper.

\subsection{Associations with other catalogues}

\subsubsection{IRDCs}

We have cross-matched the positions of the CSC associated with filaments (as
discussed in Section\,\ref{sect:csc_associations}) with the IRDC catalogue
produced by \citet{peretto2009}. We find matches between 451 CSC sources and
IRDCs and these in turn are matched with 193 of our filaments; thus 43\% of the
filaments associated with CSC sources  are also associated with IRDCs.

As mentioned in the previous subsection many of the recent searches for
filaments have used IRDCs as their starting point to identify suitable
candidates whose nature as coherent filaments was later confirmed using molecular line data. These samples are
therefore somewhat biased to relatively nearby structures that may not be
representative of the general population. Our method of starting with the dust
continuum maps has successfully recovered the majority of the filaments that
were already known, many of which were discussed in the previous subsection, and
has identified $\sim$200 similar filaments that are associated with IRDCs.
We have also identified a large number of structures that are not seen in
extinction against a strong mid-infrared background emission and so are unlikely to have been detected by the other studies. Our sample therefore offers a more global view of their Galactic distribution and physical properties.

\begin{figure}
\centering
\includegraphics[width = 0.49 \textwidth]{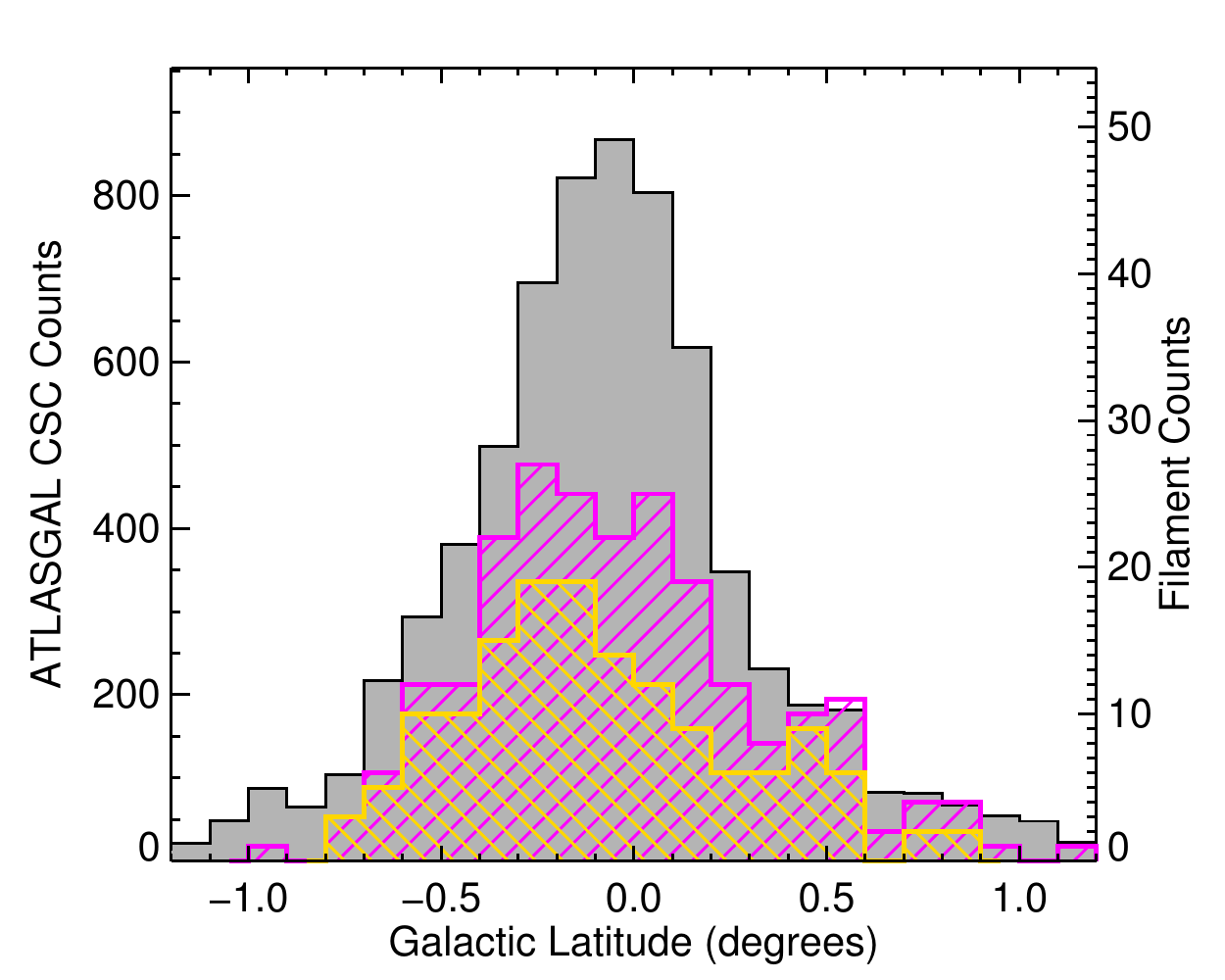} \\
\caption{Comparison of the Galactic latitude distribution of filament (magenta hatching)
with respect to that of the ATLASGAL CSC (grey; $10\degr < $|$\ell$|$ < 60\degr$).
Filaments associated with IRDCs are shown by the yellow hatching (see text for more details).
The bin size used is 0.1\degr. }
\label{fig:lat_dist}
\end{figure}

In Fig.\,\ref{fig:lat_dist} we present the numbers of sources of the whole
ATLASGAL CSC, the filaments and the filaments associated with IRDCs in
grey, magenta and yellow, respectively. Since the IRDC catalogue excludes the
innermost part of the Galactic plane (i.e. $350\degr < $|$\ell$|$ < 10\degr$)
this region has been exclude from this analysis.
The filaments have a flatter distribution compared to the ATLASGAL sources and those associated with IRDCs
are flatter still. Both the filaments and IRDCs appear to peak at more negative latitudes and
this is probably just reflecting that these samples consist of a larger
proportion of nearby structures and have a broader distribution. This gives an
indication of how far out in distance we are probing with the IRDC associated
being the most nearby structures while the whole sample of filaments, which is
not affected by extinction, is penetrating farther through the plane.

\begin{figure*}
\begin{center}
\includegraphics[width = 0.95 \textwidth]{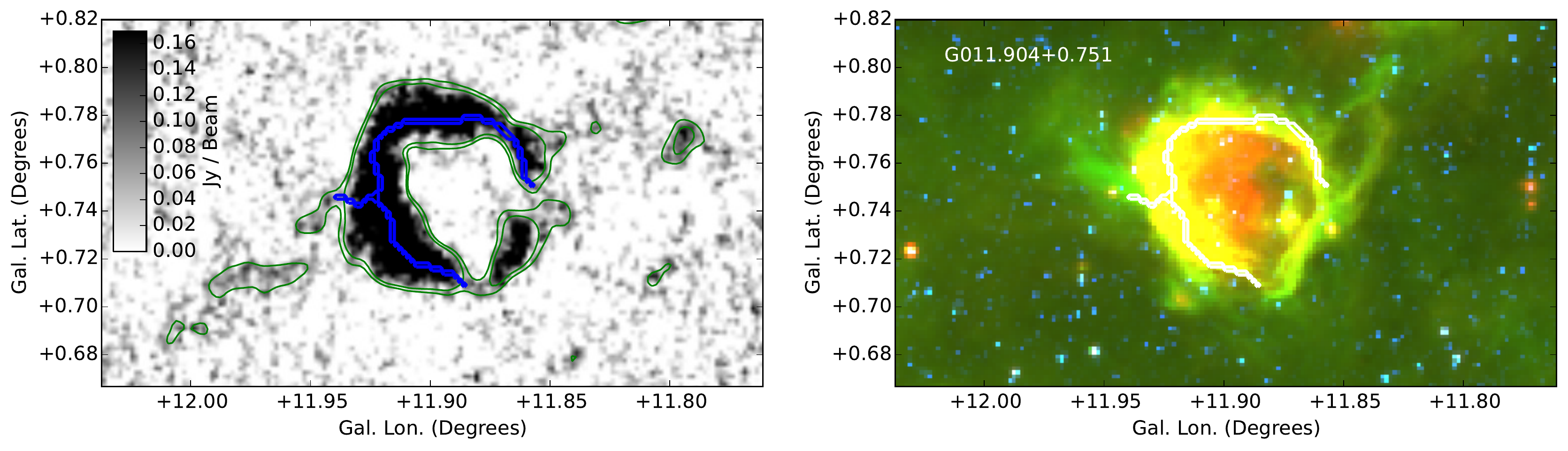}
\includegraphics[width = 0.95 \textwidth]{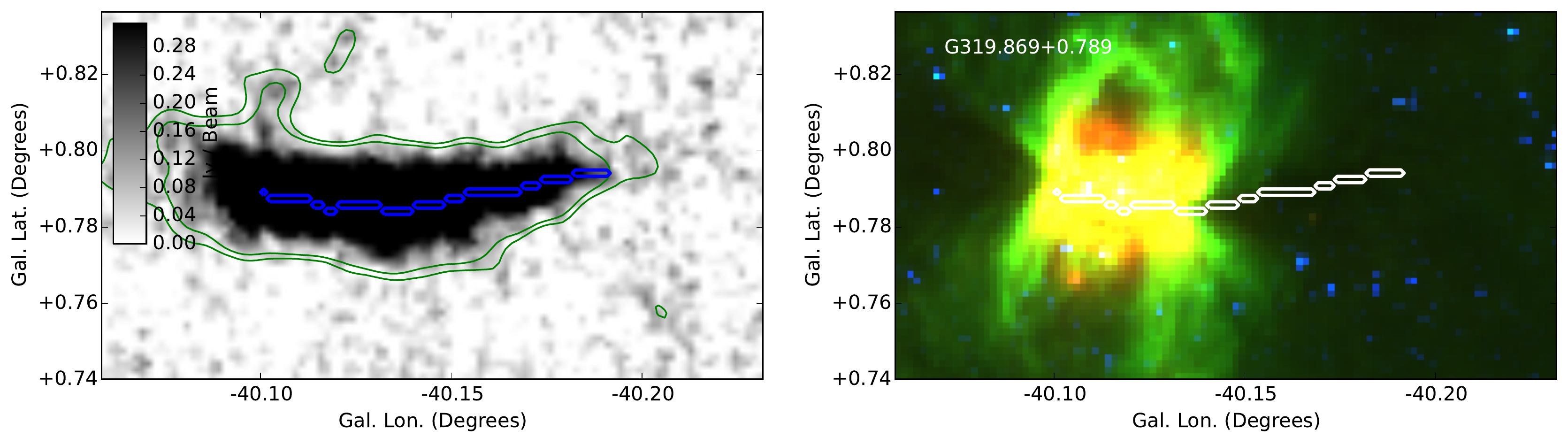}
\includegraphics[width = 0.95\textwidth]{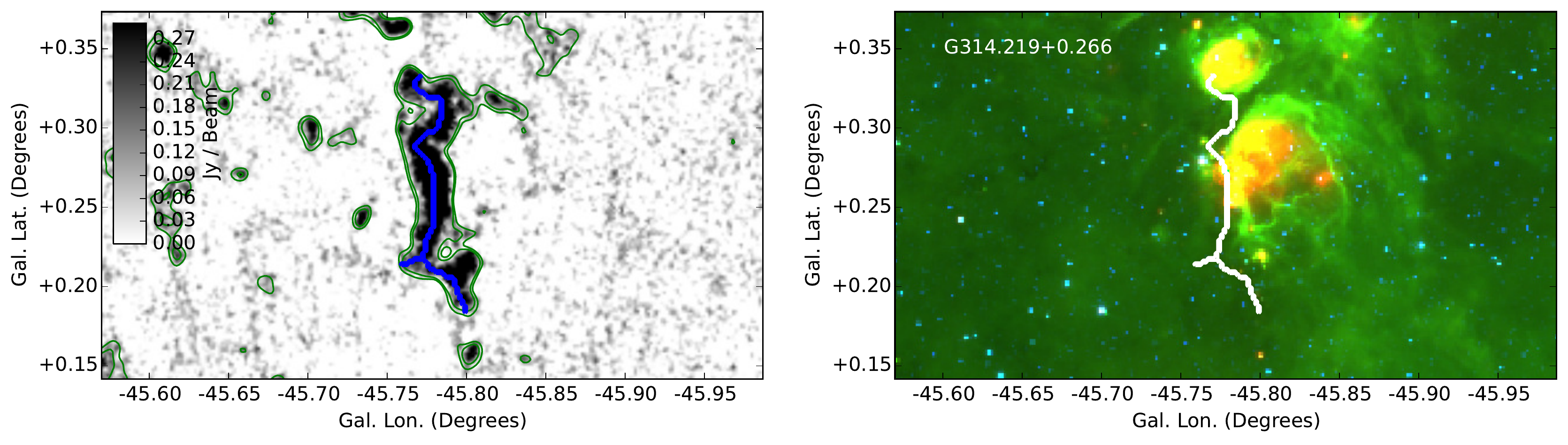}
\includegraphics[width = 0.95 \textwidth]{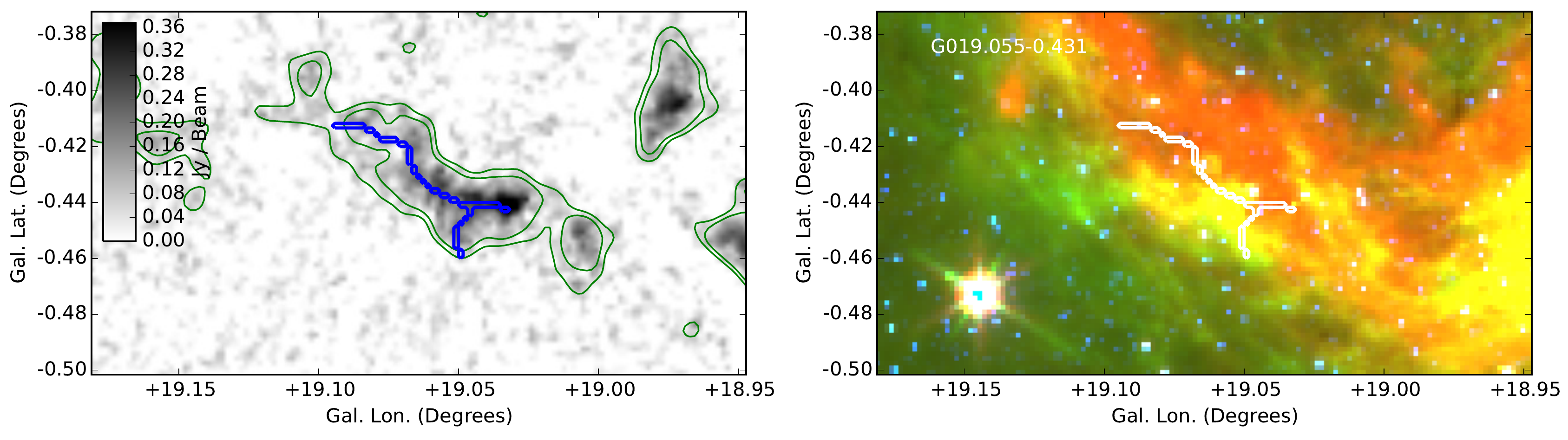}

\caption{Examples of filaments that correlate with GLIMPSE bubbles. In the upper panel we show an example of the filament tracing the dense material around the periphery of the \hii\ region; upper middle shows the filament running through the centre of a bubble; lower middle panel shows a filament weaving round two bubbles; lower panel shows a filament that is tracing a section of a much more extended bubble. For image details see caption of Fig.\,\ref{fig:structure_types}.}
\label{fig:bubbles}
\end{center}
\end{figure*}

\subsubsection{Mid-infrared bubbles}

\Disperse\ is designed to follow the peaks in the emission distribution of a
given structure and it takes no account of the morphology of the emission. As
well as identifying filamentary structures, which are generally linear
with larger aspect ratios ($>$3), it is sensitive to curved ridges of material
often found around the periphery of evolved \hii\ regions. These dense ridges
are formed from material that is swept up in front of the ionisation front as the
\hii\ regions expand. Since the processes that lead to these dense ridges and
the more linear filamentary structures are not necessarily the same it is useful
to distinguish between them.

\citet{churchwell2006} identified a large catalogue of bubbles by inspecting
mid-infrared images from the GLIMPSE survey
(\citealt{benjamin2003,churchwell2009}). This catalogue contains details for
322 partial and closed ring bubbles. We have searched for matches between these
bubbles and the filaments using a search radius twice the size of the mean
bubble radius. This has resulted in the identification of 47 filaments that are associated with bubbles,
with 6 filaments associated with two bubbles. We include the
names of the bubbles where a match has been found in
Table\,\ref{tbl:atlasgal_filaments}.

In many cases the association of these filaments with bubbles is far from clear.
In Fig.\,\ref{fig:bubbles} we show some examples of these matches, which have
been selected to illustrate some of the variety but also the ambiguities. In
some cases the filament is indeed tracing the dense material around the edge of
the \hii\ regions as one would expect (upper panel), however, in others the
filament appears to run through the middle of the bubbles (upper middle panel),
while in others the filaments appear to thread their way around two bubbles
(lower middle panel). Interestingly we find four bubbles that are associated
with 2 or more filaments; these are bubbles N2, N18, N59 and N63. These are some
of the largest bubbles identified in the Churchwell catalogue with angular radii
between 6.7 and 12 arc minutes --- the average radius of the catalogue is
$\sim$2\arcmin. It is therefore likely that in these cases the filaments are
tracing density enhanced fragments of the molecular shell surrounding the \hii\
regions (e.g. see lower panel of Fig.\,\ref{fig:bubbles}).  It is not
surprising to find many filaments on the periphery of \hii\ regions as bubbles
and shells have been found to be preferential sites of filament formation
\citep{2015A&A...580A..49I}.


\subsection{Filaments as engines of star formation}
\label{star_formation_engines}

In total, filaments and networks of filaments contribute $\sim$65\% of the total
flux contained in the detected structures. Assuming the temperature and
distance distributions are similar for different types of structures, it is
likely that they are associated with a proportional fraction of the dense gas
mass (see Section\,\ref{sect:properties_}). We found in
Section\,\ref{sect:csc_associations} that the proportion of massive star forming
clumps associated with the { filaments} is similar to their dense gas
fraction.
We also found that the filamentary structures (both networks and single
filaments) are also associated with approximately an equal proportion of the
Galactic population of MSF clumps ($\sim$67\%), suggesting a direct link between
dense mass fraction and massive star formation.

In Section\,\ref{sect:csc_associations} we found that 157 MSF clumps are
associated with 103 filaments, which is only 22\% of the filament sample. As
shown in Figs.\,\ref{fig:mass_dist} and \ref{fig:fila:comp} the filaments
associated with MSF clumps tend to be the more massive, but have a similar range
of filament lengths. For an isothermal cylinder the { balance} between
self-gravity and the internal pressure gives a { critical mass per
unit length}

\begin{equation}
M_{\rm{line, crit}} = \frac{2\sigma_{\rm v}^2}{G}=465 \left( \frac{\sigma
_{\rm v}}{{\rm{1\,km\,s^{-1}}}} \right)^2\,{\rm{M_\odot}}\,{\rm{pc}}^{-1}\;,
\end{equation}
\label{eqn:6}

\noindent where $M_{\rm{line, crit}}$ is the critical virial mass
per unit length and $\sigma_{\rm v}$ is the velocity dispersion of the gas.
A detailed analysis of the stability can be found in
\citet{2000MNRAS.311...85F}. Filaments
with $M/L$ ratios larger than the critical value are unstable and will collapse radially. Ammonia
observations of the (1,1) and (2,2) inversion transitions of a large number of
quiescent and massive star forming clumps reported line-widths ranging from
0.7-1.2\,\kms, with the larger values found towards MYSOs and \hii\ regions
(\citealt{wienen2012,urquhart2011,dunham2011}). These observations were targeted
at the peak column density regions and have an angular resolution
$\sim$30-40\arcsec, which is comparable with the widths of filaments and should
provide a reasonable estimate of their radial stability. Putting these values
into Eq.\,7 we estimate an upper and lower threshold for the critical masses of
670 and 230\,\msun\,pc$^{-1}$, respectively; these thresholds are indicated on
Fig.\,\ref{fig:fila:comp} by the dashed diagonal lines.

Comparing all of the filaments discussed in Section\,\ref{sect:gmf} we find that
nearly all of the large-scale filaments are likely to be unstable to radial
collapse and fragmentation. We also find that approximately half of the
filaments we have identified are likely to be unstable. We note, however, that
most of the filaments associated with the MSF clumps are located above this
threshold and likely to be radially collapsing. In Fig.\,\ref{fig:ml_plot} we
show the $M/L$ ratio distribution for all of the filaments as well as those
 associated with MSF clumps. The MSF associated filaments clearly
have significantly larger $M/L$ ratios compared to the rest of the population
(KS test results in a $p$-value $\ll$ 0.0013).

There is strong correlation between filaments with the largest $M/L$
ratio and the presence of massive star formation, and this appears to be fairly
consistent with the theoretically determined critical values. This also provides
a rather straightforward explanation as to why the proportion of massive star
forming filaments is relatively modest ($\sim$22\%). Furthermore, this analysis
supports a fairly simple evolutionary sequence starting with large-scale
unstable filaments that form in the spiral arms; these collapse radially and
fragment into denser filaments  and clumps, some of these remain
unstable and continue to collapse resulting in massive star formation while
others fall below the critical $M/L$ threshold and are able to resist further
collapse.

\begin{figure*}
\begin{center}
\includegraphics[width = 0.49 \textwidth]{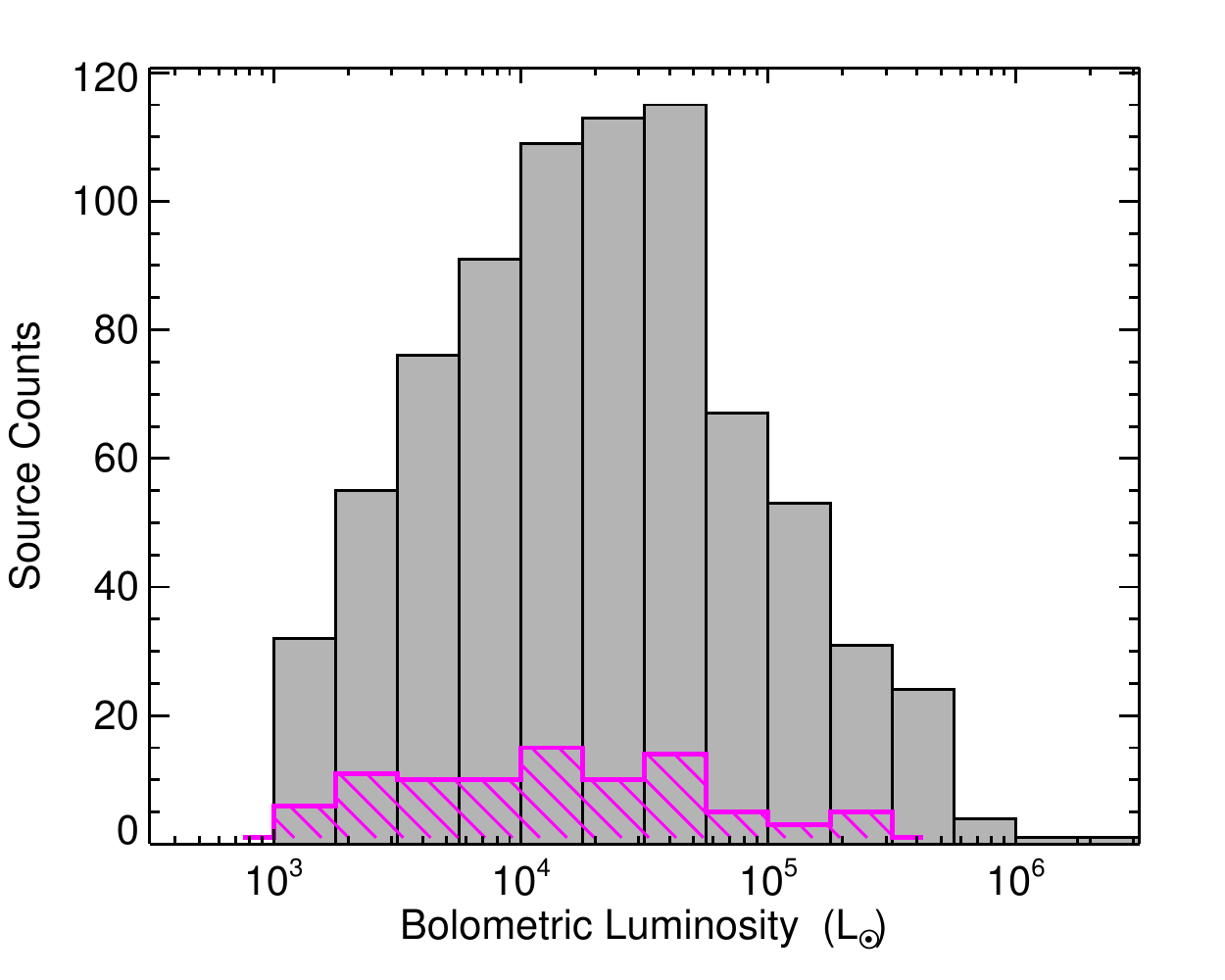}
\includegraphics[width = 0.49 \textwidth]{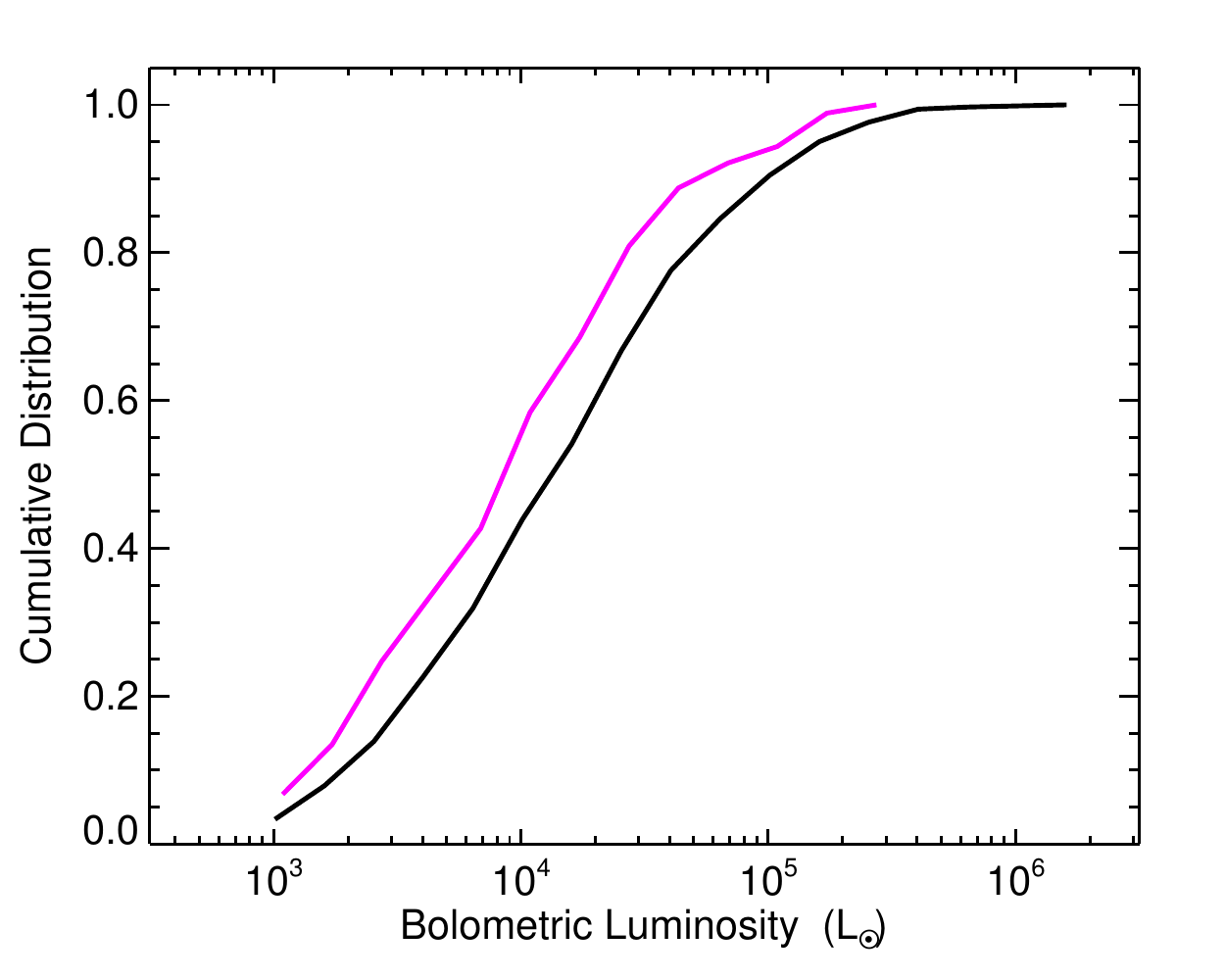}
\caption{Luminosity distribution of \hii\ regions and MYSOs associated with the filaments. The left hand panel shows the distribution of all \hii\ regions and MYSOs in the ATLASGAL region identified by the RMS survey (grey histogram) and those associated with a filament (magenta hatched histogram). The right hand panel shows the cumulative distribution of the luminosity for all massive RMS sources (black curve) and those associated with the filaments identified in this paper.}
\label{fig:lum}
\end{center}
\end{figure*}

In Fig.\,\ref{fig:lum} we compare the luminosity distribution of all MSF clumps
and those associated with filaments. In the left and right panels we present a
histogram and cumulative distribution of the two samples; these show that the
luminosities of the MSF clumps associated with the filaments cover almost the
full range of luminosities found for MYSOs and \uchii\ regions, but they have a
flatter distribution that slightly over-samples the lower luminosity end of the
distribution. However, a KS test is unable to reject the null hypothesis that
the two samples are drawn from the same parent population ($p$-value = 0.09).
This suggests that the MSF clumps associated with filaments can be
considered to be representative of the general population and that detailed
follow-up studies of this relatively small sample of nearby filaments have the
potential to provide valuable insights into how massive star forming clumps and
their large-scale molecular environments are related.

\section{Summary and conclusions}
\label{sect:summary}

In this paper we present the results of a systematic search for filamentary
structures located across the inner part of the Galaxy. This has been achieved
using the \Disperse\ algorithm
which has been used to find regions of coherent dust emission in the
ATLASGAL survey region ($300\degr < \ell < 60\degr$ and |$b$| $< 1.5\degr$). To
improve the performance of the code we first smoothed the emission maps and
decreased the resolution by a factor of 2 and used this to produce a smoothed
signal-to-noise map.
This improves the sensitivity to the weaker extended inter-clump regions
that form the links between larger structures.

 This generated a provisional catalogue of structures that was subsequently
inspected resulting in the identification of 517 filamentary structures.
Comparing the available velocity information we find 31 structures where the
velocity dispersion is greater than 10\,\kms; these
are considered to be due to line of sight blending of clumps and are removed from our
final sample. Comparison with the literature reveals that we have recovered the majority
of the filaments previously reported that are located in the ATLASGAL survey region.
 In total, we have obtained velocity information for 279 of the catalogued
filaments and have determined distances for 241 of the catalogued filaments.

Distances and velocities have been determined for many of the ATLASGAL sources
(e.g. \citealt{wienen2012, wienen2015a, urquhart2014c}) and these can be readily
assigned to the identified structures by cross matching them to
identify the dense clumps previously identified in the ATLASGAL CSC. This resulted in
associations of approximately four-fifths  of the CSC with the remainder being
associated with weaker emission that fell below the detection threshold of the
ATLASGAL CSC. We also determine the fraction of massive star formation for each
structure type by matching up clumps with methanol masers, massive YSOs and
\hii\ regions identified by \citet[][and references therein]{urquhart2014c}.

We focus on the properties of the filaments and present the most
complete view yet of their Galactic distribution and investigate their
association with the spiral arms. The main results are as follows:

\begin{enumerate}

\item We have identified a large and reliable sample of { filaments}
 located across the inner Galactic plane. This sample consists of 486
spatially and kinematically coherent filaments with a relatively even split
between the I and IV Quadrants. Most previous works have focused on filaments
located in the I Quadrant and so this catalogue is the ideal tool to begin to
explore the IV Quadrant.

\item The filaments identified have aspect ratios of 2-10, lengths between 2 and
20\,pc, widths between $\sim$0.1-2.5\,pc and masses between
$\sim$10$^{2}$-10$^{5}$\,\msun. The filaments have typical mass-to-length ratios
 of $M/L$ $\sim$\,200-2000\,\msun\,pc$^{-1}$. We find our filaments are smaller than many of
the large-scale structures ($\sim$100\,pc)  recently reported in the literature,
but are larger and significantly more massive than many of the filaments
identified by $Herschel$ studies. Our sample therefore covers an important range
of the mass-size parameter-space that is currently not well explored.

\item We find a correlation between many of the large-scale filaments and
filaments identified here, with our filaments effectively tracing the
high-column density regions of the larger structures. It is therefore likely
that many of the filaments in our sample are intimately associated much
larger-scale structures. In other words, these structures
 of different sizes should not be considered distinct, but parts of larger,
 coherent, hierarchical structures.

\item We find the { filaments studied in this work}  are approximately
aligned parallel to the Galactic mid-plane but with a slight inclination of $\sim$20\degr.
This indicated that at $\sim 10\rm\; pc$ scale probed by the filaments, the
dynamics of the Galactic disk have significant impacts on the dynamics of the
dense molecular gas.

\item Comparing the velocities of the filaments with the loci of the spiral arms
we show that they are tightly correlated with $\sim$80\% of the
filament velocities found within 10\,\kms\ of that of spiral arm. A similar
correlation is found for compact \hii\ regions and for the most massive molecular clouds in the GRS
survey. Plotting the distribution of the filaments on a 3D
map of the Milky Way we find a good correlation with the positions of the
spiral arms, and the majority of the filaments are associated with the
near side of the Scutum-Centaurus arm.


\item We find $\sim$22\% of the filaments are associated with massive star
forming clumps. Comparing the properties of the filaments associated with
massive star formation with the rest of the filament sample we find they are
significantly more massive and have higher $M/L$ ratios. Investigating the
stability of filaments we also find that the $M/L$ ratio for massive star
forming filaments are above the critical threshold and they are therefore likely
to be unstable to radial collapse. The majority of the rest of the filaments
appear to be close to equilibrium, and this provides a relatively simple
explanation for the low proportion of filaments that are associated with massive
star formation.

\item Comparing the luminosities of massive YSOs and \hii\ regions associated
with filaments with the rest of the Galactic population of these objects we find
the distributions of their luminosities are similar. This would suggest that
either a significant fraction of all massive stars form in filaments or that the
star formation is relatively insensitive to the large-scale structure.

\end{enumerate}

In this work  we provide the first Galaxy-wide sample of filamentary structures
\footnote{The full catalogue and the corresponding data files are available
at \url{http://atlasgal.mpifr-bonn.mpg.de/cgi-bin/ATLASGAL_FILAMENTS.cgi}. }. We
investigate their general properties and Galactic distribution.
We also investigate their association with the spiral arms, large-scale filaments,
IRDCs, bubbles around evolved \hii\ regions and regions of massive star
formation. This is the largest and most comprehensive catalogue of
filamentary structures in the inner Galactic plane and will provide an
excellent starting point for further, more detailed studies.

\begin{acknowledgements}

We thank the referee for careful readings and helpful comments to our paper. The
ATLASGAL project is a collaboration between the Max-Planck-Gesellschaft, the European Southern Observatory (ESO) and the Universidad de Chile. It includes
projects E-181.C-0885, E-078.F-9040(A), M-079.C-9501(A), M-081.C-9501(A) plus
Chilean data. This paper made use of information from the Red MSX Source survey
database at http://rms.leeds.ac.uk/cgi-bin/public/RMS\_DATABASE.cgi which was
constructed with support from the Science and Technology Facilities Council of
the UK. This research has made use of the SIMBAD database operated at CDS,
Strasbourg, France. This work was partially funded by the ERC Advanced
Investigator Grant GLOSTAR (247078) and was partially carried out within the
Collaborative Research Council 956, sub-project A6, funded by the Deutsche
Forschungsgemeinschaft (DFG).
Guang-Xing Li and T.\,Csengeri are supported by the Deutsche
Forschungsgemeinschaft (DFG) priority program 1573 ISM-SPP.

\end{acknowledgements}
\bibliography{atlas_filaments_new}

\begin{thebibliography}{117}
\expandafter\ifx\csname natexlab\endcsname\relax\def\natexlab#1{#1}\fi

\bibitem[{{Andr{\'e}} {et~al.}(2014){Andr{\'e}}, {Di Francesco},
  {Ward-Thompson}, {Inutsuka}, {Pudritz}, \& {Pineda}}]{andre2014}
{Andr{\'e}}, P., {Di Francesco}, J., {Ward-Thompson}, D., {et~al.} 2014,
  Protostars and Planets VI, 27

\bibitem[{{Andr{\'e}} {et~al.}(2010){Andr{\'e}}, {Men'shchikov}, {Bontemps},
  {K{\"o}nyves}, {Motte}, {Schneider}, {Didelon}, {Minier}, {Saraceno},
  {Ward-Thompson}, {di Francesco}, {White}, {Molinari}, {Testi}, {Abergel},
  {Griffin}, {Henning}, {Royer}, {Mer{\'{\i}}n}, {Vavrek}, {Attard},
  {Arzoumanian}, {Wilson}, {Ade}, {Aussel}, {Baluteau}, {Benedettini},
  {Bernard}, {Blommaert}, {Cambr{\'e}sy}, {Cox}, {di Giorgio}, {Hargrave},
  {Hennemann}, {Huang}, {Kirk}, {Krause}, {Launhardt}, {Leeks}, {Le Pennec},
  {Li}, {Martin}, {Maury}, {Olofsson}, {Omont}, {Peretto}, {Pezzuto}, {Prusti},
  {Roussel}, {Russeil}, {Sauvage}, {Sibthorpe}, {Sicilia-Aguilar}, {Spinoglio},
  {Waelkens}, {Woodcraft}, \& {Zavagno}}]{andre2010}
{Andr{\'e}}, P., {Men'shchikov}, A., {Bontemps}, S., {et~al.} 2010, \aap, 518,
  L102

\bibitem[{{Arzoumanian} {et~al.}(2011){Arzoumanian}, {Andr{\'e}}, {Didelon},
  {K{\"o}nyves}, {Schneider}, {Men'shchikov}, {Sousbie}, {Zavagno}, {Bontemps},
  {di Francesco}, {Griffin}, {Hennemann}, {Hill}, {Kirk}, {Martin}, {Minier},
  {Molinari}, {Motte}, {Peretto}, {Pezzuto}, {Spinoglio}, {Ward-Thompson},
  {White}, \& {Wilson}}]{arzoumanian2011}
{Arzoumanian}, D., {Andr{\'e}}, P., {Didelon}, P., {et~al.} 2011, \aap, 529, L6

\bibitem[{{Bania} {et~al.}(2012){Bania}, {Anderson}, \& {Balser}}]{bania2012}
{Bania}, T.~M., {Anderson}, L.~D., \& {Balser}, D.~S. 2012, \apj, 759, 96

\bibitem[{{Bania} {et~al.}(2010){Bania}, {Anderson}, {Balser}, \&
  {Rood}}]{bania2010}
{Bania}, T.~M., {Anderson}, L.~D., {Balser}, D.~S., \& {Rood}, R.~T. 2010,
  \apjl, 718, L106

\bibitem[{{Battersby} {et~al.}(2014){Battersby}, {Ginsburg}, {Bally},
  {Longmore}, {Dunham}, \& {Darling}}]{battersby2014b}
{Battersby}, C., {Ginsburg}, A., {Bally}, J., {et~al.} 2014, \apj, 787, 113

\bibitem[{{Battisti} \& {Heyer}(2014)}]{battisti2014}
{Battisti}, A.~J. \& {Heyer}, M.~H. 2014, \apj, 780, 173

\bibitem[{{Benjamin} {et~al.}(2003){Benjamin}, {Churchwell}, {Babler}, {Bania},
  {Clemens}, {Cohen}, {Dickey}, {Indebetouw}, {Jackson}, {Kobulnicky},
  {Lazarian}, {Marston}, {Mathis}, {Meade}, {Seager}, {Stolovy}, {Watson},
  {Whitney}, {Wolff}, \& {Wolfire}}]{benjamin2003}
{Benjamin}, R.~A., {Churchwell}, E., {Babler}, B.~L., {et~al.} 2003, \pasp,
  115, 953

\bibitem[{{Bertin} \& {Arnouts}(1996)}]{bertin1996}
{Bertin}, E. \& {Arnouts}, S. 1996, \aaps, 117, 393

\bibitem[{{Brand} \& {Blitz}(1993)}]{brand1993}
{Brand}, J. \& {Blitz}, L. 1993, \aap, 275, 67

\bibitem[{{Bronfman} {et~al.}(2000){Bronfman}, {Casassus}, {May}, \&
  {Nyman}}]{bronfman2000}
{Bronfman}, L., {Casassus}, S., {May}, J., \& {Nyman}, L.-{\AA}. 2000, \aap,
  358, 521

\bibitem[{{Bronfman} {et~al.}(1996){Bronfman}, {Nyman}, \&
  {May}}]{bronfman1996}
{Bronfman}, L., {Nyman}, L.-A., \& {May}, J. 1996, \aaps, 115, 81

\bibitem[{{Burkert} \& {Hartmann}(2004)}]{2004ApJ...616..288B}
{Burkert}, A. \& {Hartmann}, L. 2004, \apj, 616, 288

\bibitem[{{Burton} {et~al.}(2013){Burton}, {Braiding}, {Glueck}, {Goldsmith},
  {Hawkes}, {Hollenbach}, {Kulesa}, {Martin}, {Pineda}, {Rowell}, {Simon},
  {Stark}, {Stutzki}, {Tothill}, {Urquhart}, {Walker}, {Walsh}, \&
  {Wolfire}}]{burton2013}
{Burton}, M.~G., {Braiding}, C., {Glueck}, C., {et~al.} 2013, \pasa, 30, 44

\bibitem[{{Carey} {et~al.}(2005){Carey}, {Noriega-Crespo}, {Price}, {Padgett},
  {Kraemer}, {Indebetouw}, {Mizuno}, {Ali}, {Berriman}, {Boulanger}, {Cutri},
  {Ingalls}, {Kuchar}, {Latter}, {Marleau}, {Miville-Deschenes}, {Molinari},
  {Rebull}, \& {Testi}}]{carey2005}
{Carey}, S.~J., {Noriega-Crespo}, A., {Price}, S.~D., {et~al.} 2005, in
  Bulletin of the American Astronomical Society, Vol.~37, American Astronomical
  Society Meeting Abstracts, 1252

\bibitem[{{Churchwell} {et~al.}(2009){Churchwell}, {Babler}, {Meade},
  {Whitney}, {Benjamin}, {Indebetouw}, {Cyganowski}, {Robitaille}, {Povich},
  {Watson}, \& {Bracker}}]{churchwell2009}
{Churchwell}, E., {Babler}, B.~L., {Meade}, M.~R., {et~al.} 2009, \pasp, 121,
  213

\bibitem[{{Churchwell} {et~al.}(2006){Churchwell}, {Povich}, {Allen}, {Taylor},
  {Meade}, {Babler}, {Indebetouw}, {Watson}, {Whitney}, {Wolfire}, {Bania},
  {Benjamin}, {Clemens}, {Cohen}, {Cyganowski}, {Jackson}, {Kobulnicky},
  {Mathis}, {Mercer}, {Stolovy}, {Uzpen}, {Watson}, \&
  {Wolff}}]{churchwell2006}
{Churchwell}, E., {Povich}, M.~S., {Allen}, D., {et~al.} 2006, \apj, 649, 759

\bibitem[{{Colombo} {et~al.}(2014){Colombo}, {Meidt}, {Schinnerer},
  {Garc{\'{\i}}a-Burillo}, {Hughes}, {Pety}, {Leroy}, {Dobbs}, {Dumas},
  {Thompson}, {Schuster}, \& {Kramer}}]{colombo2014}
{Colombo}, D., {Meidt}, S.~E., {Schinnerer}, E., {et~al.} 2014, \apj, 784, 4

\bibitem[{{Contreras} {et~al.}(2013){Contreras}, {Schuller}, {Urquhart},
  {Csengeri}, {Wyrowski}, {Beuther}, {Bontemps}, {Bronfman}, {Henning},
  {Menten}, {Schilke}, {Walmsley}, {Wienen}, {Tackenberg}, \&
  {Linz}}]{contreras2013}
{Contreras}, Y., {Schuller}, F., {Urquhart}, J.~S., {et~al.} 2013, \aap, 549,
  A45

\bibitem[{{Cordes}(2004)}]{cordes2004}
{Cordes}, J.~M. 2004, in Astronomical Society of the Pacific Conference Series,
  Vol. 317, Milky Way Surveys: The Structure and Evolution of our Galaxy, ed.
  D.~{Clemens}, R.~{Shah}, \& T.~{Brainerd}, 211

\bibitem[{{Csengeri} {et~al.}(2016{\natexlab{a}}){Csengeri}, {Leurini},
  {Wyrowski}, {Urquhart}, {Menten}, {Walmsley}, {Bontemps}, {Wienen},
  {Beuther}, {Motte}, {Nguyen-Luong}, {Schilke}, {Schuller}, {Zavagno}, \&
  {Sanna}}]{2015arXiv151105138C}
{Csengeri}, T., {Leurini}, S., {Wyrowski}, F., {et~al.} 2016{\natexlab{a}},
  \aap, 586, A149

\bibitem[{{Csengeri} {et~al.}(2014){Csengeri}, {Urquhart}, {Schuller}, {Motte},
  {Bontemps}, {Wyrowski}, {Menten}, {Bronfman}, {Beuther}, {Henning}, {Testi},
  {Zavagno}, \& {Walmsley}}]{csengeri2014}
{Csengeri}, T., {Urquhart}, J.~S., {Schuller}, F., {et~al.} 2014, \aap, 565,
  A75

\bibitem[{{Csengeri} {et~al.}(2016{\natexlab{b}}){Csengeri}, {Weiss},
  {Wyrowski}, {Menten}, {Urquhart}, {Leurini}, {Schuller}, {Beuther},
  {Bontemps}, {Bronfman}, {Henning}, \& {Schneider}}]{csengeri2015a}
{Csengeri}, T., {Weiss}, A., {Wyrowski}, F., {et~al.} 2016{\natexlab{b}}, \aap,
  585, A104

\bibitem[{{Cyganowski} {et~al.}(2008){Cyganowski}, {Whitney}, {Holden},
  {Braden}, {Brogan}, {Churchwell}, {Indebetouw}, {Watson}, {Babler},
  {Benjamin}, {Gomez}, {Meade}, {Povich}, {Robitaille}, \&
  {Watson}}]{cyganowski2008}
{Cyganowski}, C.~J., {Whitney}, B.~A., {Holden}, E., {et~al.} 2008, \aj, 136,
  2391

\bibitem[{{Dame} {et~al.}(2001){Dame}, {Hartmann}, \& {Thaddeus}}]{dame2001}
{Dame}, T.~M., {Hartmann}, D., \& {Thaddeus}, P. 2001, \apj, 547, 792

\bibitem[{{Davies} {et~al.}(2012){Davies}, {Clark}, {Trombley}, {Figer},
  {Najarro}, {Crowther}, {Kudritzki}, {Thompson}, {Urquhart}, \&
  {Hindson}}]{davies2012}
{Davies}, B., {Clark}, J.~S., {Trombley}, C., {et~al.} 2012, \mnras, 419, 1871

\bibitem[{{Dempsey} {et~al.}(2013){Dempsey}, {Thomas}, \&
  {Currie}}]{dempsey2013}
{Dempsey}, J.~T., {Thomas}, H.~S., \& {Currie}, M.~J. 2013, \apjs, 209, 8

\bibitem[{{Dobbs} \& {Pringle}(2013)}]{dobbs2013}
{Dobbs}, C.~L. \& {Pringle}, J.~E. 2013, \mnras, 432, 653

\bibitem[{{Dunham} {et~al.}(2011{\natexlab{a}}){Dunham}, {Robitaille}, {Evans},
  {Schlingman}, {Cyganowski}, \& {Urquhart}}]{dunham2011}
{Dunham}, M.~K., {Robitaille}, T.~P., {Evans}, II, N.~J., {et~al.}
  2011{\natexlab{a}}, \apj, 731, 90

\bibitem[{{Dunham} {et~al.}(2011{\natexlab{b}}){Dunham}, {Rosolowsky}, {Evans},
  {Cyganowski}, \& {Urquhart}}]{dunham2011b}
{Dunham}, M.~K., {Rosolowsky}, E., {Evans}, II, N.~J., {Cyganowski}, C., \&
  {Urquhart}, J.~S. 2011{\natexlab{b}}, \apj, 741, 110

\bibitem[{{Ellsworth-Bowers} {et~al.}(2013){Ellsworth-Bowers}, {Glenn},
  {Rosolowsky}, {Mairs}, {Evans}, {Battersby}, {Ginsburg}, {Shirley}, \&
  {Bally}}]{ellsworth2013}
{Ellsworth-Bowers}, T.~P., {Glenn}, J., {Rosolowsky}, E., {et~al.} 2013, \apj,
  770, 39

\bibitem[{{Ellsworth-Bowers} {et~al.}(2015){Ellsworth-Bowers}, {Rosolowsky},
  {Glenn}, {Ginsburg}, {Evans}, {Battersby}, {Shirley}, \&
  {Svoboda}}]{bowers2015}
{Ellsworth-Bowers}, T.~P., {Rosolowsky}, E., {Glenn}, J., {et~al.} 2015, \apj,
  799, 29

\bibitem[{{Elmegreen} \& {Elmegreen}(1986)}]{1986ApJ...311..554E}
{Elmegreen}, B.~G. \& {Elmegreen}, D.~M. 1986, \apj, 311, 554

\bibitem[{{Federrath}(2016)}]{2016MNRAS.457..375F}
{Federrath}, C. 2016, \mnras, 457, 375

\bibitem[{{Fiege} \& {Pudritz}(2000)}]{2000MNRAS.311...85F}
{Fiege}, J.~D. \& {Pudritz}, R.~E. 2000, \mnras, 311, 85

\bibitem[{{Giannetti} {et~al.}(2014){Giannetti}, {Wyrowski}, {Brand},
  {Csengeri}, {Fontani}, {Walmsley}, {Nguyen Luong}, {Beuther}, {Schuller},
  {G{\"u}sten}, \& {Menten}}]{giannetti2014}
{Giannetti}, A., {Wyrowski}, F., {Brand}, J., {et~al.} 2014, \aap, 570, A65

\bibitem[{{Goldsmith} {et~al.}(2008){Goldsmith}, {Heyer}, {Narayanan}, {Snell},
  {Li}, \& {Brunt}}]{goldsmith2008}
{Goldsmith}, P.~F., {Heyer}, M., {Narayanan}, G., {et~al.} 2008, \apj, 680, 428

\bibitem[{{G{\'o}mez} \& {V{\'a}zquez-Semadeni}(2014)}]{gomez2014}
{G{\'o}mez}, G.~C. \& {V{\'a}zquez-Semadeni}, E. 2014, \apj, 791, 124

\bibitem[{{Goodman} {et~al.}(2014){Goodman}, {Alves}, {Beaumont}, {Benjamin},
  {Borkin}, {Burkert}, {Dame}, {Jackson}, {Kauffmann}, {Robitaille}, \&
  {Smith}}]{goodman2014}
{Goodman}, A.~A., {Alves}, J., {Beaumont}, C.~N., {et~al.} 2014, \apj, 797, 53

\bibitem[{{Green} \& {McClure-Griffiths}(2011)}]{green2011b}
{Green}, J.~A. \& {McClure-Griffiths}, N.~M. 2011, \mnras, 417, 2500

\bibitem[{{Green}(2009)}]{green2009}
{Green}, J.~A. a.~a. 2009, \mnras, 392, 783

\bibitem[{{Han} \& {Wielebinski}(2002)}]{han2002}
{Han}, J.-L. \& {Wielebinski}, R. 2002, \cjaa, 2, 293

\bibitem[{{Heitsch} {et~al.}(2008){Heitsch}, {Hartmann}, {Slyz}, {Devriendt},
  \& {Burkert}}]{2008ApJ...674..316H}
{Heitsch}, F., {Hartmann}, L.~W., {Slyz}, A.~D., {Devriendt}, J.~E.~G., \&
  {Burkert}, A. 2008, \apj, 674, 316

\bibitem[{{Hennebelle}(2013)}]{2013A&A...556A.153H}
{Hennebelle}, P. 2013, \aap, 556, A153

\bibitem[{{Hildebrand}(1983)}]{hildebrand1983}
{Hildebrand}, R.~H. 1983, \qjras, 24, 267

\bibitem[{{Hill} {et~al.}(2012){Hill}, {Andr{\'e}}, {Arzoumanian}, {Motte},
  {Minier}, {Men'shchikov}, {Didelon}, {Hennemann}, {K{\"o}nyves},
  {Nguyen-Luong}, {Palmeirim}, {Peretto}, {Schneider}, {Bontemps}, {Louvet},
  {Elia}, {Giannini}, {Rev{\'e}ret}, {Le Pennec}, {Rodriguez}, {Boulade},
  {Doumayrou}, {Dubreuil}, {Gallais}, {Lortholary}, {Martignac}, {Talvard}, \&
  {De Breuck}}]{2012A&A...548L...6H}
{Hill}, T., {Andr{\'e}}, P., {Arzoumanian}, D., {et~al.} 2012, \aap, 548, L6

\bibitem[{{Hill} {et~al.}(2011){Hill}, {Motte}, {Didelon}, {Bontemps},
  {Minier}, {Hennemann}, {Schneider}, {Andr{\'e}}, {Men'shchikov}, {Anderson},
  {Arzoumanian}, {Bernard}, {di Francesco}, {Elia}, {Giannini}, {Griffin},
  {K{\"o}nyves}, {Kirk}, {Marston}, {Martin}, {Molinari}, {Nguyen Luong},
  {Peretto}, {Pezzuto}, {Roussel}, {Sauvage}, {Sousbie}, {Testi},
  {Ward-Thompson}, {White}, {Wilson}, \& {Zavagno}}]{hill2011}
{Hill}, T., {Motte}, F., {Didelon}, P., {et~al.} 2011, \aap, 533, A94

\bibitem[{{Hoare} {et~al.}(2012){Hoare}, {Purcell}, {Churchwell}, {Diamond},
  {Cotton}, {Chandler}, {Smethurst}, {Kurtz}, {Mundy}, {Dougherty}, {Fender},
  {Fuller}, {Jackson}, {Garrington}, {Gledhill}, {Goldsmith}, {Lumsden},
  {Mart{\'{\i}}}, {Moore}, {Muxlow}, {Oudmaijer}, {Pandian}, {Paredes},
  {Shepherd}, {Spencer}, {Thompson}, {Umana}, {Urquhart}, \&
  {Zijlstra}}]{hoare2012}
{Hoare}, M.~G., {Purcell}, C.~R., {Churchwell}, E.~B., {et~al.} 2012, \pasp,
  124, 939

\bibitem[{{Inutsuka} {et~al.}(2015){Inutsuka}, {Inoue}, {Iwasaki}, \&
  {Hosokawa}}]{2015A&A...580A..49I}
{Inutsuka}, S.-i., {Inoue}, T., {Iwasaki}, K., \& {Hosokawa}, T. 2015, \aap,
  580, A49

\bibitem[{{Jackson} {et~al.}(2002){Jackson}, {Bania}, {Simon}, {Kolpak},
  {Clemens}, \& {Heyer}}]{jackson2003}
{Jackson}, J.~M., {Bania}, T.~M., {Simon}, R., {et~al.} 2002, ApJ, 566, L81

\bibitem[{{Jackson} {et~al.}(2010){Jackson}, {Finn}, {Chambers}, {Rathborne},
  \& {Simon}}]{jackson2010}
{Jackson}, J.~M., {Finn}, S.~C., {Chambers}, E.~T., {Rathborne}, J.~M., \&
  {Simon}, R. 2010, \apjl, 719, L185

\bibitem[{{Jackson} {et~al.}(2008){Jackson}, {Finn}, {Rathborne}, {Chambers},
  \& {Simon}}]{jackson2008}
{Jackson}, J.~M., {Finn}, S.~C., {Rathborne}, J.~M., {Chambers}, E.~T., \&
  {Simon}, R. 2008, \apj, 680, 349

\bibitem[{{Jackson} {et~al.}(2013){Jackson}, {Rathborne}, {Foster}, {Whitaker},
  {Sanhueza}, {Claysmith}, {Mascoop}, {Wienen}, {Breen}, {Herpin},
  {Duarte-Cabral}, {Csengeri}, {Longmore}, {Contreras}, {Indermuehle},
  {Barnes}, {Walsh}, {Cunningham}, {Brooks}, {Britton}, {Voronkov}, {Urquhart},
  {Alves}, {Jordan}, {Hill}, {Hoq}, {Finn}, {Bains}, {Bontemps}, {Bronfman},
  {Caswell}, {Deharveng}, {Ellingsen}, {Fuller}, {Garay}, {Green}, {Hindson},
  {Jones}, {Lenfestey}, {Lo}, {Lowe}, {Mardones}, {Menten}, {Minier}, {Morgan},
  {Motte}, {Muller}, {Peretto}, {Purcell}, {Schilke}, {Bontemps}, {Schuller},
  {Titmarsh}, {Wyrowski}, \& {Zavagno}}]{jackson2013}
{Jackson}, J.~M., {Rathborne}, J.~M., {Foster}, J.~B., {et~al.} 2013, \pasa,
  30, 57

\bibitem[{{Jackson} {et~al.}(2006){Jackson}, {Rathborne}, {Shah}, {Simon},
  {Bania}, {Clemens}, {Chambers}, {Johnson}, {Dormody}, {Lavoie}, \&
  {Heyer}}]{jackson2006}
{Jackson}, J.~M., {Rathborne}, J.~M., {Shah}, R.~Y., {et~al.} 2006, \apjs, 163,
  145

\bibitem[{{Jansson} \& {Farrar}(2012)}]{jansson2012}
{Jansson}, R. \& {Farrar}, G.~R. 2012, \apj, 757, 14

\bibitem[{{Kauffmann} {et~al.}(2008){Kauffmann}, {Bertoldi}, {Bourke}, {Evans},
  \& {Lee}}]{2008A&A...487..993K}
{Kauffmann}, J., {Bertoldi}, F., {Bourke}, T.~L., {Evans}, II, N.~J., \& {Lee},
  C.~W. 2008, \aap, 487, 993

\bibitem[{{Kim} \& {Ostriker}(2002)}]{2002ApJ...570..132K}
{Kim}, W.-T. \& {Ostriker}, E.~C. 2002, \apj, 570, 132

\bibitem[{{Kim} \& {Ostriker}(2006)}]{2006ApJ...646..213K}
{Kim}, W.-T. \& {Ostriker}, E.~C. 2006, \apj, 646, 213

\bibitem[{{Leurini} {et~al.}(2011){Leurini}, {Pillai}, {Stanke}, {Wyrowski},
  {Testi}, {Schuller}, {Menten}, \& {Thorwirth}}]{leurini2011}
{Leurini}, S., {Pillai}, T., {Stanke}, T., {et~al.} 2011, \aap, 533, A85

\bibitem[{{Li} {et~al.}(2013){Li}, {Wyrowski}, {Menten}, \&
  {Belloche}}]{li2013}
{Li}, G.-X., {Wyrowski}, F., {Menten}, K., \& {Belloche}, A. 2013, \aap, 559,
  A34

\bibitem[{{Li} {et~al.}(2014){Li}, {Goodman}, {Sridharan}, {Houde}, {Li},
  {Novak}, \& {Tang}}]{li2014}
{Li}, H.-B., {Goodman}, A., {Sridharan}, T.~K., {et~al.} 2014, Protostars and
  Planets VI, 101

\bibitem[{{Li} \& {Henning}(2011)}]{li2011}
{Li}, H.-B. \& {Henning}, T. 2011, \nat, 479, 499

\bibitem[{{Li} {et~al.}(2015){Li}, {Yuen}, {Otto}, {Leung}, {Sridharan},
  {Zhang}, {Liu}, {Tang}, \& {Qiu}}]{li2015}
{Li}, H.-B., {Yuen}, K.~H., {Otto}, F., {et~al.} 2015, \nat, 520, 518

\bibitem[{{Lumsden} {et~al.}(2013){Lumsden}, {Hoare}, {Urquhart}, {Oudmaijer},
  {Davies}, {Mottram}, {Cooper}, \& {Moore}}]{lumsden2013}
{Lumsden}, S.~L., {Hoare}, M.~G., {Urquhart}, J.~S., {et~al.} 2013, \apjs, 208,
  11

\bibitem[{{Meidt} {et~al.}(2015){Meidt}, {Hughes}, {Dobbs}, {Pety}, {Thompson},
  {Garc{\'{\i}}a-Burillo}, {Leroy}, {Schinnerer}, {Colombo}, {Querejeta},
  {Kramer}, {Schuster}, \& {Dumas}}]{meidt2015}
{Meidt}, S.~E., {Hughes}, A., {Dobbs}, C.~L., {et~al.} 2015, \apj, 806, 72

\bibitem[{{Men'shchikov}(2013)}]{2013A&A...560A..63M}
{Men'shchikov}, A. 2013, \aap, 560, A63

\bibitem[{{Men'shchikov} {et~al.}(2010){Men'shchikov}, {Andr{\'e}}, {Didelon},
  {K{\"o}nyves}, {Schneider}, {Motte}, {Bontemps}, {Arzoumanian}, {Attard},
  {Abergel}, {Baluteau}, {Bernard}, {Cambr{\'e}sy}, {Cox}, {di Francesco}, {di
  Giorgio}, {Griffin}, {Hargrave}, {Huang}, {Kirk}, {Li}, {Martin}, {Minier},
  {Miville-Desch{\^e}nes}, {Molinari}, {Olofsson}, {Pezzuto}, {Roussel},
  {Russeil}, {Saraceno}, {Sauvage}, {Sibthorpe}, {Spinoglio}, {Testi},
  {Ward-Thompson}, {White}, {Wilson}, {Woodcraft}, \&
  {Zavagno}}]{2010A&A...518L.103M}
{Men'shchikov}, A., {Andr{\'e}}, P., {Didelon}, P., {et~al.} 2010, \aap, 518,
  L103

\bibitem[{{Minier} {et~al.}(2003){Minier}, {Ellingsen}, {Norris}, \&
  {Booth}}]{minier2003}
{Minier}, V., {Ellingsen}, S.~P., {Norris}, R.~P., \& {Booth}, R.~S. 2003,
  \aap, 403, 1095

\bibitem[{{Miville-Desch{\^e}nes} {et~al.}(2010){Miville-Desch{\^e}nes},
  {Martin}, {Abergel}, {Bernard}, {Boulanger}, {Lagache}, {Anderson},
  {Andr{\'e}}, {Arab}, {Baluteau}, {Blagrave}, {Bontemps}, {Cohen},
  {Compiegne}, {Cox}, {Dartois}, {Davis}, {Emery}, {Fulton}, {Gry}, {Habart},
  {Huang}, {Joblin}, {Jones}, {Kirk}, {Lim}, {Madden}, {Makiwa}, {Menshchikov},
  {Molinari}, {Moseley}, {Motte}, {Naylor}, {Okumura}, {Pinheiro Gon{\c
  c}alves}, {Polehampton}, {Rod{\'o}n}, {Russeil}, {Saraceno}, {Schneider},
  {Sidher}, {Spencer}, {Swinyard}, {Ward-Thompson}, {White}, \&
  {Zavagno}}]{miville2010}
{Miville-Desch{\^e}nes}, M.-A., {Martin}, P.~G., {Abergel}, A., {et~al.} 2010,
  \aap, 518, L104

\bibitem[{{Moeckel} \& {Burkert}(2015)}]{2015ApJ...807...67M}
{Moeckel}, N. \& {Burkert}, A. 2015, \apj, 807, 67

\bibitem[{{Molinari} {et~al.}(2011){Molinari}, {Schisano}, {Faustini},
  {Pestalozzi}, {di Giorgio}, \& {Liu}}]{molinari2011}
{Molinari}, S., {Schisano}, E., {Faustini}, F., {et~al.} 2011, \aap, 530, A133

\bibitem[{{Molinari} {et~al.}(2010){Molinari}, {Swinyard}, {Bally}, {Barlow},
  {Bernard}, {Martin}, {Moore}, {Noriega-Crespo}, {Plume}, {Testi}, {Zavagno},
  {Abergel}, {Ali}, {Andr{\'e}}, {Baluteau}, {Benedettini}, {Bern{\'e}},
  {Billot}, {Blommaert}, {Bontemps}, {Boulanger}, {Brand}, {Brunt}, {Burton},
  {Campeggio}, {Carey}, {Caselli}, {Cesaroni}, {Cernicharo}, {Chakrabarti},
  {Chrysostomou}, {Codella}, {Cohen}, {Compiegne}, {Davis}, {de Bernardis}, {de
  Gasperis}, {Di Francesco}, {di Giorgio}, {Elia}, {Faustini}, {Fischera},
  {Fukui}, {Fuller}, {Ganga}, {Garcia-Lario}, {Giard}, {Giardino}, {Glenn},
  {Goldsmith}, {Griffin}, {Hoare}, {Huang}, {Jiang}, {Joblin}, {Joncas},
  {Juvela}, {Kirk}, {Lagache}, {Li}, {Lim}, {Lord}, {Lucas}, {Maiolo},
  {Marengo}, {Marshall}, {Masi}, {Massi}, {Matsuura}, {Meny}, {Minier},
  {Miville-Desch{\^e}nes}, {Montier}, {Motte}, {M{\"u}ller}, {Natoli}, {Neves},
  {Olmi}, {Paladini}, {Paradis}, {Pestalozzi}, {Pezzuto}, {Piacentini},
  {Pomar{\`e}s}, {Popescu}, {Reach}, {Richer}, {Ristorcelli}, {Roy}, {Royer},
  {Russeil}, {Saraceno}, {Sauvage}, {Schilke}, {Schneider-Bontemps},
  {Schuller}, {Schultz}, {Shepherd}, {Sibthorpe}, {Smith}, {Smith},
  {Spinoglio}, {Stamatellos}, {Strafella}, {Stringfellow}, {Sturm}, {Taylor},
  {Thompson}, {Tuffs}, {Umana}, {Valenziano}, {Vavrek}, {Viti}, {Waelkens},
  {Ward-Thompson}, {White}, {Wyrowski}, {Yorke}, \& {Zhang}}]{molinari2010}
{Molinari}, S., {Swinyard}, B., {Bally}, J., {et~al.} 2010, \pasp, 122, 314

\bibitem[{{Murray} \& {Rahman}(2010)}]{murray2010}
{Murray}, N. \& {Rahman}, M. 2010, \apj, 709, 424

\bibitem[{{Ossenkopf} \& {Henning}(1994)}]{ossenkopf1994}
{Ossenkopf}, V. \& {Henning}, T. 1994, \aap, 291, 943

\bibitem[{{Padoan} {et~al.}(2001){Padoan}, {Juvela}, {Goodman}, \&
  {Nordlund}}]{2001ApJ...553..227P}
{Padoan}, P., {Juvela}, M., {Goodman}, A.~A., \& {Nordlund}, {\AA}. 2001, \apj,
  553, 227

\bibitem[{{Palmeirim} {et~al.}(2013){Palmeirim}, {Andr{\'e}}, {Kirk},
  {Ward-Thompson}, {Arzoumanian}, {K{\"o}nyves}, {Didelon}, {Schneider},
  {Benedettini}, {Bontemps}, {Di Francesco}, {Elia}, {Griffin}, {Hennemann},
  {Hill}, {Martin}, {Men'shchikov}, {Molinari}, {Motte}, {Nguyen Luong},
  {Nutter}, {Peretto}, {Pezzuto}, {Roy}, {Rygl}, {Spinoglio}, \&
  {White}}]{2013A&A...550A..38P}
{Palmeirim}, P., {Andr{\'e}}, P., {Kirk}, J., {et~al.} 2013, \aap, 550, A38

\bibitem[{{Peretto} \& {Fuller}(2009)}]{peretto2009}
{Peretto}, N. \& {Fuller}, G.~A. 2009, \aap, 505, 405

\bibitem[{{Planck Collaboration} {et~al.}(2016){Planck Collaboration}, {Ade},
  {Aghanim}, {Alves}, {Arnaud}, {Arzoumanian}, {Ashdown}, {Aumont},
  {Baccigalupi}, {Banday}, {Barreiro}, {Bartolo}, {Battaner}, {Benabed},
  {Beno{\^i}t}, {Benoit-L{\'e}vy}, {Bernard}, {Bersanelli}, {Bielewicz},
  {Bock}, {Bonavera}, {Bond}, {Borrill}, {Bouchet}, {Boulanger}, {Bracco},
  {Burigana}, {Calabrese}, {Cardoso}, {Catalano}, {Chiang}, {Christensen},
  {Colombo}, {Combet}, {Couchot}, {Crill}, {Curto}, {Cuttaia}, {Danese},
  {Davies}, {Davis}, {de Bernardis}, {de Rosa}, {de Zotti}, {Delabrouille},
  {Dickinson}, {Diego}, {Dole}, {Donzelli}, {Dor{\'e}}, {Douspis}, {Ducout},
  {Dupac}, {Efstathiou}, {Elsner}, {En{\ss}lin}, {Eriksen}, {Falceta-Gon{\c
  c}alves}, {Falgarone}, {Ferri{\`e}re}, {Finelli}, {Forni}, {Frailis},
  {Fraisse}, {Franceschi}, {Frejsel}, {Galeotta}, {Galli}, {Ganga}, {Ghosh},
  {Giard}, {Gjerl{\o}w}, {Gonz{\'a}lez-Nuevo}, {G{\'o}rski}, {Gregorio},
  {Gruppuso}, {Gudmundsson}, {Guillet}, {Harrison}, {Helou}, {Hennebelle},
  {Henrot-Versill{\'e}}, {Hern{\'a}ndez-Monteagudo}, {Herranz}, {Hildebrandt},
  {Hivon}, {Holmes}, {Hornstrup}, {Huffenberger}, {Hurier}, {Jaffe}, {Jaffe},
  {Jones}, {Juvela}, {Keih{\"a}nen}, {Keskitalo}, {Kisner}, {Knoche}, {Kunz},
  {Kurki-Suonio}, {Lagache}, {Lamarre}, {Lasenby}, {Lattanzi}, {Lawrence},
  {Leonardi}, {Levrier}, {Liguori}, {Lilje}, {Linden-V{\o}rnle},
  {L{\'o}pez-Caniego}, {Lubin}, {Mac{\'{\i}}as-P{\'e}rez}, {Maino},
  {Mandolesi}, {Mangilli}, {Maris}, {Martin}, {Mart{\'{\i}}nez-Gonz{\'a}lez},
  {Masi}, {Matarrese}, {Melchiorri}, {Mendes}, {Mennella}, {Migliaccio},
  {Miville-Desch{\^e}nes}, {Moneti}, {Montier}, {Morgante}, {Mortlock},
  {Munshi}, {Murphy}, {Naselsky}, {Nati}, {Netterfield}, {Noviello}, {Novikov},
  {Novikov}, {Oppermann}, {Oxborrow}, {Pagano}, {Pajot}, {Paladini},
  {Paoletti}, {Pasian}, {Perotto}, {Pettorino}, {Piacentini}, {Piat},
  {Pierpaoli}, {Pietrobon}, {Plaszczynski}, {Pointecouteau}, {Polenta},
  {Ponthieu}, {Pratt}, {Prunet}, {Puget}, {Rachen}, {Reinecke}, {Remazeilles},
  {Renault}, {Renzi}, {Ristorcelli}, {Rocha}, {Rossetti}, {Roudier},
  {Rubi{\~n}o-Mart{\'{\i}}n}, {Rusholme}, {Sandri}, {Santos}, {Savelainen},
  {Savini}, {Scott}, {Soler}, {Stolyarov}, {Sudiwala}, {Sutton}, {Suur-Uski},
  {Sygnet}, {Tauber}, {Terenzi}, {Toffolatti}, {Tomasi}, {Tristram}, {Tucci},
  {Umana}, {Valenziano}, {Valiviita}, {Van Tent}, {Vielva}, {Villa}, {Wade},
  {Wandelt}, {Wehus}, {Ysard}, {Yvon}, \& {Zonca}}]{2016A&A...586A.138P}
{Planck Collaboration}, {Ade}, P.~A.~R., {Aghanim}, N., {et~al.} 2016, \aap,
  586, A138

\bibitem[{{Press} {et~al.}(1992){Press}, {Teukolsky}, {Vetterling}, \&
  {Flannery}}]{press1992}
{Press}, W.~H., {Teukolsky}, S.~A., {Vetterling}, W.~T., \& {Flannery}, B.~P.
  1992, {Numerical recipes in FORTRAN. The art of scientific computing}

\bibitem[{{Purcell} {et~al.}(2013){Purcell}, {Hoare}, {Cotton}, {Lumsden},
  {Urquhart}, {Chandler}, {Churchwell}, {Diamond}, {Dougherty}, {Fender},
  {Fuller}, {Garrington}, {Gledhill}, {Goldsmith}, {Hindson}, {Jackson},
  {Kurtz}, {Mart{\'{\i}}}, {Moore}, {Mundy}, {Muxlow}, {Oudmaijer}, {Pandian},
  {Paredes}, {Shepherd}, {Smethurst}, {Spencer}, {Thompson}, {Umana}, \&
  {Zijlstra}}]{purcell2013}
{Purcell}, C.~R., {Hoare}, M.~G., {Cotton}, W.~D., {et~al.} 2013, \apjs, 205, 1

\bibitem[{{Purcell} {et~al.}(2012){Purcell}, {Longmore}, {Walsh}, {Whiting},
  {Breen}, {Britton}, {Brooks}, {Burton}, {Cunningham}, {Green},
  {Harvey-Smith}, {Hindson}, {Hoare}, {Indermuehle}, {Jones}, {Lo}, {Lowe},
  {Phillips}, {Thompson}, {Urquhart}, {Voronkov}, \& {White}}]{purcell2012}
{Purcell}, C.~R., {Longmore}, S.~N., {Walsh}, A.~J., {et~al.} 2012, \mnras,
  426, 1972

\bibitem[{{Ragan} {et~al.}(2014){Ragan}, {Henning}, {Tackenberg}, {Beuther},
  {Johnston}, {Kainulainen}, \& {Linz}}]{ragan2014}
{Ragan}, S.~E., {Henning}, T., {Tackenberg}, J., {et~al.} 2014, \aap, 568, A73

\bibitem[{{Reid} {et~al.}(2014){Reid}, {Menten}, {Brunthaler}, {Zheng}, {Dame},
  {Xu}, {Wu}, {Zhang}, {Sanna}, {Sato}, {Hachisuka}, {Choi}, {Immer},
  {Moscadelli}, {Rygl}, \& {Bartkiewicz}}]{reid2014}
{Reid}, M.~J., {Menten}, K.~M., {Brunthaler}, A., {et~al.} 2014, \apj, 783, 130

\bibitem[{{Rigby} {et~al.}(2016){Rigby}, {Moore}, {Plume}, {Eden}, {Urquhart},
  {Thompson}, {Mottram}, {Brunt}, {Butner}, {Dempsey}, {Gibson}, {Hatchell},
  {Jenness}, {Kuno}, {Longmore}, {Morgan}, {Polychroni}, {Thomas}, {White}, \&
  {Zhu}}]{2016MNRAS.456.2885R}
{Rigby}, A.~J., {Moore}, T.~J.~T., {Plume}, R., {et~al.} 2016, \mnras, 456,
  2885

\bibitem[{{Roman-Duval} {et~al.}(2009){Roman-Duval}, {Jackson}, {Heyer},
  {Johnson}, {Rathborne}, {Shah}, \& {Simon}}]{roman2009}
{Roman-Duval}, J., {Jackson}, J.~M., {Heyer}, M., {et~al.} 2009, \apj, 699,
  1153

\bibitem[{{Schinnerer} {et~al.}(2013){Schinnerer}, {Meidt}, {Pety}, {Hughes},
  {Colombo}, {Garc{\'{\i}}a-Burillo}, {Schuster}, {Dumas}, {Dobbs}, {Leroy},
  {Kramer}, {Thompson}, \& {Regan}}]{schinnerer2013}
{Schinnerer}, E., {Meidt}, S.~E., {Pety}, J., {et~al.} 2013, \apj, 779, 42

\bibitem[{{Schisano} {et~al.}(2014){Schisano}, {Rygl}, {Molinari}, {Busquet},
  {Elia}, {Pestalozzi}, {Polychroni}, {Billot}, {Carey}, {Paladini},
  {Noriega-Crespo}, {Moore}, {Plume}, {Glover}, \&
  {V{\'a}zquez-Semadeni}}]{schisano2014}
{Schisano}, E., {Rygl}, K.~L.~J., {Molinari}, S., {et~al.} 2014, \apj, 791, 27

\bibitem[{{Schlingman} {et~al.}(2011){Schlingman}, {Shirley}, {Schenk},
  {Rosolowsky}, {Bally}, {Battersby}, {Dunham}, {Ellsworth-Bowers}, {Evans},
  {Ginsburg}, \& {Stringfellow}}]{schlingman2011}
{Schlingman}, W.~M., {Shirley}, Y.~L., {Schenk}, D.~E., {et~al.} 2011, \apjs,
  195, 14

\bibitem[{{Schneider} \& {Elmegreen}(1979)}]{1979ApJS...41...87S}
{Schneider}, S. \& {Elmegreen}, B.~G. 1979, \apjs, 41, 87

\bibitem[{{Schuller} {et~al.}(2009){Schuller}, {Menten}, {Contreras},
  {Wyrowski}, {Schilke}, {Bronfman}, {Henning}, {Walmsley}, {Beuther},
  {Bontemps}, {Cesaroni}, {Deharveng}, {Garay}, {Herpin}, {Lefloch}, {Linz},
  {Mardones}, {Minier}, {Molinari}, {Motte}, {Nyman}, {Reveret}, {Risacher},
  {Russeil}, {Schneider}, {Testi}, {Troost}, {Vasyunina}, {Wienen}, {Zavagno},
  {Kovacs}, {Kreysa}, {Siringo}, \& {Wei{\ss}}}]{schuller2009_full}
{Schuller}, F., {Menten}, K.~M., {Contreras}, Y., {et~al.} 2009, \aap, 504, 415

\bibitem[{{Seigar}(2005)}]{2005MNRAS.361L..20S}
{Seigar}, M.~S. 2005, \mnras, 361, L20

\bibitem[{{Shirley} {et~al.}(2013){Shirley}, {Ellsworth-Bowers}, {Svoboda},
  {Schlingman}, {Ginsburg}, {Rosolowsky}, {Gerner}, {Mairs}, {Battersby},
  {Stringfellow}, {Dunham}, {Glenn}, \& {Bally}}]{shirley2013}
{Shirley}, Y.~L., {Ellsworth-Bowers}, T.~P., {Svoboda}, B., {et~al.} 2013,
  \apjs, 209, 2

\bibitem[{{Smith} {et~al.}(2014{\natexlab{a}}){Smith}, {Glover}, {Clark},
  {Klessen}, \& {Springel}}]{smith2014a}
{Smith}, R.~J., {Glover}, S.~C.~O., {Clark}, P.~C., {Klessen}, R.~S., \&
  {Springel}, V. 2014{\natexlab{a}}, \mnras, 441, 1628

\bibitem[{{Smith} {et~al.}(2014{\natexlab{b}}){Smith}, {Glover}, \&
  {Klessen}}]{smith2014b}
{Smith}, R.~J., {Glover}, S.~C.~O., \& {Klessen}, R.~S. 2014{\natexlab{b}},
  \mnras, 445, 2900

\bibitem[{{Sousbie}(2011)}]{sousbie2011}
{Sousbie}, T. 2011, \mnras, 414, 350

\bibitem[{{Stark} \& {Lee}(2006)}]{stark2006}
{Stark}, A.~A. \& {Lee}, Y. 2006, \apjl, 641, L113

\bibitem[{{Tackenberg} {et~al.}(2014){Tackenberg}, {Beuther}, {Henning},
  {Linz}, {Sakai}, {Ragan}, {Krause}, {Nielbock}, {Hennemann}, {Pitann}, \&
  {Schmiedeke}}]{tackenberg2014}
{Tackenberg}, J., {Beuther}, H., {Henning}, T., {et~al.} 2014, \aap, 565, A101

\bibitem[{{Taylor} \& {Cordes}(1993)}]{tayor1993}
{Taylor}, J.~H. \& {Cordes}, J.~M. 1993, \apj, 411, 674

\bibitem[{{Ungerechts} \& {Thaddeus}(1987)}]{1987ApJS...63..645U}
{Ungerechts}, H. \& {Thaddeus}, P. 1987, \apjs, 63, 645

\bibitem[{{Urquhart} {et~al.}(2007){Urquhart}, {Busfield}, {Hoare}, {Lumsden},
  {Oudmaijer}, {Moore}, {Gibb}, {Purcell}, {Burton}, \&
  {Marechal}}]{urquhart_13co_south}
{Urquhart}, J.~S., {Busfield}, A.~L., {Hoare}, M.~G., {et~al.} 2007, \aap, 474,
  891

\bibitem[{{Urquhart} {et~al.}(2008){Urquhart}, {Busfield}, {Hoare}, {Lumsden},
  {Oudmaijer}, {Moore}, {Gibb}, {Purcell}, {Burton}, {Mar{\'e}chal}, {Jiang},
  \& {Wang}}]{urquhart_13co_north}
{Urquhart}, J.~S., {Busfield}, A.~L., {Hoare}, M.~G., {et~al.} 2008, \aap, 487,
  253

\bibitem[{{Urquhart} {et~al.}(2014{\natexlab{a}}){Urquhart}, {Csengeri},
  {Wyrowski}, {Schuller}, {Bontemps}, {Bronfman}, {Menten}, {Walmsley},
  {Contreras}, {Beuther}, {Wienen}, \& {Linz}}]{urquhart2014c}
{Urquhart}, J.~S., {Csengeri}, T., {Wyrowski}, F., {et~al.} 2014{\natexlab{a}},
  \aap, 568, A41

\bibitem[{{Urquhart} {et~al.}(2015){Urquhart}, {Figura}, {Moore}, {Csengeri},
  {Lumsden}, {Pillai}, {Thompson}, {Eden}, \& {Morgan}}]{urquhart2015b}
{Urquhart}, J.~S., {Figura}, C.~C., {Moore}, T.~J.~T., {et~al.} 2015, \mnras,
  452, 4029

\bibitem[{{Urquhart} {et~al.}(2014{\natexlab{b}}){Urquhart}, {Figura}, {Moore},
  {Hoare}, {Lumsden}, {Mottram}, {Thompson}, \& {Oudmaijer}}]{urquhart2014a}
{Urquhart}, J.~S., {Figura}, C.~C., {Moore}, T.~J.~T., {et~al.}
  2014{\natexlab{b}}, \mnras, 437, 1791

\bibitem[{{Urquhart} {et~al.}(2014{\natexlab{c}}){Urquhart}, {Moore},
  {Csengeri}, {Wyrowski}, {Schuller}, {Hoare}, {Lumsden}, {Mottram},
  {Thompson}, {Menten}, {Walmsley}, {Bronfman}, {Pfalzner}, {K{\"o}nig}, \&
  {Wienen}}]{urquhart2014b}
{Urquhart}, J.~S., {Moore}, T.~J.~T., {Csengeri}, T., {et~al.}
  2014{\natexlab{c}}, \mnras, 443, 1555

\bibitem[{{Urquhart} {et~al.}(2013{\natexlab{a}}){Urquhart}, {Moore},
  {Schuller}, {Wyrowski}, {Menten}, {Thompson}, {Csengeri}, {Walmsley},
  {Bronfman}, \& {K{\"o}nig}}]{urquhart2013a}
{Urquhart}, J.~S., {Moore}, T.~J.~T., {Schuller}, F., {et~al.}
  2013{\natexlab{a}}, \mnras, 431, 1752

\bibitem[{{Urquhart} {et~al.}(2011){Urquhart}, {Morgan}, {Figura}, {Moore},
  {Lumsden}, {Hoare}, {Oudmaijer}, {Mottram}, {Davies}, \&
  {Dunham}}]{urquhart2011}
{Urquhart}, J.~S., {Morgan}, L.~K., {Figura}, C.~C., {et~al.} 2011, \mnras,
  418, 1689

\bibitem[{{Urquhart} {et~al.}(2013{\natexlab{b}}){Urquhart}, {Thompson},
  {Moore}, {Purcell}, {Hoare}, {Schuller}, {Wyrowski}, {Csengeri}, {Menten},
  {Lumsden}, {Kurtz}, {Walmsley}, {Bronfman}, {Morgan}, {Eden}, \&
  {Russeil}}]{urquhart2013b}
{Urquhart}, J.~S., {Thompson}, M.~A., {Moore}, T.~J.~T., {et~al.}
  2013{\natexlab{b}}, \mnras, 435, 400

\bibitem[{{V{\'a}zquez-Semadeni} {et~al.}(2011){V{\'a}zquez-Semadeni},
  {Banerjee}, {G{\'o}mez}, {Hennebelle}, {Duffin}, \& {Klessen}}]{vazquez2011}
{V{\'a}zquez-Semadeni}, E., {Banerjee}, R., {G{\'o}mez}, G.~C., {et~al.} 2011,
  \mnras, 414, 2511

\bibitem[{{Walsh} {et~al.}(1998){Walsh}, {Burton}, {Hyland}, \&
  {Robinson}}]{walsh1998}
{Walsh}, A.~J., {Burton}, M.~G., {Hyland}, A.~R., \& {Robinson}, G. 1998,
  \mnras, 301, 640

\bibitem[{{Wang} {et~al.}(2015){Wang}, {Testi}, {Ginsburg}, {Walmsley},
  {Molinari}, \& {Schisano}}]{wang2015}
{Wang}, K., {Testi}, L., {Ginsburg}, A., {et~al.} 2015, \mnras, 450, 4043

\bibitem[{{Wang} {et~al.}(2014){Wang}, {Zhang}, {Testi}, {van der Tak}, {Wu},
  {Zhang}, {Pillai}, {Wyrowski}, {Carey}, {Ragan}, \& {Henning}}]{wang2014}
{Wang}, K., {Zhang}, Q., {Testi}, L., {et~al.} 2014, \mnras, 439, 3275

\bibitem[{{Weidner} {et~al.}(2010){Weidner}, {Bonnell}, \&
  {Zinnecker}}]{2010ApJ...724.1503W}
{Weidner}, C., {Bonnell}, I.~A., \& {Zinnecker}, H. 2010, \apj, 724, 1503

\bibitem[{{Wienen} {et~al.}(2015){Wienen}, {Wyrowski}, {Menten}, {Urquhart},
  {Csengeri}, {Walmsley}, {Bontemps}, {Russeil}, {Bronfman}, {Koribalski}, \&
  {Schuller}}]{wienen2015a}
{Wienen}, M., {Wyrowski}, F., {Menten}, K.~M., {et~al.} 2015, \aap, 579, A91

\bibitem[{{Wienen} {et~al.}(2012){Wienen}, {Wyrowski}, {Schuller}, {Menten},
  {Walmsley}, {Bronfman}, \& {Motte}}]{wienen2012}
{Wienen}, M., {Wyrowski}, F., {Schuller}, F., {et~al.} 2012, \aap, 544, A146

\bibitem[{{Williams} {et~al.}(2000){Williams}, {Blitz}, \&
  {McKee}}]{williams2000}
{Williams}, J.~P., {Blitz}, L., \& {McKee}, C.~F. 2000, Protostars and Planets
  IV, 97

\bibitem[{{Zucker} {et~al.}(2015){Zucker}, {Battersby}, \&
  {Goodman}}]{zucker2015}
{Zucker}, C., {Battersby}, C., \& {Goodman}, A. 2015, \apj, 815, 23

\end{thebibliography}

\end{document}